\documentclass[english,preprintnumbers,nofootinbib,twocolumn,superscriptaddress]{revtex4-2}

\usepackage{graphicx}
\usepackage{scrextend}
\usepackage{mathtools}
\usepackage{amssymb}
\usepackage{bbm}
\usepackage{accents}
\usepackage{verbatim}
\usepackage{hyperref}
\usepackage{xcolor}
\usepackage{stmaryrd}
\usepackage{tabulary}
\usepackage[outline]{contour}

\hypersetup{
    colorlinks,
    linkcolor={red!50!black},
    citecolor={blue!50!black},
    urlcolor={blue!80!black}
}
\usepackage[caption=false]{subfig}

\newcommand{\om} {\omega}
\newcommand{\ob} {\bar{\omega}}
\newcommand{\T} {{\scriptscriptstyle T}}

\newcommand{\V}{\mathcal{V}}
\newcommand{\Vt}{\tilde{\mathcal{V}}}
\newcommand{\Vb}{\breve{\mathcal{V}}}

\newcommand{\cB} {\boldsymbol c}
\newcommand{\ct} {\tilde{c}}
\newcommand{\cb} {\breve{c}}
\newcommand{\C}{\mathcal{C}}
\newcommand{\Ct}{\tilde{\mathcal{C}}}
\newcommand{\Cb}{\breve{\mathcal{C}}}

\newcommand{\setn} {{\mathfrak n}}
\newcommand{\setnt} {{\tilde{\mathfrak n}}}
\newcommand{\setnb} {{\breve{\mathfrak n}}}

\newcommand{\phia} {\phi_{\mathrm a}}
\newcommand{\etaa} {\eta_{\mathrm a}}
\newcommand{\varphia} {\varphi_{\mathrm a}}
\newcommand{\phib} {\phi_{\mathrm b}}
\newcommand{\etab} {\eta_{\mathrm b}}
\newcommand{\varphib} {\varphi_{\mathrm b}}
\newcommand{\phic} {\phi_{\mathrm c}}

\newcommand {\phio}{\phi^{\circ}}

\newcommand {\rhoo}{\rho^{\circ}}
\newcommand {\etao}{\eta^{\circ}}
\newcommand {\varrhoo}{\varrho^{\circ}}

\newcommand {\varphio}{\varphi^{\circ}}
\newcommand {\varphiot}{\varphi^{\tilde{\circ}}}
\newcommand {\varphiob}{\varphi^{\breve{\circ}}}
\newcommand {\varphiao}{\varphi_{\mathrm a}^{\circ}}
\newcommand {\varphibo}{\varphi_{\mathrm b}^{\circ}}
\newcommand {\phico}{\phi_{\mathrm c}^{\circ}}

\newcommand{\Gt} {\tilde{G}}
\newcommand{\Ga} {\acute{G}}
\newcommand{\Ht} {\tilde{H}}
\newcommand{\Hb} {\breve{H}}
\newcommand{\Ha} {\acute{H}}

\newcommand{\Sa} {\acute{S}}

\newcommand{\Tt} {\tilde{T}}
\newcommand{\Tb} {\breve{T}}
\newcommand{\tht} {\theta}

\newcommand{\matt}{\mathbf{t}}

\newcommand{\mats}{\mathbf{s}}
\newcommand{\matu}{\mathbf{u}}
\newcommand{{\matv}}{\mathbf{x}}
\newcommand{\matw}{\boldsymbol{\omega}}
\newcommand{\matwr}{\boldsymbol{\omega}_{r}}

\newcommand{\sigy}{\boldsymbol{\sigma}_{\!y}}
\newcommand{\sigz}{\boldsymbol{\sigma}_{\!z}}

\newcommand{\gent}{\mathbf{T}}
\newcommand{\genu}{\mathbf{U}}
\newcommand{\genz}{\mathbf{Z}_2}
\newcommand{\genxs}{\mathbf{X}_{\mathrm s}}
\newcommand{\genxa}{\mathbf{X}_\om}
\newcommand{\genxb}{\mathbf{X}'_\om}
\newcommand{\genxd}{\mathbf{X}_2}
\newcommand{\genxc}{\mathbf{X}^c_2}

\begin{document}
	
\title{Homogeneous linear intrinsic constraints in the \\stationary manifold of a $G$-invariant potential}
\author{R.~Krishnan}
\email{krisphysics@gmail.com}
\homepage{https://orcid.org/0000-0002-0707-3267}
\affiliation{Saha Institute of Nuclear Physics, 1/AF Bidhannagar, Kolkata 700064, India}  
\begin{abstract}
Given a $G$-invariant potential $\mathcal{V}$ of a scalar multiplet $\varphi$, there may exist a set of homogenous linear equations that constrain the components of a stationary point of $\mathcal{V}$ independently of the coefficients of the terms in $\mathcal{V}$. We call them homogeneous linear intrinsic constraints (HLICs). HLICs in a stationary point manifest as HLICs in the corresponding vacuum alignment of $\varphi$, which plays a central role in predictive phenomenological models. We discover that a group $\tilde{H}$ generates HLICs if the terms in $\mathcal{V}$ satisfy a condition, which we call the compatibility condition. In this paper, we also develop a procedure, which involves splitting $\mathcal{V}$ into smaller parts, to establish the existence of specific stationary points using arguments based on symmetries without the need for explicitly extremizing the potential. Using this procedure, we obtain $\tilde{H}$ as a direct product of the symmetry groups associated with the various irreducible multiplets (irreps) in $\varphi$. This results from considering the potentials of the irreps separately and verifying if the cross terms are compatible with $\tilde{H}$.\footnote{A video presentation of this paper is available \href{https://www.youtube.com/playlist?list=PLjRYJtC1E1HfEbur89bH3CT5BvMtO4FWK}{here}.}
\end{abstract}
\maketitle

A scalar field transforming under a symmetry group, and a minimum energy configuration of the field spontaneously breaking the symmetry is a recurring theme in a wide range of scenarios in physics. Examples include the Higgs, flavons and axions in particle physics, scalar dark matter fields\cite{1007.0871, 1008.4777} and inflatons\cite{hep-ph/0001333, 0805.0325, 1407.6017, 1412.8537, 1808.09601, 2108.05400} in cosmology and various fields (order parameters)\cite{GAETA2006322, MICHEL200111} in condensed matter physics. The minimum energy configuration depends not only on the symmetry group, but also on the coefficients of the invariant terms in the potential. The predictions made by a model that are claimed to be derived from the symmetries should be independent of any general change in these coefficients. In this paper, we study the coefficient-independent properties of the stationary points of the potential, which would eventually manifest as the coefficient-independent predictions of the model. The mathematical results and techniques developed here can be applied to various physical systems, in particular, to flavour models in particle physics.

Finite groups have been used extensively\cite{1510.02091, 1711.10806, 1110.6376} in building models of neutrino masses and mixing. It was observed that these models require the flavour symmetry group to be broken differently in the charged-lepton and neutrino sectors. For example, we can construct a flavour model\cite{hep-ph/0504165} resulting in tribimaximal mixing (TBM)\cite{hep-ph/0202074} using two triplets of $A_4$, say, $\phi_c$ and $\phi_\nu$ with vacuum expectation values (vevs) $\langle\phi_c\rangle\propto(1,1,1)$ and $\langle\phi_\nu\rangle\propto(1,0,0)$ which break $A_4$ to its subgroups $Z_3$ and $Z_2$, respectively. Here, $Z_3$ is the stabilizer (also known as residual symmetry group, little group or isotropy subgroup) of $\langle\phi_c\rangle$ and $Z_2$ is the stabilizer of $\langle\phi_\nu\rangle$ under $A_4$. It was shown\cite{hep-ph/0504165} that we obtain the two stabilizers and the resulting four constraints in the vevs, i.e., $\langle\phi_c\rangle_1=\langle\phi_c\rangle_2$, $\langle\phi_c\rangle_1=\langle\phi_c\rangle_3$, $\langle\phi_\nu\rangle_2=0$ and $\langle\phi_\nu\rangle_3=0$, only if the potentials of $\phi_c$ and $\phi_\nu$ are considered separately. The cross terms that contain both $\phi_c$ and $\phi_\nu$ disturb the vevs away from $\langle\phi_c\rangle\propto(1,1,1)$ and $\langle\phi_\nu\rangle\propto(1,0,0)$ spoiling the four constraints. Ref.\cite{hep-ph/0504165} as well as several others\cite{hep-ph/0512103, hep-ph/0601001, 1004.0321, 1205.3617} proposed additional mechanisms (e.g.,~introducing driving fields in a supersymmetric framework, introducing extra dimensions and branes) that forbid the cross terms to save the day. Such mechanisms essentially force the coefficients of the cross terms to vanish.

To construct beyond-TBM models that generate non-zero reactor angle as a leading order prediction, the constraints in the vevs need to be more complicated (than the four constraints we mentioned earlier, for example). This can be achieved by utilizing irreducible multiplets (irreps) of groups with a larger set of symmetry transformations like $\Delta(6n^2)$\cite{0809.0639, 1112.1340, 1201.3279, 1305.3200, 1403.4395, 1411.5845, 2011.05693} and obtaining constraints in the vevs of the irreps as a consequence of the respective stabilizers. More predictive beyond-TBM models\cite{hep-ph/0506297, 0912.1344, 1801.10197} (sometimes called indirect models\cite{1510.02091}) can be constructed if the irreps are assigned vevs having constraints that are not derivable from their respective stabilizers.  We require additional mechanisms to forbid dangerous cross terms among the various irreps in all these models, in general, and also to justify the constraints in the vevs of the irreps in indirect models, in particular. In other words, almost all models\footnote{To the author's knowledge, the only exceptions are \cite{1111.1730, 1211.5143}.} assume mechanisms to fine-tune the coefficients in the potential. 

The author aims to avoid these mechanisms and construct flavour models that are minimal extensions of the standard model implemented using scalar fields and flavour symmetries only and that lead to coefficient-independent predictions. The first step towards this goal is to understand how symmetries generate coefficient-independent constraints in the stationary points (minima) of the potential, which forms the content of this paper. This work builds on the results obtained by the author in \cite{2011.11653}. The second step is to introduce such symmetries using group extensions, which we call the framework of the auxiliary group. In a companion paper\cite{2309.11542}, the author develops this second step and constructs a beyond-TBM indirect model in a coefficient-independent way. Ref.\cite{2309.11542} builds on the research involving group extensions carried out in \cite{1006.0203, 1111.1730, 2011.11653, 1211.5143, 1901.01205, 1912.02451}.

In Section~\ref{sec:one}. we formally define a stationary manifold as a function of the coefficients in the potential $\V$. In Section~\ref{sec:two}, we split the potential into two parts: $\Vt$ and the rest of the terms. We define $\Gt$ as the symmetry group of $\Vt$, and $\Ht$ as pointwise stabilizer of the stationary manifold under $\Gt$. We introduce the compatibility condition and state that if the terms in $\V$ are compatible with $\Ht$, then $\Ht$ remains unchanged under any general change in the coefficients in $\V$. The proof of this statement is provided in Appendix~\ref{sec:appone}. Since $\Ht$ is coefficient-independent, it generates homogeneous linear intrinsic constraints (HLICs). In Section~\ref{sec:three}, we develop a step-by-step procedure in which we split $\V$ into smaller parts and obtain its stationary manifold by combining the stationary manifolds its parts. In this procedure, we obtain the various stationary manifolds by utilizing symmetry arguments based on the compatibility condition discovered in this paper and the theorems by Michel, Golubitsky and Stewart\cite{GAETA2006322, MICHEL1971, MICHEL200111, GOLUBITSKY1988}. The procedure involves the non-bifurcation scenario (Section~\ref{subsec:one}) where HLICs are conserved as well as the bifurcation scenario (Section~\ref{subsec:two}) where HLICs are generated or destroyed. In Appendix~\ref{sec:apptwo}, we provide a number of examples that give the reader a clearer view of these theoretical results. These examples also explain the results from \cite{GAETA2006322, MICHEL1971, MICHEL200111, GOLUBITSKY1988} that are prerequisites for Section~\ref{sec:three}. We recommend that the reader go through these examples along with Section~\ref{sec:three}. We also provide an example that demonstrates how HLICs are generated in 
the framework of the auxiliary group.

\subsection{Stationary manifolds of the $G$-invariant renormalizable potential}
\label{sec:one}
Consider a scalar field $\varphi$ transforming as a real, complex or quaternionic representation of a compact group $\mathbb{G}$. Compact groups consist of compact Lie groups as well as finite groups. We denote the number of real degrees of freedom (the real dimension) of $\varphi$ by $\text{dim}(\varphi)$. The real components of $\varphi$ can be denoted by $\varphi_i$ with $i=1,..,\text{dim}(\varphi)$. The invariants of the group action constructed using $\varphi$ constitute the terms in the potential. The invariant terms of mass dimension up to four are renormalizable. Let $\setn$ be the set of integers $\{1,..,|\setn|\}$ where $|\setn|$ denotes the cardinality of $\setn$. Let the number of linearly independent real invariants up to order four be $|\setn|$. The general renormalizable potential is given by
\begin{equation}\label{eq:generalpot}
\V =  \sum_{\alpha\in\setn} c_\alpha \mathcal{I}_\alpha(\varphi),
\end{equation}
where $\mathcal{I}_\alpha(\varphi)$ are the real invariants constructed using $\varphi$, and $c_\alpha$ are the corresponding real coefficients, with $\alpha$ ranging over the set of integers $\setn$. Let us consider the largest compact group under which $\varphi$ transforms as a real orthogonal matrix representation such that the potential $\V$ remains invariant. We call this group $ G$,
\begin{equation}
 G=\{g:g\in O(\text{dim}(\varphi)),\, \mathcal{I}_\alpha(g\varphi)=\mathcal{I}_\alpha(\varphi)\,\forall\,\alpha\in\setn\}.
\end{equation}
In general, this group can be larger than $\mathbb{G}$. The larger set of group transformations present in $ G$ compared to $\mathbb{G}$ are sometimes referred to as accidental symmetries. For examples with accidental symmetries, see Examples~\ref{sec:egseven}, \ref{sec:egeight}.

In the space of $\varphi$, i.e., in $\mathbbm R^{\text{dim}(\varphi)}$, let us choose a specific point, say, $\varphi=\boldsymbol\varphi$. The action of the group $G$ on $\boldsymbol\varphi$ generates a set of points which forms the orbit of $\boldsymbol\varphi$ under $G$, denoted by $G\boldsymbol\varphi$. This $G$-orbit, in general, is a union of one or more connected manifolds. One of the connected components of $G\boldsymbol\varphi$ contains the original point $\boldsymbol\varphi$. Let $S\subseteq G$ be the setwise stabilizer of this connected component. The action of elements in $S$ moves a point in this connected manifold within the manifold transitively, while the action of elements in $G$ that are not in $S$ moves the point to another connected manifold. The connected manifold that contains $\boldsymbol\varphi$ is nothing but the $S$-orbit of $\boldsymbol\varphi$, denoted by $S\boldsymbol\varphi$. Let the real dimension of $S\boldsymbol\varphi$ be $\text{dim}(S\boldsymbol\varphi)$. We may represent the continuous transformations under $S$ with the parameters $\tht={\tht}_1,..,{\tht}_{\text{dim}(S\boldsymbol\varphi)}$. These parameters can uniquely label the points in the connected manifold $S\boldsymbol\varphi$. We note that $S$ and $\text{dim}(S\boldsymbol\varphi)$ are dependent on the choice of the point $\boldsymbol\varphi$.

Let us use $c$ to denote the coefficients in the potential (\ref{eq:generalpot}), i.e., $c=c_1,..,c_{|\setn|}$. A stationary point is a point where the first derivatives of the potential vanish. Corresponding to a set of specific values for the coefficients given by $c_\alpha=\cB_\alpha\,\forall\alpha\in\setn$, i.e., $c=\cB$, let us assume\footnote{This assumption, in general, cannot be made for any arbitrary value of $\boldsymbol\varphi$. However, some specific values of $\boldsymbol\varphi$ are guaranteed to be stationary points, no matter what the values of the coefficients $c_\alpha$ are. We will study this in detail in Section~\ref{sec:three} and Appendix~\ref{sec:apptwo}.} that the potential has a stationary point at $\varphi=\boldsymbol\varphi$, i.e.,
\begin{equation}\label{eq:mapfirst}
\sum_{\alpha\in\setn}\cB_\alpha\,\partial_i \mathcal{I}_\alpha(\varphi)\big|_{\varphi=\boldsymbol\varphi} =0,
\end{equation}
where $\partial_i$ is the derivative with respect to the component $\varphi_i$. Since $G$ is the symmetry group of the potential, we obtain a $G$-orbit of stationary points $G\boldsymbol\varphi$. We call $G\boldsymbol\varphi$ `the stationary orbit corresponding to the stationary point $\boldsymbol\varphi$'. The connected part of the stationary orbit $G\boldsymbol\varphi$ that contains $\boldsymbol\varphi$ is given by the manifold $S\boldsymbol\varphi$. We call $S\boldsymbol\varphi$ `the stationary manifold corresponding to the stationary point $\boldsymbol\varphi$'.

Consider the Hessian of the potential at $\boldsymbol\varphi$,
\begin{equation}
\mathcal{H}_{ij}= \sum_{\alpha\in\setn} c_\alpha\,\partial_i \partial_j \mathcal{I}_\alpha(\varphi)\big|_{\varphi=\boldsymbol\varphi}.
\end{equation}
We may block-diagonalize $\mathcal{H}$ using an orthogonal change of basis $\varphi'= R \varphi$,
\begin{equation}\label{eq:local}
\mathcal{H}'= R\,\mathcal{H}\, R^\T = \left(\begin{matrix} 0 & 0\\
0 & \mathcal{H}'_\perp
\end{matrix}\right),
\end{equation}
where the top left block is a square matrix of dimension $\text{dim}(S\boldsymbol\varphi)$, and it involves the directions tangential to the stationary manifold $S\boldsymbol\varphi$. It vanishes because the value of the potential remains unchanged along the tangential directions. $\mathcal{H}'_\perp$ is a square matrix of dimension $(\text{dim}(\varphi)-\text{dim}(S\boldsymbol\varphi))$. We assume that $\mathcal{H}'_\perp$ is non-singular; in other words, $\V$ is a Morse-Bott function\cite{Bott, MorseBott} at $c=\cB$. 

\begin{figure}[]
\subfloat[\label{fig:onea}]{
  \includegraphics[width=0.5\columnwidth]{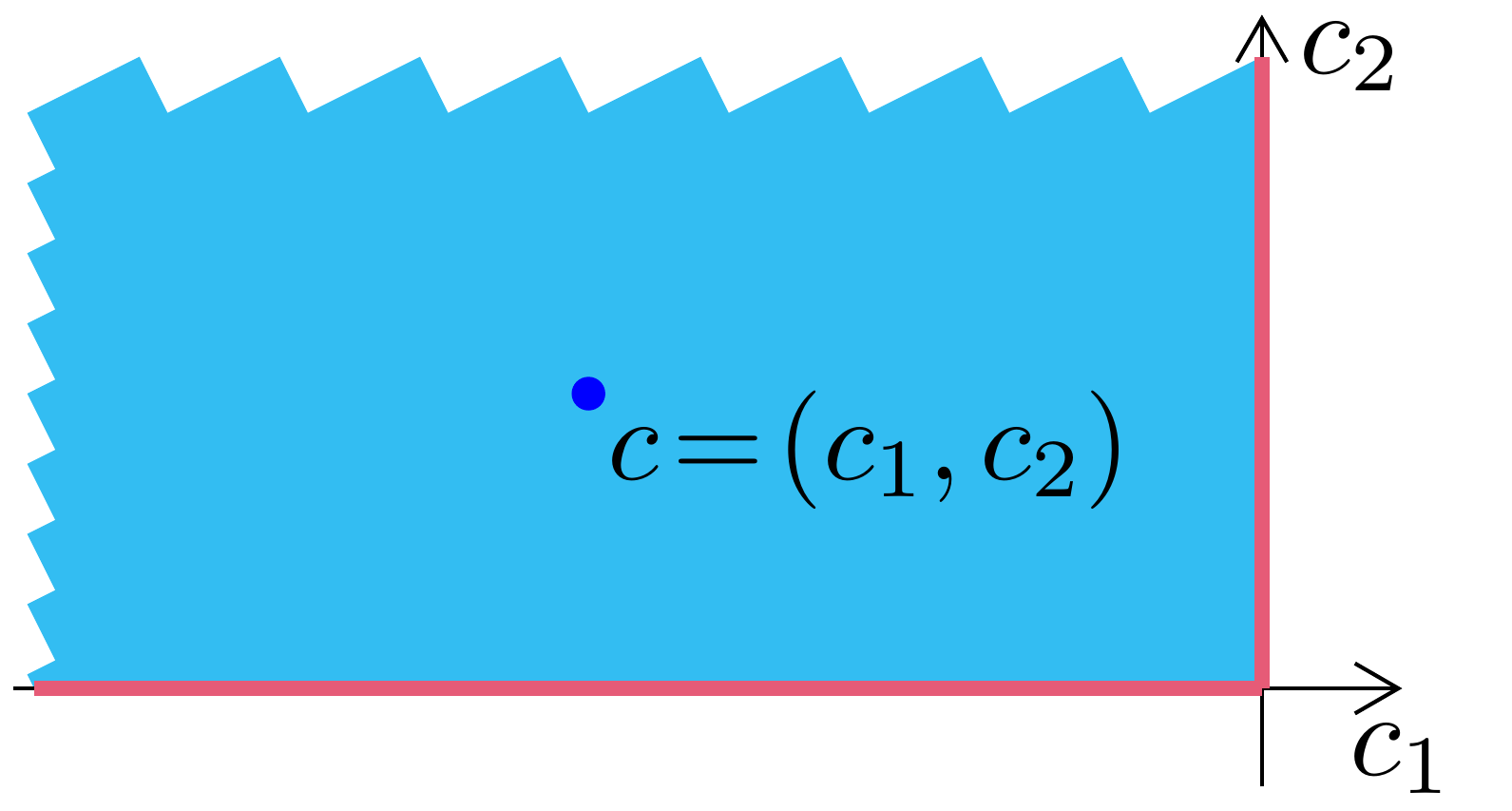}
}
\subfloat[\label{fig:oneb}]{
  \includegraphics[width=0.48\columnwidth]{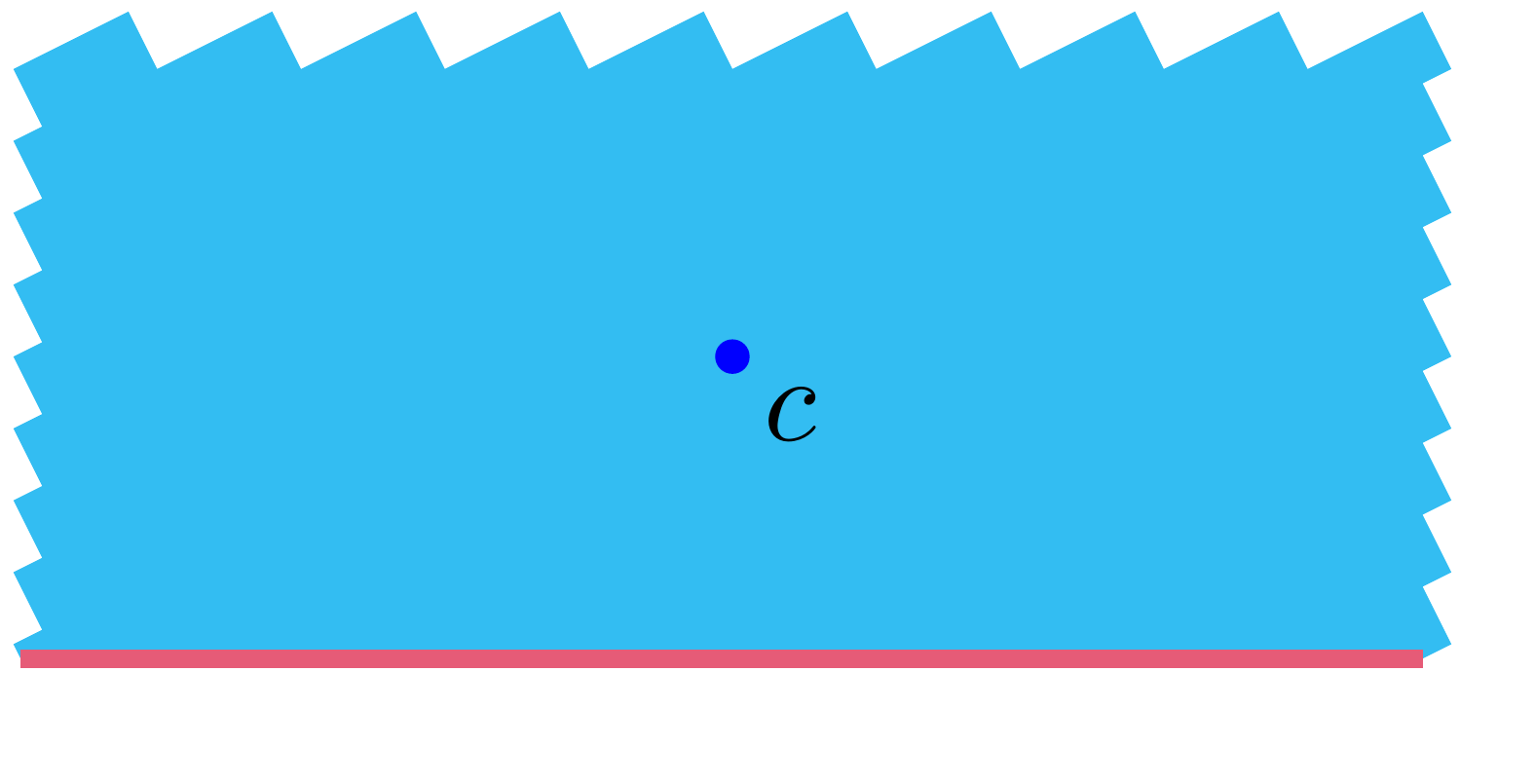}
}
\caption{(a) The coefficient space of the Higgs potential $\V=c_1 m^2 \varphi^\dagger \varphi+c_2 (\varphi^\dagger \varphi)^2$. In the blue region, the potential has the mexican-hat-like structure, which corresponds to degenerate minima forming a 3-sphere as the stationary manifold, say, $\varphio(c)$. Here, the Hessian has three vanishing eigenvalues. As $c$ approaches the $c_1$-axis, the 3-sphere $\varphio(c)$ approaches infinity, and as $c$ approaches the $c_2$-axis, $\varphio(c)$ approaches zero. The red lines indicate the boundary of the domain of $\varphio(c)$, where $\varphio(c)$ is no longer defined. (b) The  coefficient space of the potential $\V$ with $|\setn|$ coefficients. This is a schematic diagram. The stationary manifold $\varphio(c)$ is defined in the blue region $\C$, which is $|\setn|$-dimensional. The red line represents a {\it surface} of dimension less than $|\setn|$ that forms a boundary of $\C$. At the boundary, $\varphio(c)$ is no longer defined.}\label{fig:one}
\end{figure}

We want to study how the stationary manifold changes away from $S\boldsymbol\varphi$ when the coefficients are changed away from $\cB$. A change in the stationary manifold can happen in two ways. One is a smooth change that does not involve a change in its nature. The other involves a change in its nature. A typical example of the second case is the electroweak phase transition resulting from the change in the coefficients in the Higgs potential. In the space of the coefficients, the region where the Higgs potential has the well-known mexican-hat-like structure is shown in FIG~\ref{fig:onea}. At the boundary of this region, the nature of the stationary manifold changes. The theory of bifurcations\cite{GOLUBITSKY1985, GOLUBITSKY1988} describes such changes. A change in the nature of the stationary manifold is associated with a change in the sign of the eigenvalues of the Hessian. Therefore, as long as the number of vanishing eigenvalues remains the same, the change in the stationary manifold will be smooth. Currently, we are interested only in such a smooth change since our objective is to express the stationary manifold as a smooth function of $c$. 

Let $\C$ be the largest connected open neighbourhood of $\cB$ throughout which the Hessian has exactly $\text{dim}(S\boldsymbol\varphi)$ vanishing eigenvalues. A brief description of the boundary of $\C$ is provided in FIG~\ref{fig:oneb}. To ensure the existence of $\C$ in relation to $\cB$, i.e., to ensure that $\cB$ is not a point in the boundary, we need to assume that $\mathcal{H}'_\perp$ (\ref{eq:local}) is non-singular, i.e., the potential is a Morse-Bott function\cite{Bott, MorseBott}, at $\cB$. We use $\C$ as the domain to define the map $\varphio$,
\begin{gather}\label{eq:map}
\varphio\!:\C\rightarrow  S\mathbb{R}^{\text{dim}(\varphi)}\!:\,\,
\begin{split}
&\sum_{\alpha\in\setn} c_\alpha\,\partial_i \mathcal{I}_\alpha(\varphi)\big|_{\varphi=\varphio(c)(\tht)} =0 \,\,\forall \tht\\
&\varphio(\cB)=S\boldsymbol\varphi.
\end{split}
\end{gather}
This map defines $\varphio(c)$ as the stationary manifold of $\V$ expressed as a function of the coefficients $c$. It is defined in relation to the chosen stationary point $\boldsymbol\varphi$ obtained at $\cB$. In (\ref{eq:map}), $\varphio(c)(\tht)$ with $\tht=\tht_1,..,\tht_{\text{dim}(S\boldsymbol\varphi)}$ identifies individual stationary points in the manifold $\varphio(c)$. 

We may denote any point $c\in \C$ as a variation from $\cB$, i.e., $c=\cB+\delta c$. A change in the coefficients $\cB\rightarrow (\cB + \delta c)$ results in a change in the stationary manifold $\varphio(\cB)\rightarrow\varphio(\cB+\delta c)$ which satisfies
\begin{equation}\label{eq:derivative}
\sum_{\alpha\in\setn}(\cB_\alpha+\delta c_\alpha)\,\partial_i \mathcal{I}_\alpha(\varphi)\big|_{\varphi=\varphio(\cB+\delta c)(\tht)} =0 \,\,\, \forall \theta.
\end{equation}
Here, to parameterize the points in the new stationary manifold $\varphio(\cB+\delta c)$, we have used the same variables $\tht$ as we used for the case of $\varphio(\cB)$. Such a unique way of assignment between $\varphio(\cB)$ and $\varphio(\cB+\delta c)$ can be ensured by moving in the directions normal to the manifolds,
\begin{equation}\label{eq:taudef}
\text{\small$\big(\varphio\!(\cB)(\tht\!+\!\delta \tht)-\varphio\!(\cB)(\tht)\big)\!\perp\!\big(\varphio\!(\cB\!+\!\delta c)(\tht)-\varphio\!(\cB)(\tht)\big) \,\,\forall\, \delta \tht, \tht.$}
\end{equation}

In some cases, the stationary orbit $ G\boldsymbol\varphi$ will be a union of disconnected stationary points rather than a union of higher-dimensional stationary manifolds. Throughout this paper, we may use the term stationary manifold to refer to such a disconnected stationary point as well, since a point is nothing but a zero-dimensional manifold. If $\varphio(c)$ is a disconnected stationary point (instead of a higher-dimesional manifold), we do not require the parameters $\tht$. In that case, the whole of the Hessian (instead of only its part normal to the higher-dimensional manifold) is assumed to be non-singular. The Morse-Bott theory, which studies stationary manifolds, is a generalisation of the Morse theory, which was formulated earlier\cite{Bott} to study disconnected stationary points.


\subsection{Coefficient-independent symmetries of the stationary manifold}
\label{sec:two}
Given the potential $\V$ (\ref{eq:generalpot}) containing $|\setn|$ terms, we define
\begin{equation}\label{eq:pot3}
\Vt=\sum_{\alpha\in\setnt} c_\alpha \mathcal{I}_\alpha(\varphi),
\end{equation}
where $\setnt$ is a subset of $\setn$. We express $\V$ as the sum of two parts: the first part being $\Vt$ and the second part being the rest of the terms, i.e.,
\begin{equation}\label{eq:pot2}
\V=\Vt+\sum_{\alpha\in(\setn-\setnt)} c_\alpha \mathcal{I}_\alpha(\varphi).
\end{equation}
The domain of the map $\varphio$ (\ref{eq:map}) contains the point $c=\cB$. Let us assume that the second part of $\V$ (\ref{eq:pot2}) vanishes at $c=\cB$, i.e.,
\begin{equation}
\cB_\alpha=0\quad\forall\,\alpha\in(\setn-\setnt).\label{eq:cond1}
\end{equation}
Let $\Gt$ be the symmetry group of $\Vt$,
\begin{equation}
\Gt=\{g:g\in O(\text{dim}(\varphi)),\, \mathcal{I}_\alpha(g\varphi)=\mathcal{I}_\alpha(\varphi)\,\forall\,\alpha\in\setnt\}.\!\!\!
\end{equation}
We have $ G\subseteq\Gt$. Let $\Ht\subseteq\Gt$ be the pointwise stabilizer of $\varphio(\cB)$ under $\Gt$,
\begin{equation}
\Ht=\{h:h\in \Gt,\, h\,\varphio(\cB)(\tht)=\varphio(\cB)(\tht)\,\forall\,\tht\}.\!\!
\end{equation}
Let $\text{fix}(\Ht)$ be the fixed-point subspace\cite{GOLUBITSKY1988, Gaeta1994} of $\Ht$,
\begin{equation}\label{eq:fix}
\text{fix}(\Ht)=\{\varphi: h\varphi=\varphi\,\,\forall \, h\in \Ht \}.
\end{equation}

Suppose the terms that we added to $\Vt$ to obtain $\V$ (\ref{eq:pot2}), i.e., $\mathcal{I}_\alpha(\varphi)\,\forall\alpha\in(\setn-\setnt)$, satisfy the condition:
\begin{equation}
\boxed{\partial_i \mathcal{I}_\alpha(\varphi) \in \text{fix}(\Ht) \quad\forall \,\,\varphi \in \text{fix}(\Ht)}\,,\label{eq:cond2}
\end{equation}
i.e., the gradient of every newly added term calculated anywhere in the fixed-point subspace of $\Ht$ also lies in the fixed-point subspace. In Appendix~\ref{sec:appone}, we prove that the realization of this condition leads to
\begin{equation}\label{eq:key}
\varphio(c)(\tht)\in\text{fix}(\Ht)\quad\forall\,\tht, \,\,\forall\,c \in \C.
\end{equation}
The above equation can be rephrased as
\begin{equation}\label{eq:key2}
h \varphio(c)(\tht) =\varphio(c)(\tht) \quad\forall\,\tht,\,\,\forall\,\, c \in \C,\,\,h \in \Ht.
\end{equation}
In other words, $\Ht$ encapsulates the symmetries of the stationary manifold $\varphio(c)$ that are independent of the coefficients in the potential. The terms that we added to $\Vt$ to obtain $\V$ do not spoil these symmetries even though they are not, in general, invariant under $\Ht$. They have to satisfy condition (\ref{eq:cond2}) only. If a term satisfies (\ref{eq:cond2}), we say that it is {\it compatible with} $\Ht$. We call (\ref{eq:cond2}) the \mbox{\it compatibility condition}. The discovery of the compatibility condition is the main result of this paper. Compatibility is a weaker condition than invariance, i.e.,~if a term is invariant under $\Ht$, it will be compatible with $\Ht$. Please see Appendix~\ref{sec:apponeb} for proof.


Let us discuss a couple of special cases (corollaries) where the compatibility condition is satisfied by a term $\mathcal{I}_\alpha(\varphi)$. These corollaries are of practical importance in the construction of potentials in model building. Suppose we can write the term $\mathcal{I}_\alpha(\varphi)$ in the form,
\begin{equation}\label{eq:productform}
\mathcal{I}_\alpha(\varphi)=\mathcal{X}(\varphi)^{\T}\,\mathcal{Y}(\varphi),
\end{equation}
where $\mathcal{X}(\varphi)$ and $\mathcal{Y}(\varphi)$ are multiplets that are constructed with $\varphi$ and that transform under $\tilde{G}$. Let $h_{\mathcal X}$ and $h_{\mathcal{Y}}$ be the representations of $h\in \Ht$ acting on $\mathcal{X}(\varphi)$ and $\mathcal{Y}(\varphi)$, respectively,~i.e.,
\begin{equation}\label{eq:xyreps}
h_{\mathcal{X}}\mathcal{X}(\varphi)=\mathcal{X}(h \varphi),\quad h_{\mathcal{Y}}\mathcal{Y}(\varphi)=\mathcal{Y}(h\varphi).
\end{equation}
Corollary~A: If
\begin{equation}\label{eq:coroll1}
h_{\mathcal{X}} \mathcal{Y}(\varphi)=\mathcal{Y}(\varphi),\,\,\, h_{\mathcal{Y}} \mathcal{X}(\varphi)=\mathcal{X}(\varphi)\quad\forall \,\varphi\in\text{fix}(\Ht),
\end{equation}
then the compatibility condition (\ref{eq:cond2}) is satisfied. Please see Appendix~\ref{sec:apponec} for proof.

Let $\Ht_\mathcal{X}=\{h_\mathcal{X}:h\in\Ht\}$ and $\Ht_\mathcal{Y}=\{h_\mathcal{Y}:h\in\Ht\}$. In other words, let $\Ht_\mathcal{X}$ and $\Ht_\mathcal{Y}$ be the set of representation matrices of the group $\Ht$ corresponding to the multiplets $\mathcal{X}(\varphi)$ and $\mathcal{Y}(\varphi)$, respectively. If $\Ht_\mathcal{X}=\Ht_\mathcal{Y}$, condition (\ref{eq:coroll1}) will be satisfied. $\Ht_\mathcal{X}=\Ht_\mathcal{Y}$ does not imply that $h_{\mathcal{X}}= h_{\mathcal{Y}}$; the latter situation is rather trivial since $h_{\mathcal{X}}=h_{\mathcal{Y}}$ corresponds to the term (\ref{eq:productform}) being invariant under $\Ht$.
\\
\\
Corollary~B: If
\begin{equation}\label{eq:coroll2}
\mathcal{X}(\varphi)=0, \quad\mathcal{Y}(\varphi)=0\quad\forall\,\varphi\in\text{fix}(\Ht),
\end{equation}
then the compatibility condition is satisfied. We can see that (\ref{eq:coroll2}) implies (\ref{eq:coroll1}), which in turn implies (\ref{eq:cond2}). Corollary~B is the main result of \cite{2011.11653}. A number of examples that utilize corollaries~A and B are given in Appendix~\ref{sec:apptwo}.

\vspace{7mm}
\centerline{\it Homogenous linear intrinsic constraints}
\label{sec:2O}
\vspace{3mm}
We define a homogenous linear intrinsic constraint (HLIC) in the stationary manifold $\varphio(c)$ as a coefficient-independent constraint that can be expressed as a linear combination of the components of $\varphio(c)(\tht)$ being equal to zero, i.e.,
\begin{equation}\label{eq:hlic}
\kappa^\T\varphio(c)(\tht)=0\quad\forall \,c\in \C,\, \forall\tht,
\end{equation}
where $\kappa$ is a constant vector. We have shown that $\varphio(c)(\tht)$ belongs to $\text{fix}(\Ht)$ independently of $c$ and $\tht$. Therefore, the above equation is satisfied for all $\kappa$ orthogonal to $\text{fix}(\Ht)$, i.e., $\forall\,\kappa\in\text{fix}(\Ht)^\perp$ where $\text{fix}(\Ht)^\perp$ denotes the orthogonal complement of $\text{fix}(\Ht)$. We can show that $\text{fix}(\Ht)^\perp$ is nothing but the space spanned by the rows of the matrices $(h-I)$ with $h\in\Ht$.

Let $H$ be the pointwise stabilizer of $\varphio(\cB)$ under $G$. Since $H$ is a subgroup of $G$, the whole potential $\V$ remains invariant under $H$. It is well known in the mathematics literature (even though, perhaps, not presented in the same way as in this paper) that $H$ generates HLICs in $\varphio(c)(\tht)$, i.e., (\ref{eq:hlic}) is satisfied for all $\kappa\in\text{fix}(H)^\perp$. In this paper, we show that not only $H$ but also the larger group $\Ht$ (with $H\subseteq\Ht$) generates HLICs. Rather than the whole potential $\V$, only $\Vt$ is required to be invariant under $\Ht$. The terms added to $\Vt$ are required to be compatible with $\Ht$ (\ref{eq:cond2}) only. 


\subsection{Obtaining the stationary manifold of $\V$ using symmetry arguments}
\label{sec:three}

In this section, we propose to split the potential $\V$ into smaller parts containing various subsets of the whole set of invariant terms present in $\V$. The criteria followed for this spitting will become apparent as this section develops.  Given a stationary manifold of a part of $\V$, we will investigate how adding new terms to it alters the stationary manifold. The symmetries of the given stationary manifold may or may not survive the addition of the new terms. This determines whether the stationary manifold survives (non-bifurcation scenario) or is replaced (bifurcation scenario). We will learn to combine the stationary manifolds of the various parts to obtain the stationary manifold $\varphio(c)$ of the whole potential $\V$.


\subsubsection{\normalsize{The non-bifurcation scenario}}
\label{subsec:one}
We describe this scenario in relation to (\ref{eq:pot2}), i.e., we have $\Vt$ as a part of the potential and we add new terms to it to obtain $\V$. For convenience, let us use $\ct$ to denote the coefficients of the terms in $\Vt$, i.e., $\ct_\alpha=c_\alpha$ for all $\alpha\in\setnt$. Suppose we are given $\varphiot(\ct)$ as a stationary manifold of $\Vt$. Let $\Ct$ be the $|\setnt|$-dimensional domain where $\varphiot(\ct)$ is defined. We have
\begin{equation}
\sum_{\alpha\in\setnt} c_\alpha\,\partial_i \mathcal{I}_\alpha(\phi)\big|_{\phi=\varphiot(\ct)(\tht)} =0 \,\,\forall \ct\in\Ct,\,\forall \tht,
\end{equation}
where $\tht$ parametrizes the points in the stationary manifold $\varphiot(\ct)$. Here, $\tht$ represents $\text{dim}(\varphiot(\ct))$ parameters. Besides $\varphiot(\ct)$, we are given $\Ht$ as the group that generates the HLICs in $\varphiot(\ct)$, 
\begin{equation}
\Ht\varphiot(\ct)(\tht)=\varphiot(\ct)(\tht)\,\,\forall \ct\in\Ct, \forall \tht.
\end{equation}
We are also given $\Tt$ as the group that acts transitively on $\varphiot(\ct)$,
\begin{equation}
\Tt\varphiot(\ct)(\tht)=\varphiot(\ct)\,\,\forall \ct\in\Ct, \forall \tht.
\end{equation}
Towards the end of Section~\ref{sec:three}, we will explain how to obtain the stationary manifold $\varphiot(\ct)$ and the associated groups $\Ht$ and $\Tt$. For the time being, we simply assume that we are given $\varphiot(\ct)$ as a stationary manifold of the potential $\Vt$, equipped with the two groups $\Ht$ and $\Tt$. 

\begin{figure}[]
\subfloat[\label{fig:twoa}]{
  \includegraphics[width=0.5\columnwidth]{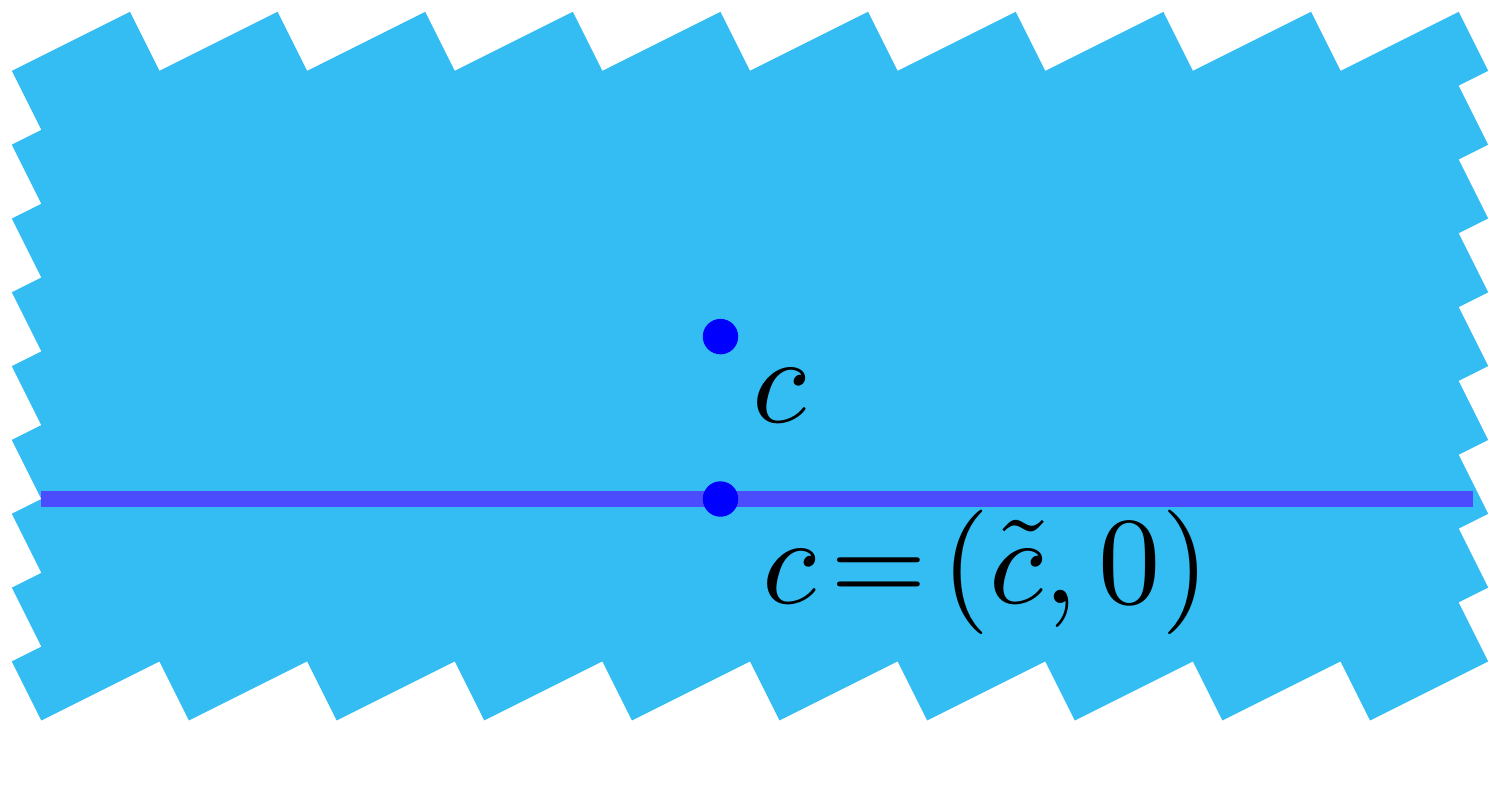}
}
\subfloat[\label{fig:twob}]{
  \includegraphics[width=0.5\columnwidth]{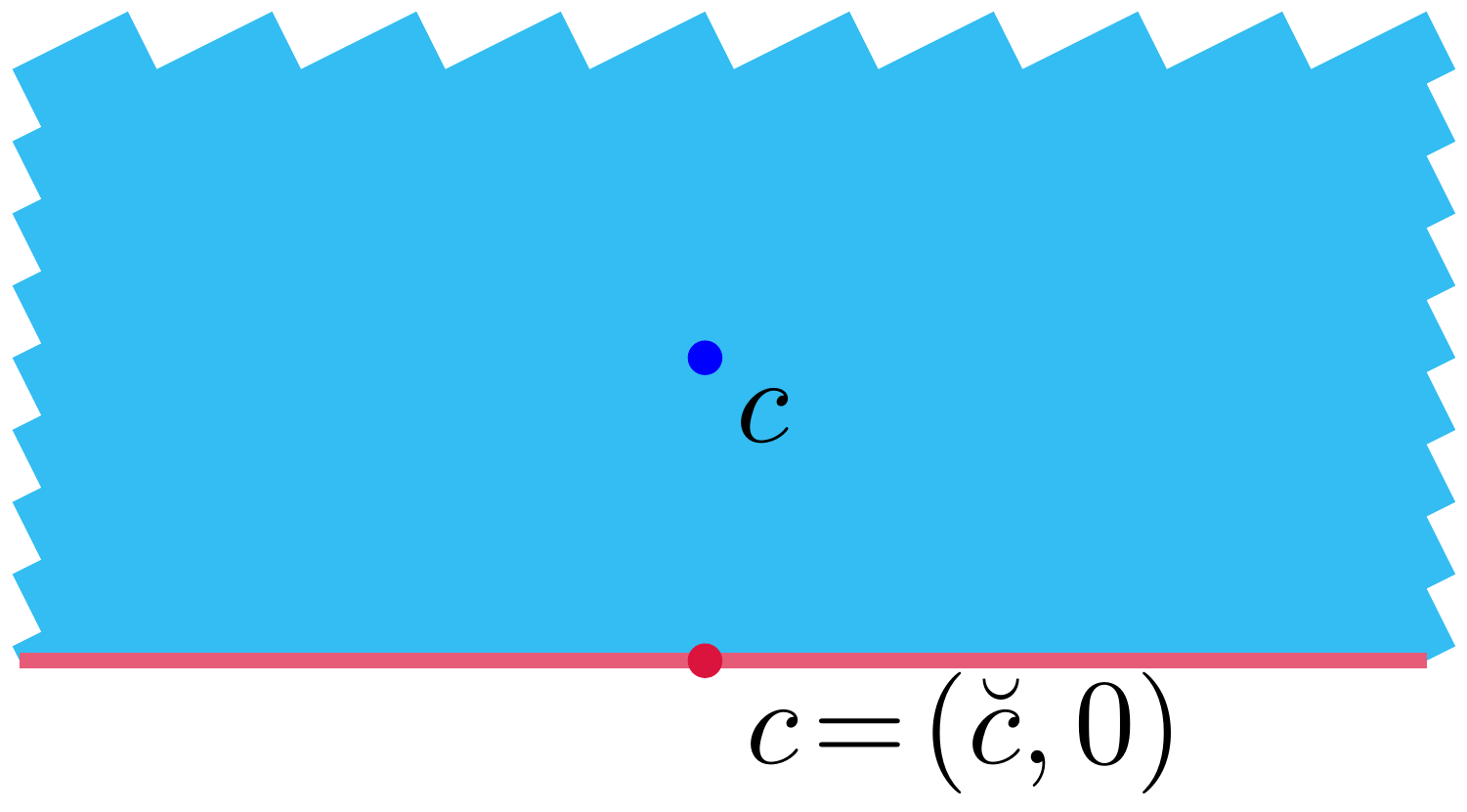}
}
\caption{(a) The non-bifurcation scenario. The stationary manifold $\varphiot(\ct)$ is defined in the $|\setnt|$-dimensional space indicated by the blue line. The stationary manifold $\varphio(c)$ is defined in the $|\setn|$-dimensional space indicated by the blue region, which includes the blue line. A point $\ct$ in the $|\setnt|$-dimensional space is represented as $(\ct,0)$ in the $|\setn|$-dimensional space. We have $\varphio(\ct,0)=\varphiot(\ct)$. (b) The bifurcation scenario. The stationary manifold $\varphiob(\cb)$ is defined in the $|\setnb|$-dimensional space indicated by the red line. The stationary manifold $\varphio(c)$ is defined in the $|\setn|$-dimensional space indicated by the blue region. The domain of $\varphiob(\cb)$ (the red line) forms one of the boundaries of the domain of $\varphio(c)$ (the blue region). Bifurcation occurs at the boundary. A point $\cb$ in the $|\setnb|$-dimensional space is represented as $(\cb,0)$ in the $|\setn|$-dimensional space. We have $\varphio(\cb,x)|_{x\rightarrow0}\subseteq\varphiob(\cb)$.}\label{fig:one}
\end{figure}

The existence of $\varphiot(\ct)$ implies that $\Vt$ is a Morse-Bott function at $\ct\in\Ct$. We obtain $\V$ by adding the terms $\mathcal{I}_\alpha(\varphi)$ with $\alpha\in(\setn-\setnt)$ to $\Vt$. Let us use $c=(\ct,0)$ to concisely denote the values of $c$ that correspond to $c_\alpha=\ct_\alpha$ for all $\alpha\in\setnt$ and $c_\alpha=0$ for all $\alpha\in(\setn-\setnt)$. The potential $\V$ will be a Morse-Bott function at $c=(\ct,0)\,\forall \ct\in\Ct$ {\it if $\Tt$ remains unbroken under the addition of the new terms and if the newly added terms are compatible with} $\Ht$. This scenario implies that there exists a stationary manifold of $\V$, say, $\varphio(c)$, such that it satisfies $\varphio(\ct,0)=\varphiot(\ct)$ and its domain $\C$ encompasses $\Ct$. $\Ct$ forms a $|\setnt|$-dimensional {\it surface} inside the $|\setn|$-dimensional region $\C$, with $c=(\ct,0)$ denoting the points in this surface. This scenario is depicted in FIG~\ref{fig:twoa}. 

The introduction of the new terms does not change the nature of the stationary manifold. $\varphiot(\ct)$ is nothing but $\varphio(c)$ given at a lower dimensional surface within the domain of $\varphio(c)$. As stated earlier, even if the added terms break $\Ht$ to a smaller group $H$, the HLICs continue to be generated by $\Ht$, i.e., the HLICs remain unaffected in this scenario. To conclude, given $\varphiot(\ct)$ equipped with $\Ht$ and $\Tt$, the non-bifurcation scenario guarantees the existence of $\varphio(c)$ also equipped with $\Ht$ and $\Tt$.

\subsubsection{\normalsize{The bifurcation scenario}}
\label{subsec:two}
To distinguish the bifurcation scenario from the previous one, we adopt a notation where we replace the tilde with a breve, i.e., we start with the potential $\Vb$,
\begin{equation}
\Vb=\sum_{\alpha\in\setnb} c_\alpha\,\mathcal{I}_\alpha(\varphi),\label{eq:pot4}
\end{equation}
and add new terms to it to obtain $\V$ (\ref{eq:generalpot}),
\begin{equation}\label{eq:bifscenario}
\V=\Vb+\sum_{\alpha\in(\setn-\setnb)} c_\alpha\,\mathcal{I}_\alpha(\varphi),
\end{equation}
where  $\setnb$ is a subset of $\setn$. For convenience, we denote the coefficients in $\Vb$ with $\cb$, i.e., $\cb_\alpha=c_\alpha\,\,\forall\alpha\in\setnb$. Let $\varphiob(\cb)$ be a stationary manifold of the potential $\Vb$ defined for the $|\setnb|$-dimensional domain $\Cb$. We have
\begin{equation}
\sum_{\alpha\in\setnb} c_\alpha\,\partial_i \mathcal{I}_\alpha(\varphi)\big|_{\varphi=\varphiob(\cb)(\tht)} =0 \,\,\forall \cb\in\Cb,\,\forall \tht,
\end{equation}
where $\tht$ parametrizes the points in the manifold $\varphiob(\cb)$.  Here, $\tht$ represents $\text{dim}(\varphiob(\cb))$ parameters. Let a group $\Hb$ generate the HLICs in $\varphiob(\cb)$, and a group $\Tb$ act transitively on $\varphiob(\cb)$,
\begin{align}
\Hb\varphiob(\cb)(\tht)&=\varphiob(\cb)(\tht)\,\,\forall \cb\in\Cb, \forall \tht,\\
\Tb\varphiob(\cb)(\tht)&=\varphiob(\cb)\,\,\forall \cb\in\Cb, \forall \tht.
\end{align}
In the following paragraph, we describe the procedure to obtain $\varphiob(\cb)$, $\Hb$ and $\Tb$.

If $\varphi$ is an irreducible multiplet, we define $\Vb$ as
\begin{equation}\label{eq:irreppot}
\Vb=c_1 m^2 |\varphi|^2+c_2 |\varphi|^4,
\end{equation}
where $|\varphi|$ is the norm, i.e., $|\varphi|^2=\varphi^\dagger\varphi$.  In (\ref{eq:irreppot}), we introduce $m$ with mass dimension one to ensure that $c_1$ is dimensionless. We have $\cb=(c_1, c_2)$. We define $\Cb$ as the region where $c_1<0$ and $c_2>0$. For $\cb\in\Cb$, we obtain an $n$-sphere with $n=\text{dim}(\varphi)-1$ as the stationary manifold (minima) $\varphiob(\cb)$. Since $\varphiob(\cb)$ does not possess any HLIC, $\Hb$ is trivial. Since the orthogonal group $O(\text{dim}(\varphi))$ acts transitively on the $n$-sphere $\varphiob(\cb)$, we have $\Tb=O(\text{dim}(\varphi))$. Thus we obtain
\begin{equation}\label{eq:irrepcase}
\begin{split}
&\varphiob(\cb)=n\text{-sphere},\\
&\Hb=1, \quad\Tb=O(\text{dim}(\varphi)).
\end{split}
\end{equation}
On the other hand, if $\varphi$ is reducible, we consider it as being composed of multiplets (which themselves may or may not be reducible). For example, let $\varphi$ be composed of two multiplets $\phi$ and $\varrho$, i.e., $\varphi=(\phi,\varrho)^\T$. In this case, we define $\Vb$ as
\begin{equation}
\Vb=\V_\phi+\V_\varrho.
\end{equation}
where $\V_\phi$ and $\V_\varrho$ are potentials constructed with $\phi$ and $\varrho$, respectively. Let $\phio(c_\phi)$ and $\varrhoo(c_\varrho)$ be the stationary manifolds of $\V_\phi$ and $\V_\varrho$ where $c_\phi$ and $c_\varrho$ represent the coefficients in $\V_\phi$ and $\V_\varrho$, respectively. Let $\Ht_\phi$ and $\Ht_\varrho$ generate the HLICs in $\phio(c_\phi)$ and $\varrhoo(c_\varrho)$, and $\Tt_\phi$ and $\Tt_\varrho$ act transitively on  $\phio(c_\phi)$ and $\varrhoo(c_\varrho)$, respectively. We combine $\phio(c_\phi)$ and $\varrhoo(c_\varrho)$ to obtain $\varphiob(\cb)$ the stationary manifold of $\Vb$,
\begin{equation}\label{eq:repcase}
\begin{split}
&\varphiob(\cb)=(\phio(c_\phi),\varrhoo(c_\varrho))^\T,\\
&\Hb=\Ht_\phi\times\Ht_\varrho, \quad\Tb=\Tt_\phi\times\Tt_\varrho.
\end{split}
\end{equation}
The domain $\Cb$ of $\varphiob(\cb)$ is nothing but $\Cb_\phi\times\Cb_\varrho$ where $\Cb_\phi$ and $\Cb_\varrho$ are the domains of $\phio(c_\phi)$ and $\varrhoo(c_\varrho)$, respectively. In this paper, we always use $\varphi$ to denote the full multiplet. If $\varphi$ is reducible, the various objects associated with its constituent multiplets are denoted with subscripts, e.g.,~$\V_\phi$, $c_\phi$, $\Ht_\phi$ etc. On the other hand, we do not use subscripts when describing the objects associated with $\varphi$, e.g.,~$\Vb$, $\cb$, $\Hb$ etc. To summarise, we obtained $\varphiob(\cb)$ as a stationary manifold of the potential $\Vb$, equipped with the two groups $\Hb$ and $\Tb$ (\ref{eq:irrepcase}), (\ref{eq:repcase}).

We want to obtain a stationary manifold of $\V$ (\ref{eq:bifscenario}) in relation to $\varphiob(\cb)$ resulting from bifurcation. This scenario corresponds to $\V$ not being a Morse-Bott function at $c=(\cb,0)$, where $(\cb,0)$ denotes $c_\alpha=\cb_\alpha$ for all $\alpha\in\setnb$ and $c_\alpha=0$ for all $\alpha\in(\setn-\setnb)$. The set of points $(\cb,0)$ with $\cb\in\Cb$ forms a $|\setnb|$-dimensional surface (in the $|\setn|$-dimensional space of $c$) at which the bifurcation occurs. The stationary manifold $\varphiob(\cb)$ is defined only at this surface, outside of which new stationary manifolds emerge. We name one such stationary manifold $\varphio(c)$, with the points in it denoted by $\varphio(c)(\tht)$. Its $|\setn|$-dimensional domain $\C$ has the $|\setnb|$-dimensional surface $\Cb$ as one of its boundaries. This scenario is shown in FIG~\ref{fig:twob}. Based on the HLICs in the newly obtained stationary manifold $\varphio(c)$, we classify the bifurcation scenario into two cases\footnote{We do not claim that all bifurcation scenarios fall into one of these two cases. The aim of Section~\ref{sec:three} is not a comprehensive analysis of bifurcation. Rather, we are proposing a convenient procedure to identify stationary manifolds in a variety of settings applicable in physics that involve discrete symmetries. This procedure requires that the groups that generate HLICs such as $\Hb$ and $\Ht$ are finite.}: {\it bifurcation leads to the generation of new HLICs}; {\it bifurcation leads to the destruction of the pre-existing HLICs}.

\vspace{3mm}
\centerline{\it Bifurcation generates HLICs}
\vspace{1.5mm}

Let $\Ga$ be the subgroup of $\Tb$ under which the newly added terms remain invariant,
\begin{equation}
\Ga=\{g:g\in\Tb, \,\,\mathcal{I}_\alpha(g\phi)=\mathcal{I}_\alpha(\phi)\,\,\forall\alpha\in(\setn-\setnb)\}.
\end{equation}
We assume that $\Ga$ is a proper subgroup of $\Tb$, i.e., the added terms break $\Tb$. To identify $\varphio(c)$, we need to study the stratification\cite{MICHEL200111} of $\varphiob(\cb)$ under $\Ga$. Let $\mathcal M$ be a connected part of a $\Ga$-orbit in a maximal symmetry stratum in $\varphiob(\cb)$. Michel's theorem\cite{MICHEL1971, MICHEL200111} and Golubitsky's and Stewart's reduction lemma\cite{GAETA2006322, GOLUBITSKY1988} guarantee the existence of a stationary manifold of $\V$ in relation to $\mathcal M$. We identify this guaranteed stationary manifold as $\varphio(c)$. We have $\varphio(\cb,x)|_{x\rightarrow0}=\mathcal{M}$.

Let $\Ha$ be the pointwise stabilizer of $\mathcal{M}$ under $\Ga$,
\begin{equation}\label{eq:bfpnt}
\Ha=\{h:h\in\Ga,\,\, h\varphi=\varphi\,\,\forall\varphi\in\mathcal{M}\}.
\end{equation}
The HLICs in $\varphio(c)$ are generated by 
\begin{equation}\label{eq:hlicgen}
\Ht=\Ha\times\Hb.
\end{equation}
Let $\Sa$ be the setwise stabilizer of $\mathcal{M}$ under $\Ga$,
\begin{equation}\label{eq:bfset}
\Sa=\{s:s\in\Ga,\,\, s\varphi\in\mathcal{M}\,\,\forall\varphi\in\mathcal{M}\}.
\end{equation}
The pointwise stabilizer ($\Ha$) forms a normal subgroup of the setwise stabilizer ($\Sa$). We define $\Tt$ as the quotient,
\begin{equation}\label{eq:bftrn}
\Tt=\Sa/\Ha.
\end{equation}
$\Tt$ acts transitively\footnote{Both $\Sa$ and $\Tt$ act transitively in $\varphio(c)$. We define $\Tt$ by taking the quotient because we do not want the group defined for the transitive action to contain the HLIC-generating symmetries.} in $\varphio(c)$. 

Since $\Tb$ is broken, the dimension of $\mathcal{M}$, and subsequently the dimension of $\varphio(c)$, which is obtained in relation to $\mathcal{M}$, will be less than that of $\varphiob(\cb)$. Hence, the number of parameters represented by $\tht$ in $\varphio(c)(\tht)$ will be less than that in $\varphiob(\cb)(\tht)$. To conclude, we started with the stationary manifold $\varphiob(\cb)$ equipped with $\Hb$ and $\Tb$ and obtained $\varphiob(\cb)$ equipped with $\Ht$ (\ref{eq:hlicgen}) and $\Tt$  (\ref{eq:bftrn}). Here, bifurcation generates HLICs since $\Ht$ (\ref{eq:hlicgen}) is larger than $\Hb$\footnote{In situations where $\Hb$ is nontrivial, we need to verify that the newly added terms are compatible with $\Ha\times\Hb$ to ensure the validity of our analysis. Also, $\Ha$ is nontrivial in most situations. An exception is when the whole of $\varphiob(\cb)$ forms a single maximal symmetry stratum under $\Ga$. Another type of exception appears in Appendices D 3 and D 4 of the companion paper\cite{2309.11542}. In these exceptions, where $\Ha=1$, the bifurcation does not generate HLICs.}.

\vspace{2mm}
\centerline{\it Bifurcation destroys HLICs}
\vspace{1.5mm}

Here, we assume that the newly added terms remain invariant under $\Tb$. Therefore, the whole of $\varphiob(\cb)$ remains as a single connected $\Tb$-orbit even after the addition of the new terms. Since $\varphiob(\cb)$ also forms a maximal symmetry stratum (under $\Tb$), we have $\mathcal{M}=\varphiob(\cb)$. Suppose the new terms are not compatible with $\Hb$. This leads to bifurcation. As in the previous case, we are guaranteed to obtain a stationary manifold of $\V$ in relation to $\mathcal M$, say, $\varphio(c)$, with $\varphio(\cb,x)|_{x\rightarrow0}=\mathcal{M}$. 

The HLICs in $\varphio(c)$ are generated by the subgroup of $\Hb$ under which the newly added terms are compatible, i.e., they are generated by
\begin{equation}\label{eq:hlicgen2}
\Ht\!=\!\{h\!:h\!\in\!\Hb,\, \partial_i \mathcal{I}_\alpha(\varphi)\!\in \text{fix}(\Ht) \,\,\forall \,\varphi \in \text{fix}(\Ht), \alpha\in(\setn-\setnb)\}.
\end{equation}
The group of transitive action in $\varphio(c)$ is nothing but $\Tb$ itself, i.e.,
\begin{equation}\label{eq:bftrn2}
\Tt=\Tb.
\end{equation}
The stationary manifolds $\varphio(c)$ and $\varphiob(\cb)$ have the same dimension, i.e., the number of parameters represented by $\tht$ in $\varphio(c)(\tht)$ and $\varphiob(\cb)(\tht)$ are the same. To conclude, we obtained $\varphiob(\cb)$, $\Ht$ (\ref{eq:hlicgen2}) and $\Tt$ (\ref{eq:bftrn2}) given $\varphiob(\cb)$, $\Hb$ and $\Tb$ as in the previous case. Here, however, bifurcation destroys HLICs since $\Ht$ (\ref{eq:hlicgen2}) is a proper subgroup of $\Hb$.

In (\ref{eq:bifscenario}), we added new terms to $\Vb$ to obtain the whole potential $\V$. In some situations, we utilize a two-step process to construct $\V\,$:  at first, adding terms to $\Vb$ to obtain $\Vt$, and then adding more terms to $\Vt$ to obtain $\V$,
\begin{align}
\Vt&=\Vb+\sum_{\alpha\in(\setnt-\setnb)} c_\alpha\,\mathcal{I}_\alpha(\varphi),\label{eq:step1}\\
\V&=\Vt+\sum_{\alpha\in(\setn-\setnt)} c_\alpha\,\mathcal{I}_\alpha(\varphi).\label{eq:step2}
\end{align}
The addition of terms in (\ref{eq:step1}) refers to the bifurcation scenario. Accordingly, we follow the procedure described in Section~\ref{subsec:two}, and obtain the stationary manifold of $\Vt$ as $\varphiot(\ct)$ along with the groups $\Ht$ and $\Tt$. Then we move on to add more terms as given in (\ref{eq:step2}) which refers to the non-bifurcation scenario. We utilize $\varphiot(\ct)$, $\Ht$ and $\Tt$ obtained earlier and follow the procedure described in  Section~\ref{subsec:one} to obtain the stationary manifold of $\V$ as $\varphio(c)$ with the groups $\Ht$ and $\Tt$ preserved. 

\begin{figure}[]
\subfloat[\label{fig:irrep}]{
 \includegraphics[width=0.25\textwidth]{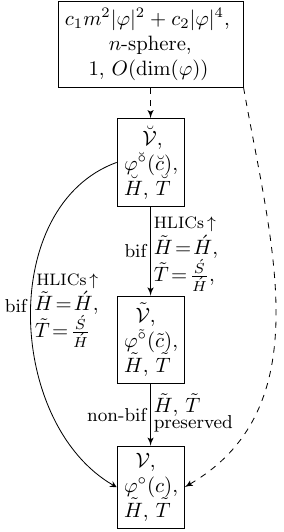}}
\subfloat[\label{fig:rerep}]{
 \includegraphics[width=0.25\textwidth]{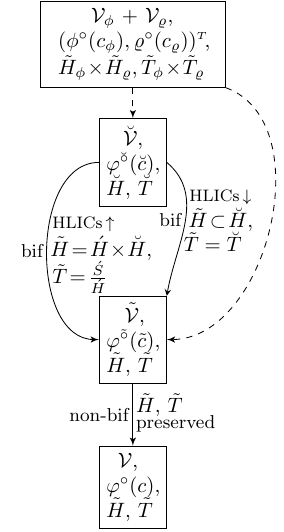}}
\caption{(a) and (b) describe the steps to obtain the stationary manifolds, with $\varphi$ being an irrep and a reducible representation, respectively. The dotted arrows denote assignment, e.g., the short arrow in (a) denotes (\ref{eq:irreppot}),(\ref{eq:irrepcase}). The solid arrows denote the bifurcation (bif) and the non-bifurcation (non-bif) scenarios. We start with a potential, its stationary manifold and the associated groups at the top of (a) and (b) and obtain $\V$, $\varphio(c)$, $\Ht$ and $\Tt$ at the bottom via one of several routes.}\label{fig:rep}
\end{figure}

FIG.~\ref{fig:rep} summarises the procedure developed in this section. The left route and middle route in FIG.~\ref{fig:irrep} represent the splitting of $\V$ as given in (\ref{eq:bifscenario}) and (\ref{eq:step1}), (\ref{eq:step2}), respectively, for $\varphi$ being an irrep. The right route represents the situation where no more terms are added to $\Vb$ (\ref{eq:irreppot}) so that $\Vb$ is equal to $\V$. FIG.~\ref{fig:rerep} describes the scenarios where $\varphi$ is a reducible representation. The left and middle routes correspond to (\ref{eq:step1}), (\ref{eq:step2}), where HLICs are generated and destroyed, respectively. The right route shows the splitting of $\V$ that involves the non-bifurcation scenario only. At the top of FIG.~\ref{fig:rerep}, we have used $\phio(c_\phi)$ and $\varrhoo(c_\varrho)$ as inputs. If a multiplet, say, $\phi$, is an irrep, we follow the steps in FIG.~\ref{fig:irrep} to obtain $\phio(c_\phi)$. On the other hand, if a multiplet, say, $\varrho$, is reducible we consider its constituent multiplets, say, $\varrho=(\rho,\eta)^\T$. We follow the steps in FIG.~\ref{fig:rerep} with $\rhoo(c_\rho)$ and $\etao(c_\eta)$ as inputs, and obtain $\varrhoo(c_\varrho)$. We continue this procedure recursively until $\varphi$ is split into its constituent irreps, say, $\varphi=(\phi,\rho,...)^\T$. In this way, we construct the whole potential $\V$, which includes the potentials of all the irreps as well as all possible cross terms among the irreps, and obtain the stationary manifold $\varphio(c)$ of $\V$. The resulting $\Ht$, which generates the HLICs in $\varphio(c)$, will be the direct product of the HLIC-generating groups of the irreps. We demonstrate this procedure with a few simple examples in Appendix~\ref{sec:apptwo}. For a flavour model where this procedure is adopted, please see the companion paper\cite{2309.11542}. 



\subsection{Summary}
\label{sec:four}

We consider a multiplet $\varphi$ transforming under a compact group $\mathbb G$. Its general renormalizable potential $\V$ consists of all independent invariant terms $\mathcal{I}_\alpha(\varphi)$ up to the fourth order and the corresponding coefficients $c_\alpha$, i.e., $\V=\sum_\alpha c_\alpha \mathcal{I}_\alpha(\varphi)$. Since we do not include higher-order terms in $\V$, its symmetry group, $G$, can be larger than $\mathbb G$, i.e., $\mathbb{G}\subseteq G$. We express a stationary manifold of $\V$ as a function of the coefficients, namely, $\varphio(c)$, with a domain $\mathcal C$. Even though a change in $c$ causes a change in $\varphio(c)$, its pointwise stabilizer, $H\subseteq G$, remains independent of $c$, i.e., $h\varphio(c)=\varphio(c) \,\,\forall \,h\in H, c\in \mathcal{C}$. This implies that there exists a set of homogenous linear equations that constrain the components of $\varphio(c)$ in a coefficient-independent way. We call them homogeneous linear intrinsic constraints (HLICs). These constraints can be given as $\kappa^\T\varphio(c)=0\,\forall\,c\in\C$, where $\kappa$ belongs to the orthogonal complement of the fixed-point subspace of $H$, i.e., $\kappa\in\text{fix}(H)^\perp$. In modeling a physical system with discrete symetries, we often assign one of the stationary points (minima) of $\V$ as the expectation value $\langle\varphi\rangle$ via spontaneous symmetry breaking. The HLICs in $\varphio(c)$ leads to constraints on the components of $\langle\varphi\rangle$ that manifest as coefficient-independent predictions of the model.

In \cite{2011.11653}, we showed that $\varphio(c)$ can contain HLICs beyond those generated by $H$, i.e., we can have $\kappa^\T\varphio(c)=0\,\,\forall\,\,c\in\C$ even for $\kappa\not\in\text{fix}(H)^\perp$. The current work formalizes and generalizes the results from \cite{2011.11653}. We analyze the potential $\V$ by splitting it, resulting in the part $\Vt$ having the symmetry group $\Gt$ with $G\subseteq\Gt$. We define $\Ht$ as the pointwise stabilizer of $\varphio(c)$ under $\Gt$. We discover that $\kappa^\T\varphio(c)=0 \,\,\forall\,\,c\in\C,\, \kappa\in \text{fix}(\Ht)^\perp$ if the terms in $\V$ satisfy the {\it compatibility condition}\,: $\partial_i \mathcal{I}_\alpha(\varphi) \in \text{fix}(\Ht) \,\forall\,\varphi \in \text{fix}(\Ht)^\perp$. The compatibility condition is more general than the requirement of invariance, i.e., the terms $\mathcal{I}_\alpha(\varphi)$ that are compatible with $\Ht$ need not be invariant under it. To easily verify the compatibility condition, we provide two corollaries which are its special cases.

We develop a step-by-step procedure to identify $\varphio(c)$ by splitting $\V$ into several parts. First, we obtain the minima of the parts of $\V$ constructed with the irreps in $\varphi$ as $n$-spheres. More terms are added to these parts leading to bifurcation and emergence of new stationary manifolds. We utilize the theorems by Michel, Golubitsky and Stewart to identify these stationary manifolds. We may further add terms that satisfy the compatibility condition resulting in the non-bifurcation scenario. We combine the stationary manifolds of the various irreps and add the cross terms, which again falls under the bifurcation or the non-bifurcation scenarios. The bifurcation scenario can create new HLICs or destroy pre-existing ones, while the non-bifurcation scenario preserves HLICs. By including all renormalizable terms, we obtain the whole potential $\V$ and its stationary manifold $\varphio(c)$. Following this procedure, we construct $\Ht$ as the direct product of the stabilizers corresponding to each irrep, with all cross terms being compatible with $\Ht$. In situations where $\varphio(c)$ contains more HLICs than those generated by $H$, i.e.,~where $\text{fix}(H)^\perp$ forms a subspace of $\text{fix}(\Ht)^\perp$, our procedure helps to determine $\Ht$.  More importantly, in general, it serves as a way to handle the stabilizers corresponding to each irrep separately so that we do not have to determine $G$ and $H$.

To demonstrate this procedure, we provide a number of examples in Appendix~\ref{sec:apptwo}. In the flavour physics literature, authors resort to explicit extremization to obtain the minima. By contrast, our procedure utilizes theorems on symmetries well-established in the mathematics literature. One of the aims of this paper is to make these theorems more accessible and usable to physicists. Our examples provide a comprehensive study of these theorems. They also demonstrate the various ways driving fields can be utilized in our procedure. Finally, we provide an example where we construct an {\it effective} irrep by taking the tensor product of several {\it elementary} \mbox{irreps}. Under a set of auxiliary generators, the elementary \mbox{irreps} transform nontrivially while the effective irrep remains invariant. We show that the auxiliary symmetries, though {\it hidden} from the effective irrep, can play a role in determining its HLICs. A realistic flavour model that invokes the framework of the auxiliary group and utilizes the mathematical procedure developed in this paper is constructed in the companion paper\cite{2309.11542}.

The non-renormalizable terms, which are of order higher than four, are suppressed. The truncation of the potential at a given order may increase symmetries in two ways: it produces accidental symmetries, i.e.,~enlarges $\mathbb{G}$ to $ G$, and it leads to $\Ht$ being larger than $H\subseteq G$. The higher-order terms that are not compatible with $\Ht$ will perturb the HLICs generated by $\Ht$. In this paper, we have ignored the higher-order terms since their effect is small. On the other hand, we may utilize the non-renormalizable terms, as an alternative to driving fields, to break the accidental continuous symmetries in the potential. In that case, we need to consider the non-renormalizable terms hierarchically, e.g.,~we utilize the fifth order terms and ignore the sixth, and verify that the terms we utilize, e.g.,~the fifth order terms, are compatible with $\Ht$. Such scenarios can also be analysed using our mathematical precedure even though we have not done so in this paper.

\vspace{5mm}
\centerline{\bf Acknowledgments} 
\vspace{2mm}
I am grateful to Sujatha Ramakrishnan for the many stimulating discussions. I sincerely thank Ambar Ghosal and Debasish Majumdar for their help and support. I gratefully acknowledge the seemless administrative assistance provided by Pourjok Majumder at SINP, Kolkata. I am indebted to Srubabati Goswami for lending a helping hand when most in need. 

\onecolumngrid
\vspace{8mm}
\noindent\makebox[\linewidth]{\resizebox{0.5\linewidth}{1pt}{$\bullet$}}\bigskip
\twocolumngrid

\appendix

\onecolumngrid
\vspace{-4mm}

\section{Proofs}

\subsection{To prove that (\ref{eq:cond2}) leads to (\ref{eq:key})}
\label{sec:appone}

The partial derivatives of the potential (\ref{eq:pot2}) vanishes at the stationary manifold (\ref{eq:map}). At $c=\cB$, we obtain
\begin{equation}\label{eq:map2}
\Big(\sum_{\alpha\in\setnt}\cB_\alpha\,\partial_i \mathcal{I}_\alpha(\varphi)+\sum_{\alpha\in(\setn-\setnt)} 0\,\partial_i \mathcal{I}_\alpha(\varphi)\Big)\Big|_{\varphi=\varphio(\cB)(\tht)} =0 \,\,\, \forall \tht.
\end{equation}
Since a change in the coefficients, $\cB\rightarrow \cB+\delta c$, causes a change in the manifold, $\varphio(\cB)\rightarrow \varphio(\cB+\delta c)$, we have
\begin{equation}\label{eq:derivativeI}
\Big(\sum_{\alpha\in\setnt}(\cB_\alpha+\delta c_\alpha)\,\partial_i \mathcal{I}_\alpha(\varphi)+\sum_{\alpha\in(\setn-\setnt)}\delta c_\alpha\,\partial_i \mathcal{I}_\alpha(\varphi)\Big)\Big|_{\varphi=\varphio(\cB+\delta c)(\tht)} =0 \,\,\, \forall \tht.
\end{equation}
To obtain the stationary manifold at $\cB+\delta c$, we use the Taylor expansion in several variables,
\begin{equation}\label{eq:taylor}
\varphio(\cB+\delta c)(\tht)=\sum_{|t|=0}^\infty \frac{\partial^t \varphio(\cB)(\tht)}{t!}\delta c^{t},
\end{equation}
where $t$ denotes a $|\setn|$-tuple having whole numbers $t_\alpha$ as its elements and
\begin{equation}
|t|=\sum_{\alpha\in\setn} t_\alpha\,, \quad t!=\prod_{\alpha\in\setn}t_\alpha!\,,\quad
\partial^t = \prod_{\alpha\in\setn}\frac{\partial^{t_\alpha}\,\,\,}{\partial c_\alpha^{t_\alpha}}\,, \quad\delta c^t =\prod_{\alpha\in\setn}\delta c_\alpha^{t_\alpha}.
\end{equation}
Here we have adopted the {\it multi-index} notation of Taylor expansion described in \cite{Folland1999}. Also, $\partial^t \varphio(\cB)(\tht)$ is an abuse of notation which really denotes $\partial^t \varphio(c)(\tht)|_{c=\cB}$. Applying expansion (\ref{eq:taylor}) on (\ref{eq:derivativeI}), we obtain
\begin{equation}\label{eq:derivativetaylor}
\Big(\sum_{\alpha\in\setnt}(\cB_\alpha+\delta c_\alpha)\,\partial_i \mathcal{I}_\alpha(\varphi)+\sum_{\alpha\in(\setn-\setnt)}\delta c_\alpha\,\partial_i \mathcal{I}_\alpha(\varphi)\Big)\Big|_{\varphi=\Sigma_w+\overline{\Sigma}_w} =0,
\end{equation}
where
\begin{equation}\label{eq:AB}
\Sigma_w=\sum_{|t|=0}^{w} \frac{\partial^t \varphio(\cB)(\tht)}{t!}\delta c^{t}, \quad \overline{\Sigma}_w= \sum_{|t|=w+1}^\infty \frac{\partial^t \varphio(\cB)(\tht)}{t!}\delta c^{t},
\end{equation}
with $w$ being any whole number. Taylor expanding (\ref{eq:derivativetaylor}) around $\Sigma_w$, we obtain
\begin{equation}\label{eq:derivativetaylor2}
\begin{split}
&\sum_{\alpha\in\setnt}(\cB_\alpha+\delta c_\alpha)\,\Big( \partial_i \mathcal{I}_\alpha(\varphi)\big|_{\varphi=\Sigma_w} +  \partial_i\partial_j \mathcal{I}_\alpha(\varphi)\big|_{\varphi=\Sigma_w}\sum_{|t|=w+1}^\infty \frac{\partial^t {\varphio}_{\!\!\!j}(\cB)(\tht)}{t!}\delta c^t +\mathcal{HO}\Big)\\
&+\sum_{\alpha\in(\setn-\setnt)}\delta c_\alpha\,\Big( \partial_i \mathcal{I}_\alpha(\varphi)\big|_{\varphi=\Sigma_w} +  \partial_i\partial_j \mathcal{I}_\alpha(\varphi)\big|_{\varphi=\Sigma_w}\sum_{|t|=w+1}^\infty \frac{\partial^t {\varphio}_{\!\!\!j}(\cB)(\tht)}{t!}\delta c^t +\mathcal{HO}\Big)=0  \quad \forall\tht,
\end{split}
\end{equation}
where $\mathcal{HO}$ are higher-order terms containing $\delta c^{t}$ with $|t|\geq2(w+1)$.

Carrying out a transformation $\varphi'= h\varphi$, where $h\in\Ht$, in (\ref{eq:derivativeI}), we obtain
\begin{equation}\label{eq:temp}
\Big(\sum_{\alpha\in\setnt}(\cB_\alpha+\delta c_\alpha)\,h^\T_{\,\,ij}\,\partial'_j \mathcal{I}_\alpha(h^\T\varphi')+\sum_{\alpha\in(\setn-\setnt)}\delta c_\alpha\,h^\T_{\,\,ij}\,\partial'_j \mathcal{I}_\alpha(h^\T\varphi')\Big)\Big|_{{\varphi'}=h\varphio(\cB+\delta c)(\tht)} =0 \,\,\, \forall \tht,
\end{equation}
where $\partial'_j$ are the derivatives with respect to $\varphi'_j$. Multiplying the LHS of the above equation with $h$, then simplifying using the relation $\mathcal{I}_\alpha(h^\T\varphi')=\mathcal{I}_\alpha(\varphi')\,\forall\,\alpha\in\setnt$, and finally replacing $\varphi'$ with $\varphi$, we obtain
\begin{equation}\label{eq:derivativetrof}
\Big(\sum_{\alpha\in\setnt}(\cB_\alpha+\delta c_\alpha)\,\partial_i \mathcal{I}_\alpha(\varphi)+\sum_{\alpha\in(\setn-\setnt)}\delta c_\alpha\,\partial_i \mathcal{I}_\alpha(h^\T\varphi)\Big)\Big|_{{\varphi}=h\varphio(\cB+\delta c)(\tht)} =0 \,\,\, \forall \tht.
\end{equation}
Substituting (\ref{eq:taylor}) and (\ref{eq:AB}) in the above equation, we obtain 
\begin{equation}\label{eq:derivativetrof}
\Big(\sum_{\alpha\in\setnt}(\cB_\alpha+\delta c_\alpha)\,\partial_i \mathcal{I}_\alpha(\varphi)+\sum_{\alpha\in(\setn-\setnt)}\delta c_\alpha\,\partial_i \mathcal{I}_\alpha(h^\T\varphi)\Big)\Big|_{{\varphi}=h\Sigma_w+h\overline{\Sigma}_w} =0\quad \forall\tht,
\end{equation}
which leads to
\begin{equation}\label{eq:derivativetaylor3}
\begin{split}
&\sum_{\alpha\in\setnt}(\cB_\alpha+\delta c_\alpha)\,\Big( \partial_i \mathcal{I}_\alpha(\varphi)\big|_{\varphi=h\Sigma_w} +  \partial_i\partial_j \mathcal{I}_\alpha(\varphi)\big|_{\varphi=h\Sigma_w}h_{jk}\sum_{|t|=w+1}^\infty \frac{\partial^t {\varphio}_{\!\!\!k}(\cB)(\tht)}{t!}\delta c^t +\mathcal{HO}\Big)\\
&+\sum_{\alpha\in(\setn-\setnt)}\delta c_\alpha\,\Big( \partial_i \mathcal{I}_\alpha(h^\T\varphi)\big|_{\varphi=h\Sigma_w} +  \partial_i\partial_j \mathcal{I}_\alpha(h^\T\phi)\big|_{\varphi=h\Sigma_w}h_{jk}\sum_{|t|=w+1}^\infty \frac{\partial^t {\varphio}_{\!\!\!k}(\cB)(\tht)}{t!}\delta c^t +\mathcal{HO}\Big)=0  \,\, \forall\tht,
\end{split}
\end{equation}
similar to (\ref{eq:derivativetaylor2}).

Our proof utilizes the method of induction. For that purpose, let us assume
\begin{equation}\label{eq:condition}
h \Sigma_w = \Sigma_w \quad \forall\tht.
\end{equation}
Given (\ref{eq:condition}) and (\ref{eq:cond2}), we subtract (\ref{eq:derivativetaylor2}) from (\ref{eq:derivativetaylor3}) to obtain
\begin{equation}
\begin{split}
&\sum_{\alpha\in\setnt}(\cB_\alpha+\delta c_\alpha)\,\partial_i\partial_j \mathcal{I}_\alpha(\varphi)\big|_{\varphi=\Sigma_w}(h_{jk}-\delta_{jk})\Big(\sum_{|t|=w+1}^\infty \frac{\partial^t {\varphio}_{\!\!\!k}(\cB)(\tht)}{t!}\delta c^t +\mathcal{HO}\Big)\\
&+\sum_{\alpha\in(\setn-\setnt)}\delta c_\alpha\,\left(\partial_i\partial_j \mathcal{I}_\alpha(h^\T\varphi)\big|_{\varphi=\Sigma_w}h_{jk}-\partial_i\partial_j \mathcal{I}_\alpha(\varphi)\big|_{\varphi=\Sigma_w}\delta_{jk}\right)\Big(\sum_{|t|=w+1}^\infty \frac{\partial^t {\varphio}_{\!\!\!k}(\cB)(\tht)}{t!}\delta c^t +\mathcal{HO}\Big)=0\,\, \forall \tht.
\end{split}
\end{equation}
Every term in this equation contains products of $\delta c_\alpha$ with order $w+1$ or higher. Extracting terms of order $w+1$, we obtain
\begin{equation}
\sum_{\alpha\in\setnt}\cB_\alpha\partial_i\partial_j \mathcal{I}_\alpha(\varphi)\big|_{\varphi=\Sigma_w} \, (h_{jk}-\delta_{jk})\,\partial_{t} {\varphio}_{\!\!\!k}(\cB)(\tht)=0 \quad \forall  \,\,|t|=w+1,\,\,\tht.
\end{equation}
The above equation can be rewritten as a matrix equation,
\begin{equation}\label{eq:matpro}
\mathcal{H}\,\,(h-I)\,\partial_{t} \varphio(\cB)(\tht) = 0 \quad \forall  \,\,|t|=w+1,\,\,\forall\tht,
\end{equation}
where $\mathcal{H}$ is the Hessian matrix calculated at $\varphi=\Sigma_w$. The change of vector $\varphio(\cB+\delta c)(\tht)-\varphio(\cB)(\tht)$ is always along the directions normal to the stationary manifold $\varphio(\cB)$ since we are keeping $\tht$ fixed, i.e., the tangential components of $\partial_{t} \varphio(\cB)(\tht)$ vanish. Since the Hessian has a non-singular block in the space normal to the stationary manifold, (\ref{eq:matpro}) leads to
\begin{equation}\label{eq:hder}
h\,\partial^t {\varphio}(\cB)(\theta)=\partial^t {\varphio}(\cB)(\tht)\quad \forall  \,\,|t|=w+1,\,\,\forall\tht.
\end{equation}
By definition, we have 
\begin{equation}\label{eq:step}
\Sigma_{w+1}=\Sigma_w+\sum_{|t|=w+1}\frac{\partial^t \varphio(\cB)(\tht)}{t!}\delta c^{t}.
\end{equation}
From (\ref{eq:step}), (\ref{eq:hder}) and assuming (\ref{eq:condition}),we obtain
\begin{equation}
h \Sigma_{w+1} = \Sigma_{w+1}\quad \forall\tht.
\end{equation}
Thus we have proved that 
\begin{equation}\label{eq:implies}
\text{if}\quad h \Sigma_w = \Sigma_w \,\, \forall\tht,\quad\text{then}\quad h \Sigma_{w+1} = \Sigma_{w+1}\,\, \forall\tht.
\end{equation}

By definition, $\Ht$ is the pointwise stabilizer of the stationary manifold $\varphio(\cB)$. Since $\Sigma_0=\varphio(\cB)(\tht)$, we obtain
\begin{equation}\label{eq:stab0}
h \Sigma_0 = \Sigma_0\quad\forall\,h\in \Ht,\,\,\forall\tht.
\end{equation}
(\ref{eq:stab0}) and (\ref{eq:implies}) lead to
\begin{equation}
h \Sigma_\infty = \Sigma_\infty\quad\forall\,h\in \Ht,\,\,\forall\tht.
\end{equation}
$\Sigma_\infty$ is nothing but $\varphio(\cB+\delta c)(\tht)$. Thus we conclude that $\Ht$ is the pointwise stabilizer of $\varphio(c)$ not only at $c=\cB$ but also at all values of $c\in \C$.

\vspace{8mm}
\twocolumngrid

\subsection{To prove that invariance under \texorpdfstring{$\tilde{H}$}{H} leads to (\ref{eq:cond2})}
\label{sec:apponeb}

Carrying out a transformation $\varphi'= h\varphi$, where $h\in\Ht$, on the gradient of $\mathcal{I}_\alpha(\varphi)$, we obtain
\begin{equation}\label{eq:apponeb1}
\partial \mathcal{I}_\alpha(\varphi) = h^\T\,\partial' \mathcal{I}_\alpha(h^\T\varphi').
\end{equation}
Replacing $\varphi'$ with $\varphi$ and multiplying the LHS and RHS of the above equation with $h$, we obtain
\begin{equation}\label{eq:apponeb2}
h\partial \mathcal{I}_\alpha(\varphi) = \partial \mathcal{I}_\alpha(h^\T\varphi).
\end{equation}
If $\mathcal{I}_\alpha(\varphi)$ is invariant under $\Ht$, we have $\mathcal{I}_\alpha(h^\T\varphi)=\mathcal{I}_\alpha(\varphi)$. In that case, (\ref{eq:apponeb2}) becomes
\begin{equation}\label{eq:apponeb3}
h\partial \mathcal{I}_\alpha(\varphi) = \partial \mathcal{I}_\alpha(\varphi),
\end{equation}
i.e.,~the gradient of the term lies in the fixed-point subspace of $\Ht$, which implies that condition (\ref{eq:cond2}) is satisfied.
\vspace{-1mm}

\subsection{To prove that (\ref{eq:coroll1}) leads to (\ref{eq:cond2})}
\label{sec:apponec}

We start with (\ref{eq:apponeb2}) and expand its RHS using (\ref{eq:productform}),
\begin{equation}\label{eq:apponec1}
\begin{split}
h\partial \mathcal{I}_\alpha(\varphi) &= \partial \mathcal{I}_\alpha(h^\T\varphi)\\
&= \partial \big(\mathcal{X}(h^\T\varphi)^\T\,\mathcal{Y}(h^\T\varphi)\big)\\
&= \partial \mathcal{X}(h^\T\varphi)^\T\mathcal{Y}(h^\T\varphi)+ \mathcal{X}(h^\T\varphi)^\T\partial\mathcal{Y}(h^\T\varphi)
\end{split}
\end{equation}
Given (\ref{eq:xyreps}) and $\varphi\in\text{fix}(\Ht)$, the above equation becomes
\begin{equation}\label{eq:apponec2}
h\partial \mathcal{I}_\alpha(\varphi) =\partial \mathcal{X}(\varphi)^\T h_{\mathcal{X}} \mathcal{Y}(\varphi)+\mathcal{X}(\varphi)^\T h_{\mathcal{Y}}^\T\partial \mathcal{Y}(\varphi).
\end{equation}
Applying (\ref{eq:coroll1}) on (\ref{eq:apponec2}), we obtain
\begin{equation}
\begin{split}
h\partial \mathcal{I}_\alpha(\varphi) &= \partial \mathcal{X}(\varphi)^\T \mathcal{Y}(\varphi)+\mathcal{X}(\varphi)^\T\partial \mathcal{Y}(\varphi)\\
&=\partial \mathcal{I}_\alpha(\varphi) \quad\forall \,\,\varphi\in\text{fix}(\Ht),
\end{split}
\end{equation}
i.e.,~the gradient of the term calculated anywhere in $\text{fix}(\Ht)$ also lies in $\text{fix}(\Ht)$, which implies that condition (\ref{eq:cond2}) is satisfied.

\onecolumngrid
\vspace{5mm}

\section{Examples}
\label{sec:apptwo}
\vspace{2mm}

\twocolumngrid

All our examples are based on the dihedral group $D_6$, the symmetric group $S_4$ and groups closely related to them. In our discussion, we utilize the following matrices as their generators:
\begin{equation}
\begin{split}
&\matwr=\!\left(\begin{matrix} -\frac{1}{2} & -\frac{\sqrt{3}}{2}\\
\frac{\sqrt{3}}{2} & -\frac{1}{2}
\end{matrix}\right)\!,\,
\sigz=\!\left(\begin{matrix} 1 & 0\\
0 & -1
\end{matrix}\right)\!,\,i\sigy=\!\left(\begin{matrix} 0 & 1\\
-1 & 0
\end{matrix}\right)\!,\!\!\\
&\mathbf s_3= \left(\begin{matrix} 1 & 0 & 0\\
0 & -1 & 0\\
0 & 0 & -1
\end{matrix}\right)\!, \,\, \mathbf t_3 = \left(\begin{matrix} 0 & 1 & 0\\
0 & 0 & 1\\
1 & 0 & 0
\end{matrix}\right)\!, \,\,\mathbf u_3 = -\left(\begin{matrix} 1 & 0 & 0\\
0 & 0 & 1\\
0 & 1 & 0
\end{matrix}\right)\!.
\end{split}
\end{equation}
We explain the prerequisites such as the stratification of a manifold under group action and the theorems by Michel, Golubitsky and Stewart in Example~\ref{sec:egone} using $D_6$ (the symmetry group of an equilateral triangle) as the example. To better understand these prerequisites, the reader may also go through the first part of \ref{sec:egfour} where we utlise the group $S_4\times Z_2$ (the symmetry group of a cube). Even though the examples in this appendix involve rather small groups, the procedure demonstrated here applies to finite groups in general and, in particular, to those utilized in flavour physics such as $\Delta(6n^2)$.

\subsection{The symmetry group of an equilateral triangle\\ $\mathbb{G}=D_{6}$}
\label{sec:egone}

\begin{figure}[]
\subfloat[\label{fig:D6a}]{%
  \includegraphics[width=0.48\columnwidth]{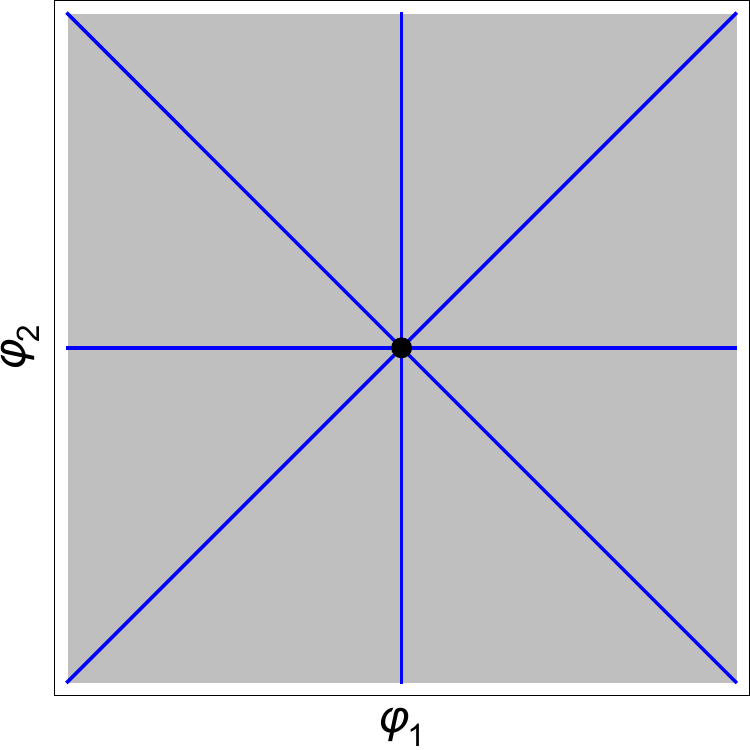}%
}\hspace{0.1cm}
\subfloat[\label{fig:D6b}]{%
  \includegraphics[width=0.48\columnwidth]{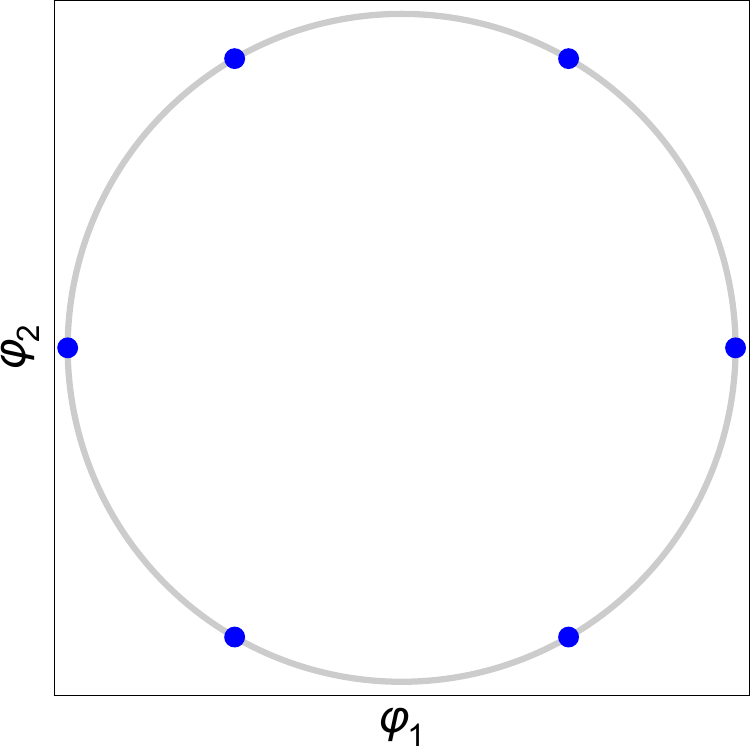}%
}
\caption{(a) Stratification of $\mathbbm R^2$ under $D_6$ from Example~\ref{sec:egone}.\hspace{1cm} (b) Stratification of circle (\ref{eq:circle}) under $\Ga=D_6$ from Example~\ref{sec:egone}.}\label{fig:D6}
\end{figure}

We begin with the dihedral group $D_6$ and a field $\varphi=({\varphi}_1,{\varphi}_2)$ transforming as its irreducible doublet. We use
\begin{equation}\label{eq:D6gen}
\mathbf T \equiv\matwr, \quad \mathbf U \equiv \sigz.
\end{equation}
as the generators of the doublet representation of $D_6$. This representation acts on the real plane $\mathbb{R}^2$. The stratification of $\mathbb{R}^2$ under $D_6$ is shown in FIG~\ref{fig:D6a}. There are three strata: the origin (black), the rays (blue) and the rest of the plane (grey). The origin forms an orbit (under $D_6$) with just one point. It remains invariant under the action of every element of the group, i.e., its stabilizer is the whole group. No other point with such a stabilizer exists, hence the origin forms a stratum with just one point. Any three points equidistant from the origin and lying in the rays that are at angles $0$, $\frac{2\pi}{3}$ and $-\frac{2\pi}{3}$ form an orbit under $D_6$. Similarly, a set of three equidistant points lying in the rays at angles $\pi$, $-\frac{\pi}{3}$ and $\frac{\pi}{3}$ also form an orbit. The stabilizers of such three points are the $Z_2$ subgroups generated by $\sigz$, $\sigz\matwr$ and $\sigz\matwr^2$, respectively. These subgroups are all conjugate to each other, i.e., $\matwr(\sigz)\matwr^\T=\sigz\matwr$, $\matwr(\sigz\matwr)\matwr^\T=\sigz\matwr^2$. In other words, the stabilizers of all points in the rays are equal up to conjugation. Therefore, the blue region forms a stratum. Any orbit in the grey region consists of a set of six points. The stabilizers of all such points are trivial. Hence the whole grey region forms a stratum. 

Let us discuss some mathematical definitions and general facts related to strata. The stabilizers of all points in an orbit are equal up to conjugation, i.e., they form a conjugacy class of stabilizers. All orbits with the same conjugacy class of stabilizers form a stratum. The manifold on which the group acts stratifies under the group action, i.e., the manifold can be expressed as a union of strata which are disjoint sets. Two strata $S_1$ and $S_2$ are neighbouring strata if points in one of them, say, $S_2$, are the limit points of the other, say, $S_1$. In that case, the stabilizer of every point in $S_1$ will be a subgroup of the stabilizer of every point in $S_2$ up to conjugation. In other words, a stratum constituted by the limit points of another stratum will have a higher class of symmetries than the other. The union of $S_1$ and all its neighbouring strata with higher symmetries than $S_1$ (such as $S_2$) form the closure of $S_1$. The stratum with the lowest class of symmetries is called the generic stratum. The generic stratum is open and dense, so the whole manifold forms its closure. It is the only stratum whose dimension is equal to that of the manifold. If none of the neighbouring strata has higher symmetries than a given stratum, it is called a maximal symmetry stratum. We may have more than one maximal symmetry stratum in a stratified manifold. If and only if a stratum is closed, it is a maximal symmetry stratum. In FIG~\ref{fig:D6a}, we have one maximal symmetry stratum, i.e.,~the black stratum. The grey stratum is the generic stratum. The black stratum is closed. The closure of the blue stratum is the union of the blue and the black strata. The closure of the grey stratum is $\mathbb{R}^2$.

It is possible to establish the existence of stationary points (or manifolds) of a potential invariant under a group by simply studying the stratification of the space under the group action. Such an approach is quite helpful as it does not require explicitly extremizing the potential. Michel's theorem\cite{MICHEL200111, MICHEL1971} forms the basis of this approach. It is given by: {\it Every maximal symmetry stratum is guaranteed to contain at least one stationary orbit.} Case~a) {\it If the maximal symmetry stratum contains a finite number of orbits, all of them are stationary.} An equivalent statement is: {\it If an orbit is isolated in its stratum, it is stationary.}   Case~b) {\it If the maximal symmetry stratum contains an infinite (continuous) set of orbits, at least two stationary orbits are guaranteed to be present in every connected component of the stratum.} As corollaries to this theorem, we can make the following statements. Case~a) {\it For a potential $\V$ constructed with $\varphi$ and invariant under $G$, a point $\boldsymbol \varphi$ (or a connected part of a $G$-orbit, say, $S\boldsymbol \varphi$) having the stabilizer (or the pointwise stabilizer) $H\subseteq G$ is guaranteed to be a stationary point (or a stationary manifold), if no point with stabilizer $H$ exists infinitesimally close to $\boldsymbol \varphi$ (or $S\boldsymbol \varphi$).} The black point in FIG~\ref{fig:D6a} is such a guaranteed stationary point. We describe the case of a guaranteed higher-dimensional stationary manifold in Example~\ref{sec:egseven}. Case~b) {\it If all points (or connected parts of $G$-orbits) in a connected closed subspace have the same stabilizer (or pointwise stabilizer) $H\subseteq G$, and no point with stabilizer $H$ exists infinitesimally close to it, then at least two stationary points (or manifolds) are guaranteed to be present in it.} For such a case, see Example~\ref{sec:egfive}.

Now, let us return to our discussion on $D_6$ and the doublet $\varphi$ transforming under it. The general renormalizable potential constructed with $\varphi$ is given by
\begin{equation}\label{eq:dhpot}
\V= c_1 m^2 |\varphi|^2+ c_2 |\varphi|^4 + c_3 m\, \varphi^{2\alpha\,\T}\varphi,
\end{equation}
where $\varphi^{2\alpha}=({\varphi}^2_1-{\varphi}^2_2, -2{\varphi}_1{\varphi}_2)^\T$ is a doublet which transforms in the same way as $\varphi$. The symmetry group of $\V$ is $G=D_6$. We have $\setn=\{1,2,3\}$ and $c=(c_1,c_2,c_3)$. The origin, FIG~\ref{fig:D6a}, has higher symmetries than every point in its neighbourhood. Therefore, by Michel's theorem, a general potential constructed with $D_6$ symmetry must have a stationary point at the origin. It can be verified that the potential (\ref{eq:dhpot}) has a stationary point at the origin for all values of the coefficients $c$. Nevertheless, the origin is not very interesting phenomenologically. 

Let us split $\V$ (\ref{eq:dhpot}) into two parts,
\begin{align}
\Vb&= c_1 m^2 |\varphi|^2+ c_2 |\varphi|^4,\label{eq:dhpotb1}\\
\V&=\Vb+ c_3  m\, \varphi^{2\alpha\,\T}\varphi,\label{eq:dhpotb2}
\end{align}
in accordance with the left route in FIG~\ref{fig:irrep}. We have $\setnb=\{1,2\}$. The symmetry group of $\Vb$ is $O(2)$. Let $\Cb$, with elements $\cb=(c_1,c_2)$, be the subset of $\mathbb{R}^2$ satisfying the conditions $c_1<0$ and $c_2>$. The stationary manifold (minima) of $\Vb$ is a circle which we express as the map $\varphiob$ with the domain $\Cb$,
\begin{equation}\label{eq:circle}
\varphiob(\cb)(\theta)=r(\cb)(\cos \theta, \sin \theta)^\T.
\end{equation}
Here, $r(\cb)$, as a function of $\cb=(c_1, c_2)$, denotes the radius of the circle. The Hessian becomes singular at $c_1=0$ as well as $c_2=0$, and hence they form the boundary of the domain $\Cb$. We may rewrite $\Vb$ (\ref{eq:dhpotb1}) as
\begin{equation}\label{eq:dhpotb3}
\Vb = k_1(\cb) (|\varphi|^2- r(\cb)^2)^2-k_1(\cb)r(\cb)^4,
\end{equation}
where $k_1(\cb)$ is positive. This form of $\Vb$ manifestly shows that the minima of the potential is a circle of radius $r(\cb)$. By comparing (\ref{eq:dhpotb3}) with (\ref{eq:dhpotb1}), we can obtain the expressions for $r(\cb)$ and $k_1(\cb)$ as functions of $\cb=(c_1, c_2)$. We call $r(\cb)$ and $k_1(\cb)$ arbitrary constants. 

According to (\ref{eq:irrepcase}), we have
\begin{equation}
\Hb=1, \quad\Tb=O(2).
\end{equation}
To summarise, we have got $\varphiob(\cb)$, $\Hb$ and $\Tb$. Introducing the third term (\ref{eq:dhpotb2}) breaks $\Tb=O(2)$ to $ \Ga=D_6$. This results in bifurcation.

Let us consider the stratification of the circle (\ref{eq:circle}) under $\Ga=D_6$, as shown in FIG~\ref{fig:D6b}. The six blue points constitute one stratum, and the rest of the circle (the grey arcs) constitutes the generic stratum. The six points form two orbits, which are isolated in the blue stratum. Hence, the blue stratum is a maximal symmetry stratum containing a finite number of orbits. Therefore, Michel's theorem\footnote{We apply Michel's theorem on the circle after removing the radial dependence, as explained in Ref.\cite{GAETA2006322}. This is possible because of the reduction lemma by Golubitsky and Stewart\cite{GOLUBITSKY1988}, with the help of which we can apply Michel's theorem on a submanifold.} guarantees the existence of stationary points of $\V$ (\ref{eq:dhpotb2}) in relation to the blue points. 

Cosider a blue point, say, $\mathcal{M}=\varphiob(\cb)(\frac{2\pi}{3})$. This point breaks $ \Ga=D_6$ to its pointwise stabilizer (\ref{eq:bfpnt}) $\Ha=Z_2$ generated by $\sigz\matwr$. Since $\mathcal{M}$ is a point, as opposed to a higher-dimensional manifold, its setwise stabilizer (\ref{eq:bfset}) is equal to its pointwise stabilizer, i.e., $\Sa=Z_2$.  According to the left route in FIG~\ref{fig:irrep}, we have $\Ht=\Ha$ and $\Tt=\Sa/\Ha$, i.e., we obtain $\Ht=Z_2$ and $\Tt=1$. With this information, we obtain the stationary point of $\V$ in relation to $\mathcal{M}$ as
\begin{equation}\label{eq:d6min}
\varphio(c)=r(c)\big(\cos\frac{2\pi}{3},\sin\frac{2\pi}{3}\big)^\T,
\end{equation}
where the radius $r(c)$ is a function of $c=(c_1, c_2, c_3)$. The domain of $\varphio(c)$, i.e., $\C$, has the structure as given in FIG~\ref{fig:twob}, i.e., $c_3=0$ corresponds to the red line forming the boundary of $\C$. The HLIC in $\varphio(c)$ is given by $\left(-\sin\frac{2\pi}{3},\cos\frac{2\pi}{3}\right)\varphio(c)=0$ which can be obtained from the equation $\sigz\matwr\varphio(c)=\varphio(c)$. The orthogonal complement of the fixed-point subspace of $\Ht$, i.e., $\text{fix}(Z_2)^\perp$, is nothing but the one-dimensional space spanned by the vector $\left(-\sin\frac{2\pi}{3},\cos\frac{2\pi}{3}\right)^\T$. 

We may rewrite $\V$ (\ref{eq:dhpotb2}) as
\begin{equation}\label{eq:dhpotb4}
\begin{split}
\V &= k_1(c) (|\varphi|^2- r(c)^2)^2-k_1(c)r(c)^4\\
&\qquad\qquad\qquad\qquad+ k_2(c)|\varphi^{2\alpha}-r(c)\varphi|^2,
\end{split}
\end{equation}
with $k_1(c)$ and $k_2(c)$ positive, to manifestly show that the stationary point (\ref{eq:d6min}) can be obtained as a minimum. $\V$ (\ref{eq:dhpotb4}) is bounded from below. At its minimum, we have $|\varphi|^2=r(c)^2$ and $\varphi^{2\alpha}=r(c)\varphi$. These two equations have a discrete set of solutions, and (\ref{eq:d6min}) is one among them. The functional dependence of the arbitrary constants $r(c)$, $k_1(c)$ and $k_2(c)$ on $c=(c_1,c_2,c_3)$ can be explicitely obtained by comparing (\ref{eq:dhpotb4}) with (\ref{eq:dhpotb2}).

\subsection{The symmetry group of a regular hexagon\\ $\mathbb{G}=D_{12}$}
\label{sec:egtwo}

Consider the group $\mathbb{G}=D_{12}$, which is isomorphic to $D_{6}\times Z_2$, generated by 
\begin{equation}\label{eq:reppi3}
\mathbf T \equiv \matwr, \quad \mathbf U \equiv \sigz, \quad \mathbf Z_2\equiv -1.
\end{equation}
Let $\phi=(\phi_1, \phi_2)^\T$ be a field transforming under the above representation. $D_{12}$ has an unfaithful representation that corresponds to 
\begin{equation}\label{eq:rep2pi3}
\mathbf T \equiv 1, \quad \mathbf U \equiv 1, \quad \mathbf Z_2\equiv -1.
\end{equation}
Let $\eta$ be a singlet transforming under the above representation. We combine $\phi$ and $\eta$ into a reducible multiplet and call it $\varphi$, i.e., $\varphi=(\phi,\eta)^\T$. The general fourth order potential constructed with $\varphi$ is given by
\begin{equation}\label{eq:dhpot2}
\begin{split}
\V&=c_1 m^2 |\phi|^2+ c_2 |\phi|^4 + c_3 m^2 \eta^2+ c_4 \eta^4 \\
&\qquad\qquad\qquad\qquad\quad+ c_5 \phi^{2\alpha\,\T}\phi\eta+ c_6 |\phi|^2\eta^2.
\end{split}
\end{equation}
Here, $\phi^{2\alpha}=\left(\phi^2_1-\phi^2_2, -2{\phi}_1{\phi}_2\right)^{\!\T}$ transforms in the same way as $\phi\eta$. The symmetry group of $\V$ is $G=D_{12}$. We split the potential in the following way
\begin{align}
\V_\phi&=c_1 m^2 |\phi|^2+ c_2 |\phi|^4,\label{eq:eg21}\\
\V_\eta&=c_3 m^2 \eta^2+ c_4 \eta^4,\\
\Vb&=\V_\phi+\V_\eta,\label{eq:eg23}\\
\Vt&=\Vb+c_5\phi^{2\alpha\,\T}\phi\eta,\label{eq:eg24}\\
\V&=\Vt+c_6 |\phi|^2\eta^2.\label{eq:eg25}
\end{align}
We define $\setn_\phi=\{1,2\}$, $\setn_\eta=\{3,4\}$, $\setnb=\{1,..,4\}$, $\setnt=\{1,..,5\}$, $\setn=\{1,..,6\}$, $c_\phi=(c_1,c_2)$, $c_\eta=(c_3,c_4)$, $\cb=(c_1,..,c_4)$, $\ct=(c_1,..,c_5)$ and $c=(c_1,..,c_6)$.

The construction of $\V_\phi$ is in accordance with the right route in FIG~\ref{fig:irrep}. It has the symmetry group $O(2)_\phi$. We obtain its minima as a circle,
\begin{equation}\label{eq:minb}
\phio(c_\phi)(\theta)=r_{\!\phi}(c_\phi)(\cos\theta,\sin\theta)^\T,
\end{equation}
with the domain, say, $\C_\phi$ given by $c_1<0$ and $c_2>0$. We have $\Ht_\phi=1$ and $\Tt_\phi=O(2)_\phi$. We may rewrite $\V_\phi$ as 
\begin{equation}\label{eq:eg27}
\V_\phi = k_1(c_\phi) (|\phi|^2- r_{\!\phi}(c_\phi)^2)^2-k_1(c_\phi)r_{\!\phi}(c_\phi)^4,
\end{equation}
to make the minima manifest. Comparing (\ref{eq:eg27}) with (\ref{eq:eg21}), we can calculate the functional dependence of the arbitrary constants $r_{\!\phi}(c_\phi)$ and $k_1(c_\phi)$ on $c_\phi=(c_1,c_2)$. 

\begin{figure}[]
\subfloat[\label{fig:D12}]{%
  \includegraphics[width=0.48\columnwidth]{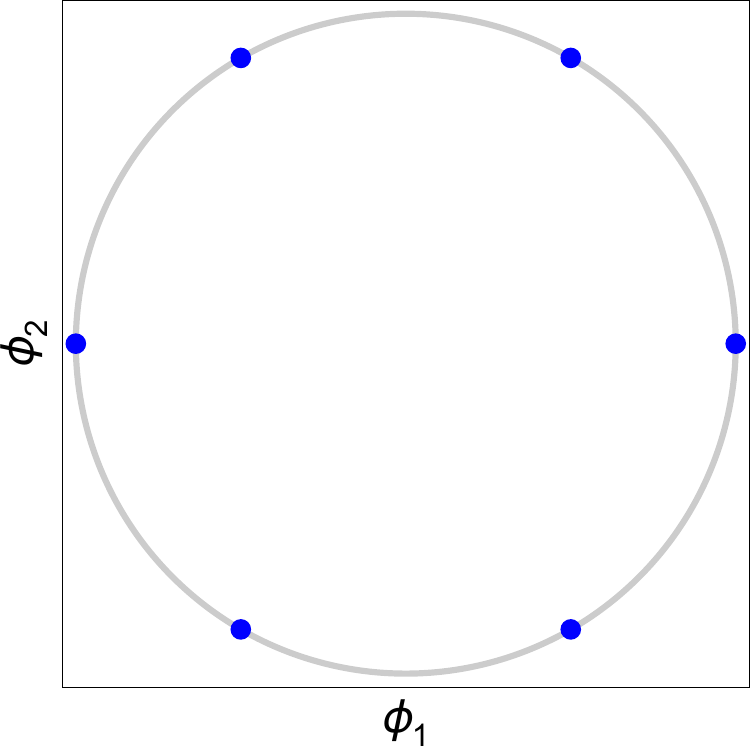}%
}\hspace{0.1cm}
\subfloat[\label{fig:D24}]{%
  \includegraphics[width=0.48\columnwidth]{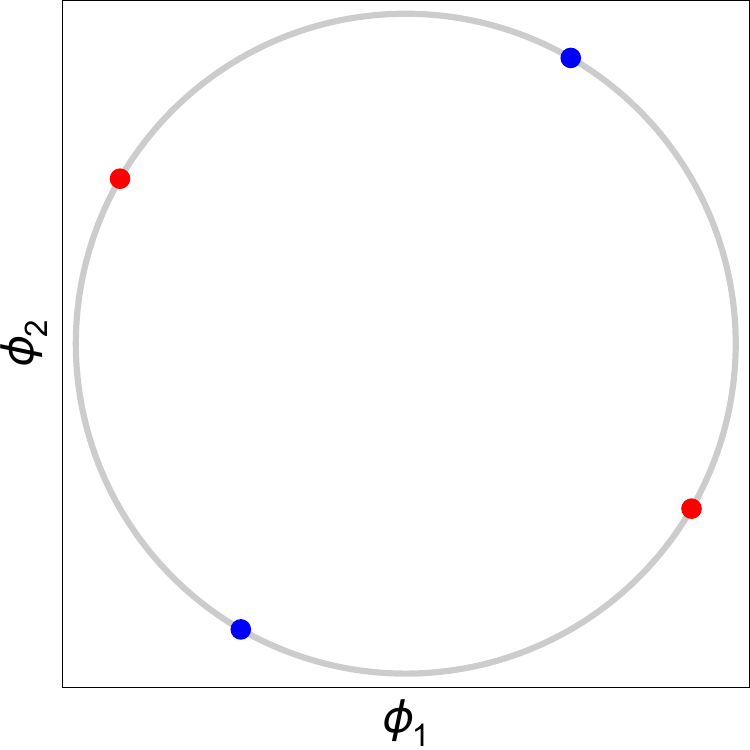}%
}
\caption{(a) Stratification of circle (\ref{eq:dh12circle}) under $\Ga=D_{6\phi}$ from  Example~\ref{sec:egtwo}. (b) Stratification of circle (\ref{eq:dh24circle}) under $\Ga=(Z_{2}'\times Z_{2}'')_\phi$ from  Example~\ref{sec:egthree}.}\label{fig:Dhigh}
\end{figure}

The symmetry group of $\V_\eta$ is $Z_2$. One of its minima is
\begin{equation}
\etao(c_\eta)=-r_{\!\eta}(c_\eta),
\end{equation}
with the domain $\C_\eta$ given by $c_3<0$ and $c_4>0$. We have $\Ht_\eta=1$ and $\Tt_\eta=1$. We may rewrite $\V_\eta$ as
\begin{equation}\label{eq:dh12potb}
\V_\eta= k_2(c_\eta) (\eta^2- r_{\!\eta}(c_\eta)^2)^2-k_2(c_\eta)r_{\!\eta}(c_\eta)^4.
\end{equation}

The construction of $\Vb$ (\ref{eq:eg23}), $\Vt$ (\ref{eq:eg24}) and $\V$ (\ref{eq:eg25}) follows the left route in FIG~\ref{fig:rerep}. For $\Vb$, we obtain the minima as the circle,
\begin{equation}\label{eq:dh12circle}
\varphiob(\cb)(\theta)=\left(r_{\!\phi}(c_\phi)(\cos\theta,\sin\theta), -r_{\!\eta}(c_\eta)\right)^\T,
\end{equation}
where the domain of $\varphiob(\cb)$ is $\Cb=\C_\phi\times\C_\eta$. We have $\Hb=\Ht_\phi\times \Ht_\eta=1$ and $\Tb= \Tt_\phi\times \Tt_\eta=O(2)_\phi$. 

We obtain $\Vt$ by adding the $5^\text{th}$ term to $\Vb$. This breaks $\Tb=O(2)_\phi$ to $\Ga=D_{6\phi}$ generated by the actions of $\matwr$ and $\sigz$ on $\phi$. This causes bifurcation and stratifies the circle (\ref{eq:dh12circle}), as shown in FIG~\ref{fig:D12}. The grey arcs form the generic stratum. The six blue points constitute the only maximal symmetry stratum. It contains two orbits of three points each. Symmetry arguments (Michel's theorem and Golubitsky and Stewart's reduction lemma) guarantee the existence of stationary points of $\Vt$ in relation to these points. Consider a point, say, $\mathcal{M}=\varphiob(\cb)(\frac{\pi}{3})$. The point $\mathcal{M}$ remains invariant under $\matw_r\sigz$, which generates a $Z_2$ group, say, $Z_{2\phi}$, i.e., $\mathcal{M}$ breaks $\Ga=D_{6\phi}$ to $\Ha=Z_{2\phi}$. We have $\Ht=\Ha\times\Hb=Z_{2\phi}$. Since $\Sa=\Ha=Z_{2\phi}$, we obtain $\Tt=\Sa/\Ha=1$. Therefore, in relation to $\mathcal{M}$, we obtain the stationary point,
\begin{equation}\label{eq:eg2mintilde}
\varphiot(\ct)=\big(r_{\!\phi}(\ct)(\cos\frac{\pi}{3},\sin\frac{\pi}{3}), -r_{\!\eta}(\ct)\big)^\T.
\end{equation}
The HLIC in $\varphiot(\ct)$ is given by $\text{fix}(\Ht)^\perp=\text{fix}(Z_{2\phi})^\perp$ which is the space spanned by $((-\sin\frac{\pi}{3},\cos\frac{\pi}{3}),0)^\T$. 

Rewriting $\Vt$ (\ref{eq:eg24}) as
\begin{equation}\label{eq:eg2Vtilde}
\begin{split}
\Vt&=k_1(\ct) (|\phi|^2- r_{\!\phi}(\ct)^2)^2-k_1(\ct)r_{\!\phi}(\ct)^4\\
&\quad +k_2(\ct) (\eta^2- r_{\!\eta}(\ct)^2)^2-k_2(\ct)r_{\!\eta}(\ct)^4\\
&\quad +k_3(\ct)\big|\phi^{2\alpha}\!\!-\frac{r_{\!\phi}(\ct)}{r_{\!\eta}(\ct)}\phi\eta\big|^2\!\!+\frac{k_3(\ct)}{2}\big(|\phi|^2\!\!-\frac{r_{\!\phi}(\ct)^2}{r_{\!\eta}(\ct)^2}\eta^2\big)^2,
\end{split}
\end{equation}
with $k_1(\ct)$,  $k_2(\ct)$ and  $k_3(\ct)$ positive, makes (\ref{eq:eg2mintilde}) manifestly a minimum of $\Vt$. By comparing (\ref{eq:eg2Vtilde}) with (\ref{eq:eg24}), we can obtain the five arbitrary constants $r_{\!\phi}(\ct)$, $r_{\!\eta}(\ct)$, $k_1(\ct)$,  $k_2(\ct)$ and  $k_3(\ct)$ as functions of five coefficeints $\ct$.

We obtain $\V$ (\ref{eq:eg25}) by adding the $6^\text{th}$ term to $\Vt$. The $6^\text{th}$ term is the product of norms. It is an invariant and hence compatible with $\Ht$. Also, $\Tt$ is trivial. Therefore, we are in the non-bifurcation scenario. The resulting minimum for the whole potential $\V$ is given by
\begin{equation}\label{eq:eg2min}
\varphio(c)=\big(r_{\!\phi}(c)(\cos\frac{\pi}{3},\sin\frac{\pi}{3}), -r_{\!\eta}(c)\big)^\T.
\end{equation}
$\Ht=\!Z_{2\phi}$ and $\Tt=\!1$ are preserved. We may rewrite $\V$ as
\begin{equation}\label{eq:eg2V}
\begin{split}
\V&=k_1(c) (|\phi|^2- r_{\!\phi}(c)^2)^2-k_1(c)r_{\!\phi}(c)^4\\
&\quad +k_2(c) (\eta^2- r_{\!\eta}(c)^2)^2-k_2(c)r_{\!\eta}(c)^4\\
&\quad +k_3(c)\big|\phi^{2\alpha}\!\!-\frac{r_{\!\phi}(c)}{r_{\!\eta}(c)}\phi\eta\big|^2\!\!+k_4(c)\big(|\phi|^2\!\!-\frac{r_{\!\phi}(c)^2}{r_{\!\eta}(c)^2}\eta^2\big)^2,
\end{split}
\end{equation}
to make the minimum (\ref{eq:eg2min}) manifest. The six arbitrary constants $r_{\!\phi}(c)$, $r_{\!\eta}(c)$, $k_1(c)$,..,$k_4(c)$ can be obtained in terms of the six coefficients $c$ by comparing (\ref{eq:eg2V}) with (\ref{eq:eg25}).

We assert that all stationary manifolds presented in this paper can be obtained as minima. To make these minima manifest, we can always rewrite the potentials as bounded from below in terms of the arbitrary constants, similar to (\ref{eq:eg27}), (\ref{eq:dh12potb}), (\ref{eq:eg2Vtilde}) and (\ref{eq:eg2V}). The number of arbitrary constants will be equal to the number of coefficients, and we can always obtain their functional dependence on the coefficients\footnote{In some cases, the solutions may not be in closed-form. Nevertheless, this is not an issue because we are not interested in their functional dependence on the coefficients. Even though the arbitrary constants that are not $k$'s appear in the expressions of the stationary manifolds, e.g.,~$r_{\!\phi}(c)$, $r_{\!\eta}(c)$ appear in $\varphio(c)$ (\ref{eq:eg2min}), the predictions claimed to be based on symmetries must be based on the HLICs only, which are coefficient-independent.} by comparing the potential with its rewritten form. In the rest of this paper, to keep our discussion short, we will rewrite the potentials using arbitrary constants only for the cases that we find interesting.

In constructing $\V$ (\ref{eq:dhpot2}), we used $\eta$ as a driving field. Driving fields were originally proposed in a supersymmetric flavour model\cite{hep-ph/0504165} as fields with $R$-charge equal to two. Since we do not wish to assume a supersymmetric framework, our definition of a driving field differs from that of \cite{hep-ph/0504165}. If a field does not appear in the fermion mass terms constructed at the leading order, we call it a driving field. It is introduced to help us in obtaining the desired minima for the other fields that do appear in the mass terms. The flavour model constructed in the companion paper\cite{2309.11542} extensively uses such driving fields. In the current work, we introduce driving fields to obtain the desired minima for the other fields that interest us, e.g.,~the driving field $\eta$ helps to obtain the minimum for $\phi$ (\ref{eq:eg2min}). Without $\eta$, the general potential constructed with $\phi$ will be nothing but $\V_\phi$ (\ref{eq:eg21}), whose symmetry group is $O(2)_\phi$. The introduction of $\eta$ serves to break this accidental continuous symmetry to $D_{6\phi}$. We use driving fields in a similar way in Examples~\ref{sec:egthree} and \ref{sec:egeight}, where they help us by breaking the accidental continuous symmetries. Besides this, they can help us to obtain the minima in two other ways, which we will discuss in Example~\ref{sec:egfour}.

\subsection{The symmetry group of a regular dodecagon\\ $\mathbb{G}=D_{24}$}
\label{sec:egthree}

Consider the group $\mathbb{G}=D_{24}$ generated by 
\begin{equation}\label{eq:reppi3}
\mathbf T \equiv \matwr, \quad \mathbf U \equiv \sigz, \quad \mathbf Z_2\equiv i\sigy.
\end{equation}
Let $\phi=(\phi_1, \phi_2)^\T$ be a field transforming under the above representation. As driving fields, we use a doublet $\rho=(\rho_1, \rho_2)^\T$ and a singlet $\eta$. They transform as the unfaithful representations of $D_{24}$ given by
\begin{align}\label{eq:rep2pi3}
&\mathbf T \equiv \matw_r, \quad \mathbf U \equiv \sigz, \quad \mathbf Z_2 \equiv -1,\quad\text{and}\\
&\mathbf T \equiv 1, \quad\,\,\,\, \mathbf U \equiv 1, \,\,\,\,\,\,\,\,\, \mathbf Z_2 \equiv -1,
\end{align}
respectively. The combined multiplet of $\rho$ and $\eta$ is named $\varrho$, i.e., $\varrho=(\rho,\eta)^\T$, and the combined multiplet of $\phi$ and $\varrho$ is named $\varphi$, i.e., $\varphi=(\phi,\varrho)^\T=(\phi,\rho,\eta)^\T$. We construct the various potentials involving these multiplets in the following way,
\begin{align}
\V_\phi&=c_1 m^2 |\phi|^2+ c_2 |\phi|^4,\label{eq:eg31}\\
\begin{split}\label{eq:eg32}
\V_\varrho&=c_3 m^2 |\rho|^2+ c_4 |\rho|^4 + c_5 m^2 \eta^2+ c_6 \eta^4\\
&\qquad\qquad\qquad\qquad\quad+ c_7 \rho^{2\alpha\,\T}\rho\eta+ c_8 |\rho|^2\eta^2,
\end{split}\\
\Vb&=\V_\phi+\V_\varrho,\label{eq:eg33}\\
\Vt&=\Vb+c_9 \,m\,\phi^{2\alpha\,\T}\rho,\label{eq:eg34}\\
\V&=\Vt+c_{10} |\phi|^2|\rho|^2+c_{11} |\phi|^2\eta^2,\label{eq:eg35}
\end{align}
where $\rho^{2\alpha}\!=\!({\rho}^2_1-{\rho}^2_2,-2{\rho}_1{\rho}_2)^{\T}$\!, $\phi^{2\alpha}\!=\!({\phi}^2_1-{\phi}^2_2, -2{\phi}_1{\phi}_2)^\T$. $\V$ (\ref{eq:eg35}) is the general renormalizable potential constructed with $\varphi$. For the purpose of this discussion, we use the notations: $c_\phi=(c_1,c_2)$, $c_\varrho=(c_3,..,c_8)$, $\cb=(c_1,..,c_8)$, $\ct=(c_1,..,c_9)$ and $c=(c_1,..,c_{11})$.

We have constructed $\V_\phi$ (\ref{eq:eg31}) in accordance with the right route in FIG~\ref{fig:irrep}. $\V_\phi$ has the symmetry group $O(2)_\phi$. We obtain its minima as a circle,
\begin{equation}
\phio(c_\phi)(\theta)=r_{\!\phi}(c_\phi)(\cos\theta,\sin\theta)^\T.
\end{equation}
We have $\Ht_\phi=1$ and $\Tt_\phi=O(2)_\phi$. $\V_\varrho$ (\ref{eq:eg32}) has the same form as the potential $\V$ (\ref{eq:eg25}) obtained in Example~\ref{sec:egtwo}. Therefore, for $\V_\varrho$, we may assign the minimum having the same form as (\ref{eq:eg2min}), i.e.,
\begin{equation}
\varrhoo(c_\varrho)=\big(r_{\!\rho}(c_\varrho)(\cos\frac{\pi}{3},\sin\frac{\pi}{3}), -r_{\!\eta}(c_\varrho)\big)^\T,
\end{equation}
with $\Ht_\varrho=Z_{2\rho}$ generated by $\rho\rightarrow\matw_r\sigz\rho$, and $\Tt_\varrho=1$.

The construction of $\Vb$ (\ref{eq:eg33}), $\Vt$ (\ref{eq:eg34}) and $\V$ (\ref{eq:eg35}) follows the left route in FIG~\ref{fig:rerep}. The minima of $\Vb$ is the circle,
\begin{equation}\label{eq:dh24circle}
\begin{split}
\varphiob(\cb)(\theta)&=\big(r_{\!\phi}(c_\phi)(\cos\theta,\sin\theta), \\
&\qquad\quad r_{\!\rho}(c_\varrho)(\cos\frac{\pi}{3},\sin\frac{\pi}{3}), -r_{\!\eta}(c_\varrho)\big)^{\!\T}\!\!.
\end{split}
\end{equation}
We have $\Hb=\Ht_\phi\times\Ht_\varrho=Z_{2\rho}$ and $\Tb=\Tt_\phi\times \Tt_\varrho=O(2)_\phi$. We obtain $\Vt$ by adding the $9^\text{th}$ term $\phi^{2\alpha\,\T}\rho$ (\ref{eq:eg34}). This term breaks $\Tb$ to $\Ga=(Z_{2}'\times Z_{2}'')_\phi$, where the actions of the generators of $Z_{2}'$ and $Z_{2}''$ are given by $\phi\rightarrow\matwr\sigz\phi$ and $\phi\rightarrow-\phi$, respectively. This leads to bifurcation. The circle (\ref{eq:dh24circle}) gets stratified under $\Ga=(Z_{2}'\times Z_{2}'')_\phi$, as shown in FIG~\ref{fig:D24}. We have three strata: blue, red and grey. The grey arcs form the generic stratum. The blue and red points are situated at angles $\frac{\pi}{3}+\frac{n\pi}{2}$. The stabilizers of the blue and red points are the $Z_2$ groups generated by $\matwr\sigz$ and $-\matwr\sigz$, respectively. These $Z_2$ groups are not conjugate to each other. Hence, the blue and red points form separate strata. These four points form two orbits isolated in their respective strata. Therefore, by symmetry arguments, stationary points of $\Vt$ must exist in relation to these points. 

We may choose one among them as $\mathcal M$, say, a red point $\mathcal{M}=\varphiob(\cb)(-\frac{\pi}{6})$. This choice breaks $\Ga=(Z_{2}'\times Z_{2}'')_\phi$ to, say, $Z_{2\phi}$, which is generated by the group action $\phi\rightarrow-\matwr\sigz\phi$. We have $\Ha=\Sa=Z_{2\phi}$. We obtain $\Ha\times\Hb=Z_{2\phi}\times Z_{2\rho}$. Since $\Hb$ is nontrivial, we need to verify if the term $\phi^{2\alpha\,\T}\rho$ is compatible with $\Ha\times\Hb=Z_{2\phi}\times Z_{2\rho}$. The generators of $Z_{2\phi}$ and $Z_{2\rho}$ acting on $\phi$ are $-\matwr\sigz$ and $1$, respectively, and acting on $\rho$ are $1$ and $\matwr\sigz$, respectively. Corresponding to $-\matwr\sigz$ and $1$ acting on $\phi$, we obtain $\matwr\sigz$ and $1$ acting on the multiplet $\phi^{2\alpha}$. This implies that the set of representation matrices of $Z_{2\phi}\times Z_{2\rho}$ acting on both $\phi^{2\alpha}$ and $\rho$ are the same, i.e., $\{1, \matwr\sigz\}$. Therefore, by corollary A (\ref{eq:coroll1}), the term $\phi^{2\alpha\,\T}\rho$ is compatible with $Z_{2\phi}\times Z_{2\rho}$. Thus, we obtain $\Ht=\Ha\times\Hb=Z_{2\phi}\times Z_{2\rho}$. We also have $\Tt=\Sa/\Ha=1$. To conclude, in relation to $\mathcal M$, we are guaranteed to obtain the stationary point,
\begin{equation}
\begin{split}
\varphiot(\ct)&=\big(r_{\!\phi}(\ct)(\cos(-\frac{\pi}{6}),\sin(-\frac{\pi}{6})),\\
&\qquad\qquad \quad r_{\!\rho}(\ct)(\cos\frac{\pi}{3},\sin\frac{\pi}{3}), -r_{\!\eta}(\ct)\big)^{\!\T}\!,
\end{split}
\end{equation}
for the potential $\Vt$ (\ref{eq:eg34}). The HLICs in $\varphiot(\ct)$ are given by the space $\text{fix}(Z_{2\phi}\times Z_{2\rho})^\perp$, which is the $2$-dimensional space spanned by the vectors $\left((\sin\frac{\pi}{6},\cos\frac{\pi}{6}),(0,0),0\right)^\T$ and $\left((0,0),(-\sin\frac{\pi}{3},\cos\frac{\pi}{3}), 0\right)^\T$.

Finally, we add the $10^\text{th}$ and $11^\text{th}$ terms to $\Vt$ to obtain the whole potential $\V$ (\ref{eq:eg35}). These two terms are invariants and hence compatible. Therefore, we are guaranteed to obtain the stationary point of $\V$,
\begin{equation}
\begin{split}
\varphio(c)&=\big(r_{\!\phi}(c)(\cos(-\frac{\pi}{6}),\sin(-\frac{\pi}{6})),\\
&\qquad\qquad\quad r_{\!\rho}(c)(\cos\frac{\pi}{3},\sin\frac{\pi}{3}), -r_{\!\eta}(c)\big)^{\!\T}\!,
\end{split}
\end{equation}
with $\Ht=Z_{2\phi}\times Z_{2\rho}$, $\Tt=1$ and the HLICs as obtained earlier.

\subsection{The symmetry group of a cube\\ $\mathbb{G}=S_{4}\times Z_2$}
\label{sec:egfour}

Consider the symmetry group of a cube, $S_4\times Z_2$, generated by
\begin{equation}\label{eq:s4gens}
\mathbf S\equiv \mathbf s_3, \quad \mathbf T\equiv \mathbf t_3, \quad \mathbf U\equiv -\mathbf u_3, \quad \mathbf Z_2 \equiv -1.
\end{equation}
$S_4$ has been used widely for model building in neutrino physics. Ref.~\cite{hep-ph/0508231, hep-ph/0602244, 0811.0345, 2003.00506} are a few examples where a real basis of $S_4$, as generated by $\mathbf s_3$, $\mathbf t_3$ and $\mathbf u_3$, was used. We construct the general renormalizable potential using  $\varphi=({\varphi}_1,{\varphi}_2,{\varphi}_3)^{\T}$ transforming as the triplet (\ref{eq:s4gens}),
\begin{equation}\label{eq:s4pot2}
\V= c_1 m^2 |\varphi|^2+ c_2 |\varphi|^4 + c_3 |\varphi^{2\alpha}|^2,
\end{equation}
where $\varphi^{2\alpha}=(-2{\varphi}_1^2+{\varphi}_2^2+{\varphi}_3^2, \sqrt{3}{\varphi}_2^2-\sqrt{3}{\varphi}_3^2)^\T$ and $|\varphi^{2\alpha}|^2=\varphi^{2\alpha\,\T}\varphi^{2\alpha}$. The doublet $\varphi^{2\alpha}$ transforms as 
\begin{equation}\label{eq:s4d6gens}
\mathbf S\equiv 1, \quad \mathbf T\equiv \mathbf \matw_r, \quad \mathbf U\equiv \sigz, \quad \mathbf Z_2 \equiv 1,
\end{equation}
which is isomorphic to the doublet of $D_6$ (\ref{eq:D6gen}). 

\begin{figure}[]
\subfloat[\label{fig:S41}]{%
  \includegraphics[width=0.5\columnwidth]{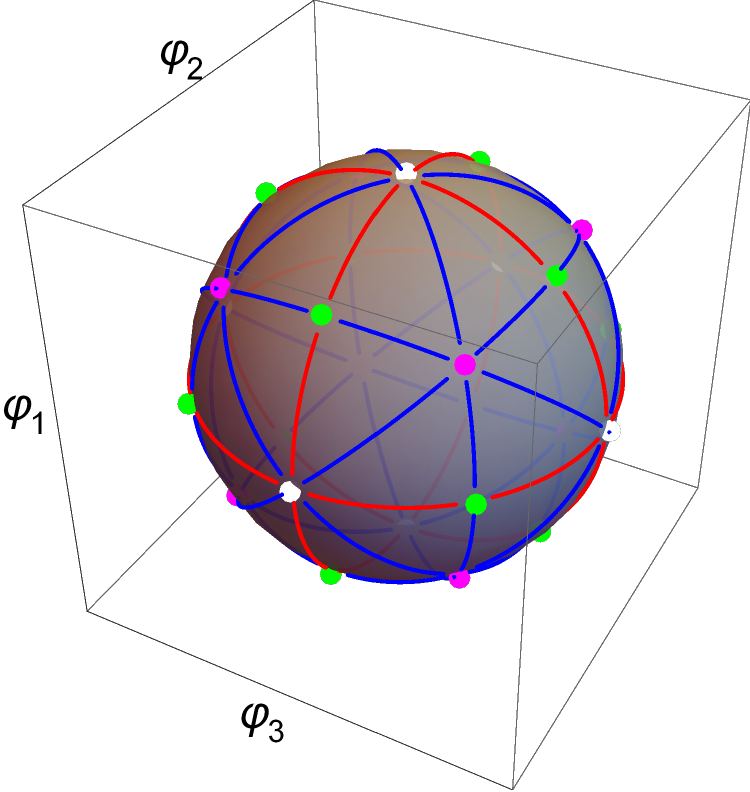}%
}
\subfloat[\label{fig:S42}]{%
  \includegraphics[width=0.5\columnwidth]{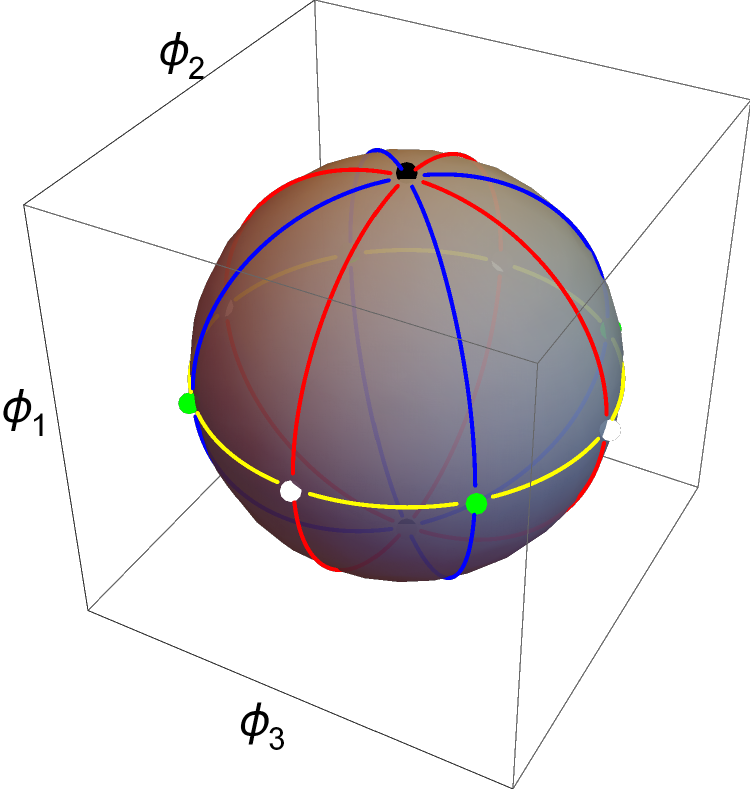}%
}
\caption{(a) Stratification of $2$-sphere (\ref{eq:eg4sphere}) under $\Ga=S_4\times Z_2$ from  Example~\ref{sec:egfour}. (b) Stratification of $2$-sphere (\ref{eq:eg4asphere}) under $\Ga=(Z_{2}\times Z'_{2}\times Z''_{2}\rtimes Z_{2}''')_\phi$ from  Example~\ref{sec:egfoura}.}\label{fig:S4}
\end{figure}

Following the left route in FIG~\ref{fig:irrep}, we split the potential as
\begin{align}
\Vb&=c_1 m^2 |\varphi|^2+ c_2 |\varphi|^4,\label{eq:eg41}\\
\V&=\Vb+c_3 |\varphi^{2\alpha}|^2.\label{eq:eg42}
\end{align}
We define $\cb=(c_1, c_2)$ and $c=(c_1, c_2, c_3)$. The symmetry group of $\Vb$ is $O(3)$. In the domain $c_1<0$, $c_2>0$, we obtain the minima of $\Vb$ as a 2-sphere,
\begin{equation}\label{eq:eg4sphere}
\varphiob(\cb)(\tht)=r(\cb)(\sin \theta_1 \cos \theta_2, \sin \theta_1 \sin \theta_2, \cos \theta_1)^\T \!,\!\!
\end{equation}
where $\tht$ consists of the parameters $\theta_1$ and $\theta_2$. We have $\Hb=1$ and $\Tb=O(3)$. 

The third term (\ref{eq:eg42}) breaks $\Tb=O(3)$ to $ \Ga=S_4\times Z_2$ causing bifurcation. The stratification of the $2$-sphere (\ref{eq:eg4sphere}) under the action of $S_4\times Z_2$ is shown in FIG~\ref{fig:S41}. We have the grey generic stratum. There are two kinds of one-dimensional strata (blue and red), which form arcs in the sphere. The blue stratum includes points such as $\propto(a,b,b)$, and the red stratum includes points such as $\propto(0,a,b)$. We also have points that form orbits isolated in their strata. These include points such as $\propto(1,1,1)$ (pink stratum), $\propto(0,1,1)$ (green stratum) and $\propto(1,0,0)$ (white stratum) corresponding to the cube's vertices, edge centres and face centres, respectively. By symmetry arguments, these points are guaranteed to be stationary points of $\V$ (\ref{eq:eg42}). 

Consider a pink point, say,
\begin{equation}\label{eq:cubem1}
\mathcal{M}=\frac{r(\cb)}{\sqrt{3}}(1,1,1).
\end{equation}
$\mathcal{M}$ remains invariant under $\matt_3\equiv\mathbf T$ and $-\matu_3\equiv\mathbf U$. The matrices $\mathbf t_3$ and $-\mathbf u_3$ generate the permutations of three objects, i.e., they generate the group $S_3$, which is isomorphic to $D_6$. This group forms the pointwise and the setwise stabilizer of $\mathcal{M}$ under $\Ga$, i.e., $\Ha=\Sa=D_6$. Therefore, we obtain $\Ht=\Ha=D_6$, and $\Tt=\Sa/\Ha=1$. The guaranteed stationary point of $\V$ in relation to $\mathcal{M}$ is
\begin{equation}\label{eq:cubemin1}
\varphio(c)=\frac{r(c)}{\sqrt{3}}\left(1,1,1\right)^\T.
\end{equation}
The HLICs in $\varphio(c)$ correspond to $\text{fix}(\Ht)^\perp=\text{fix}(D_6)^\perp$, which is the space spanned by $\frac{1}{\sqrt{6}}(2,-1,-1)^{\T}$ and $\frac{1}{\sqrt{2}}(0,1,-1)^{\T}$. 

$\V$ (\ref{eq:eg42}) can be rewritten as 
\begin{equation}\label{eq:cubenewpot1}
\V=k_1(c) (|\varphi|^2- r(c)^2)^2-k_1(c)r(c)^4+k_2(c)|\varphi^{2\alpha}|^2.
\end{equation}
This expression is bounded from below and attains a minimum at (\ref{eq:cubemin1}). By comparing (\ref{eq:cubenewpot1}) with (\ref{eq:eg42}), we can obtain the arbitrary constants $r(c)$, $k_1(c)$ and $k_2(c)$ as functions of $c=(c_1, c_2, c_3)$.

Instead of (\ref{eq:cubem1}), we may choose a white point as $\mathcal{M}$, say,
\begin{equation}\label{eq:cubem2}
\mathcal{M}=r(\cb)(1,0,0). 
\end{equation}
$\mathcal{M}$ (\ref{eq:cubem2}) remains invariant under $\mats_3\equiv\mathbf S$, $-\matt_3\mats_3\matt_3^2\equiv\mathbf Z_2\mathbf {TST}^2$ and $-\matu_3\equiv\mathbf U$. The matrices $\mathbf s_3$, $-\mathbf t_3\mathbf s_3\mathbf t_3^2$ and $-\mathbf u_3$ individually generate $Z_2$ groups, say, $Z_2'$, $Z_2''$ and $Z_2'''$, respectively. When taken together, they generate $Z_2'\times Z_2''\rtimes Z_2'''$, i.e., $\mathcal{M}$ (\ref{eq:cubem2}) breaks $\Ga=S_4\times Z_2$ to $\Ha=\Sa=Z_2'\times Z_2''\rtimes Z_2'''$. This leads to $\Ht=Z_2'\times Z_2''\rtimes Z_2'''$ and $\Tt=1$. Corresponding to (\ref{eq:cubem2}), we obtain the guaranteed stationary point,
\begin{equation}\label{eq:cubemin2}
\varphio(c)=r(c)\left(1,0,0\right)^\T.
\end{equation}
The HLICs in $\varphio(c)$ (\ref{eq:cubemin2}) are given by $\text{fix}(Z_2'\times Z_2''\rtimes Z_2''')^\perp$, which is the space spanned by $(0,1,0)^{\T}$ and $(0,0,1)^{\T}$. In this case, we may rewrite the potential as
\begin{equation}\label{eq:cubenewpot2}
\V=k_1(c) (|\varphi|^2- r(c)^2)^2-k_1(c)r(c)^4+k_2(c)|\varphi^{2\beta}|^2,
\end{equation}
where $\varphi^{2\beta}\!=\!({\varphi}_2 {\varphi}_3,{\varphi}_3 {\varphi}_1,{\varphi}_1 {\varphi}_2)^\T$ and $|\varphi^{2\beta}|^2\!=\!\varphi^{2\beta\,\T}\varphi^{2\beta}$, so that (\ref{eq:cubemin2}) is manifestly a minimum. By comparing (\ref{eq:cubenewpot2}) with (\ref{eq:eg42}), we can obtain the arbitrary constants $r(c)$, $k_1(c)$ and $k_2(c)$ in (\ref{eq:cubenewpot2}) as functions of $c=(c_1, c_2, c_3)$. Note that there are only two independent quartic invariants. We used $|\varphi|^4$ and $|\varphi^{2\alpha}|^2$ in (\ref{eq:eg42}), (\ref{eq:cubenewpot1}), while we have used $|\varphi|^4$ and $|\varphi^{2\beta}|^2$ in (\ref{eq:cubenewpot2}). 

Symmetry arguments guarantees that $\V$ (\ref{eq:eg42}) has stationary points corresponding to the green stratum also. However, we can show that no domain in the coefficient-space exists where the green stationary points form minima. Rather, they always appear as saddle points. In this paper, we first obtain an $n$-sphere as the minima and then break the orthogonal group to obtain stationary manifolds with respect to the $n$-sphere. Hence, the resulting stationary manifolds can be either minima or saddle points. In situations where every domain in the coefficient space corresponds to the stationary manifolds being saddle points only, it is still possible to obtain them as minima if we introduce driving fields. This is the second way the driving fields can be used, i.e., to change the nature of the stationary manifolds from being saddle points to minima. We will demonstrate this in relation to the green stratum in Example~\ref{sec:egfoura}. 

Stationary manifolds may or may not be present in non-maximal symmetry strata. To establish their presence in these strata, we need to explicitly extremize the potential in a case-by-case manner since Michel's theorem is not applicable here. For example, by explicitly extremising $\V$ (\ref{eq:eg42}), we can show that no domain exists in the coefficient space that leads to stationary points in the blue, red or grey strata. Nevertheless, in Example~\ref{sec:egfourb}, we obtain a minimum in the blue stratum by introducing a driving field. In general, we can obtain minima in non-maximal symmetry strata by introducing driving fields, which is the third way we can use them.

\subsubsection{Obtaining the green stationary points as minima}
\label{sec:egfoura}

Let the triplet $\phi=(\phi_1, \phi_2, \phi_3)^\T$ transform as (\ref{eq:s4gens}). To obtain minima at the green points, we use a driving field $\rho=(\rho_1, \rho_2, \rho_3)^\T$, which transforms as 
\begin{equation}
\mathbf S\equiv \mathbf s_3, \quad \mathbf T\equiv \mathbf t_3, \quad \mathbf U\equiv \mathbf u_3, \quad \mathbf Z_2 \equiv -1,
\end{equation}
We name the combined multiplet $\varphi$, i.e., $\varphi=(\phi,\rho)^\T$. We construct the potentials in the following way,
\begin{align}
\V_\phi&=c_1 m^2 |\phi|^2+ c_2 |\phi|^4,\label{eq:eg4a1}\\
\V_\rho&=c_3 m^2 |\rho|^2+ c_4 |\rho|^4 + c_5 |\rho^{2\alpha}|^2,\label{eq:eg4a2}\\
\Vb&=\V_\phi+\V_\rho,\label{eq:eg4a3}\\
\Vt&=\Vb+c_6|\phi^{2\alpha}|^2+c_7 \phi^{2\alpha\,\T}\rho^{2\alpha},\label{eq:eg4a4}\\
\V&=\Vt+c_8 \phi^{2\beta\,\T}\rho^{2\beta}+c_9|\phi|^2|\rho|^2,\label{eq:eg4a5}
\end{align}
where $\phi^{2\alpha}=(-2{\phi}_1^2+{\phi}_2^2+{\phi}_3^2, \sqrt{3}{\phi}_2^2-\sqrt{3}{\phi}_3^2)^\T$, $\rho^{2\alpha}=(-2{\rho}_1^2+{\rho}_2^2+{\rho}_3^2, \sqrt{3}{\rho}_2^2-\sqrt{3}{\rho}_3^2)^\T\!\!,$\, $\phi^{2\beta}\!=\!({\phi}_2{\phi}_3, {\phi}_3{\phi}_1, {\phi}_1{\phi}_2)^\T$ and $\rho^{2\beta}=({\rho}_2{\rho}_3, {\rho}_3{\rho}_1, {\rho}_1{\rho}_2)^\T$. $\V$ is the general renormalizable potential constructed using $\phi$ and $\rho$. We define $c_\phi=(c_1,c_2)$, $c_\rho=(c_3,c_4,c_5)$, $\cb=(c_1,..,c_5)$, $\ct=(c_1,..,c_7)$ and $c=(c_1,..,c_9)$.

$\V_\phi$ (\ref{eq:eg4a1}) is constructed in accordance with the right route in FIG~\ref{fig:irrep}. The minima of $\V_\phi$ is a $2$-sphere,
\begin{equation}
\phio(c_\phi)(\tht)=r_{\!\phi}(c_\phi)(\sin \theta_1 \cos \theta_2, \sin \theta_1 \sin \theta_2, \cos \theta_1)^\T\!,
\end{equation}
with $\Ht_\phi=1$ and $\Tt_\phi=O(3)_\phi$. The structure of $\V_\rho$ is the same as that of (\ref{eq:eg42}). Therefore, we obtain 
\begin{equation}
\rhoo(c_\rho)=r_{\!\rho}(c_\rho)\left(1,0,0\right)^\T,
\end{equation}
as one of its minima with $\Ht_\rho=(Z'_{2}\times Z''_{2}\rtimes Z_{2}''')_\rho$ generated by $\mats_3\equiv\mathbf S$, $-\matt_3\mats_3\matt_3^2\equiv\mathbf Z_2\mathbf {TST}^2$ and $-\matu_3\equiv\mathbf Z_2\mathbf U$, and with $\Tt_\rho=1$. 

The construction of $\Vb$ (\ref{eq:eg4a3}), $\Vt$ (\ref{eq:eg4a4}) and $\V$ (\ref{eq:eg4a5}) is in accordance with the left route in FIG~\ref{fig:rerep}. We obtain the $2$-sphere,
\begin{equation}\label{eq:eg4asphere}
\begin{split}
\varphiob(\cb)(\tht)&=\big(r_{\!\phi}(c_\phi)(\sin \theta_1 \cos \theta_2, \sin \theta_1 \sin \theta_2, \cos \theta_1)\\
&\qquad\qquad\qquad \qquad  r_{\!\rho}(c_\rho)\left(1,0,0\right)\big)^\T,
\end{split}
\end{equation}
as the minima of $\Vb$ with $\Hb=\Ht_\phi\times \Ht_\rho=(Z'_{2}\times Z''_{2}\rtimes Z_{2}''')_\rho$ and $\Tb=\Tt_\phi\times \Tt_\rho=O(3)_\phi$. 

We add the $6^\text{th}$ and $7^\text{th}$ terms to $\Vb$ to obtain $\Vt$ (\ref{eq:eg4a4}). These terms break $\Tb=O(3)_\phi$ to $\Ga=(Z_{2}\times Z'_{2}\times Z''_{2}\rtimes Z_{2}''')_\phi$ generated by $-1$, $-\mathbf s_3$, $-\mathbf t_3\mathbf s_3\mathbf t_3^2$ and $-\mathbf u_3$ acting on $\phi$. The $2$-sphere $\varphiob(\cb)(\tht)$ (\ref{eq:eg4asphere}) gets stratified under $(Z_{2}\times Z'_{2}\times Z''_{2}\rtimes Z_{2}''')_\phi$, as shown in FIG.~\ref{fig:S42}. Let us consider the following isolated green point in the $2$-sphere,
\begin{equation}
\mathcal{M}=\big(\frac{r_{\!\phi}(c_\phi)}{\sqrt{2}}(0,1,1),\,r_{\!\rho}(c_\rho)\left(1,0,0\right)\big)^\T.
\end{equation}
Its stabilizer under $\Ga$ is the Klein four-group generated by $-\mathbf s_3$ and $-\mathbf u_3$ acting on $\phi$, i.e., $\Ha=\Sa=(Z'_{2}\times Z'''_{2})_\phi$. We have $\Ha\times\Hb=(Z'_{2}\times Z'''_{2})_\phi\times (Z'_{2}\times Z''_{2}\rtimes Z_{2}''')_\rho$. The $6^\text{th}$ term $|\phi^{2\alpha}|^2$ is compatible with $\Ha\times\Hb$ since it remains invariant under it. To verify the compatibility of the $7^\text{th}$ term $\phi^{2\alpha\,\T}\rho^{2\alpha}$, let us examine the generators of $\Ha\times\Hb$ acting on $\phi$ and $\rho$. The generators of $Z'_{2\phi}$, $Z'''_{2\phi}$, $Z'_{2\rho}$, $Z''_{2\rho}$ and $Z_{2\rho}'''$ acting on $\phi$ are $-\mats_3$, $-\matu_3$, $1$, $1$ and $1$, respectively, and acting on $\rho$ are $1$, $1$, $\mats_3$, $-\matt_3\mats_3\matt_3^2$ and $-\matu_3$, respectively. Corresponding to the multiplet $\phi^{2\alpha}$, they are given by $1$, $\sigz$, $1$, $1$ and $1$, respectively, and corresponding to the multiplet $\rho^{2\alpha}$, they are given by $1$, $1$, $1$, $1$ and $\sigz$, respectively. In other words, the set of representation matrices of $\Ha\times\Hb$ acting on both $\phi^{2\alpha}$ and $\rho^{2\alpha}$ are the same, i.e., $\{1, \sigz \}$. Therefore, by corollary A, the $7^\text{th}$ term $\phi^{2\alpha\,\T}\rho^{2\alpha}$ is compatible with $\Ha\times\Hb$. Thus we obtain $\Ht=(Z'_{2}\times Z'''_{2})_\phi\times (Z'_{2}\times Z''_{2}\rtimes Z_{2}''')_\rho$. We also have $\Tt=\Sa/\Ha=1$. Hence, we obtain the guaranteed stationary point of $\Vt$ in relation to $\mathcal M$ as
\begin{equation}\label{eq:cubemin3}
\varphiot(\ct)=\big(\frac{r_{\!\phi}(\ct)}{\sqrt{2}}(0,1,1),\,r_{\!\rho}(\ct)\left(1,0,0\right)\big)^\T.
\end{equation}
The HLICs in $\varphiot(\ct)$ correspond to $\text{fix}(\Ht)^\perp$, the space spanned by $((1,0,0),(0,0,0))^\T$, $(\frac{1}{\sqrt{2}}(0,1,-1),(0,0,0))^\T$, $((0,0,0),(0,1,0))^\T$ and $((0,0,0),(0,0,1))^\T$.

To show that the green stationary point (\ref{eq:cubemin3}) is indeed a minimum, we rewrite $\Vt$ (\ref{eq:eg4a4}) in the form
\begin{equation}
\begin{split}
\Vt&=k_1(\ct) (|\phi|^2- r_{\!\phi}(\ct)^2)^2-k_1(\ct)r_{\!\phi}(\ct)^4\\
&\quad +k_2(\ct) (|\rho|^2\!- r_{\!\rho}(\ct)^2)^2-k_2(\ct)r_{\!\rho}(\ct)^4+k_3(\ct)|\rho^{2\beta}|^2\\
&\quad +k_4(\ct)\big|\phi^{2\alpha}+\frac{r_{\!\phi}(\ct)^2}{2r_{\!\rho}(\ct)^2}\rho^{2\alpha}\big|^2+k_5(\ct)(\phi^\T\rho)^2\!\!\!\\
&\quad +\frac{k_5(\ct)}{2}\frac{r_{\!\rho}(\ct)^2}{r_{\!\phi}(\ct)^2}\big(|\phi|^2-\frac{r_{\!\phi}(\ct)^2}{r_{\!\rho}(\ct)^2}|\rho|^2\big)^2+\frac{k_5(\ct)}{2}|(\phi\rho)^\alpha|^2,
\end{split}
\end{equation}
where $(\phi\rho)^\alpha\!=\!(-2\phi_1\rho_1+\phi_2\rho_2+\phi_3\rho_3, \sqrt{3}\phi_2\rho_2-\sqrt{3}\phi_3\rho_3)^\T$\!. This expression is bound from below. It contains seven arbitrary constants $r_{\!\phi}(\ct)$, $r_{\!\rho}(\ct)$, $k_1(\ct), .., k_5(\ct)$, each of which is a function of the seven coefficients $\ct=(c_1, .., c_7)$.

Finally, we add the $8^\text{th}$ and $9^\text{th}$ terms to $\Vt$ to obtain $\V$ (\ref{eq:eg4a5}). We can show that the $8^\text{th}$ term, $\phi^{2\beta\,\T}\rho^{2\beta}=\phi_2\phi_3\rho_2\rho_3+\phi_3\phi_1\rho_3\rho_1+\phi_1\phi_2\rho_1\rho_2$, does not satisfy corollories A and B of the compatibility condition. Therefore, we have to verify whether it satisfies the general condition, i.e., whether $\partial_i (\phi^{2\beta\,\T}\rho^{2\beta}) \in \text{fix}(\Ht) \quad\forall \,\,\varphi \in \text{fix}(\Ht)$. We have $\partial_i (\phi^{2\beta\,\T}\rho^{2\beta})=((\phi_3\rho_3\rho_1+\phi_2\rho_1\rho_2,\phi_3\rho_2\rho_3+\phi_1\rho_1\rho_2, \phi_2\rho_2\rho_3+\phi_1\rho_3\rho_1), (\phi_3\phi_1\rho_3+\phi_1\phi_2\rho_2, \phi_2\phi_3\rho_3+\phi_1\phi_2\rho_1, \phi_2\phi_3\rho_2+\phi_3\phi_1\rho_1))^\T$. The general form of a vector in the subspace $\text{fix}(\Ht)$ is $(\frac{a}{\sqrt{2}}(0,1,1),b(1,0,0))^\T$. At such a vector, $\partial_i (\phi^{2\beta\,\T}\rho^{2\beta})$ becomes $((0,0,0),(0,0,0))^\T$ which also lies in the subspace $\text{fix}(\Ht)$. Therefore, the $8^\text{th}$ term satisfies the compatibility condition. The $9^\text{th}$ is also compatible since it is an invariant. Therefore, we obtain the minimum of the potential $\V$ (\ref{eq:eg4a5}) as
\begin{equation}\label{eq:greenpoint}
\varphio(c)=\big(\frac{r_{\!\phi}(c)}{\sqrt{2}}(0,1,1),\,r_{\!\rho}(c)\left(1,0,0\right)\big)^\T,
\end{equation}
with $\Ht=(Z'_{2}\times Z'''_{2})_\phi\times (Z'_{2}\times Z''_{2}\rtimes Z_{2}''')_\rho$ and $\Tt=1$. The alignment $\phi=\frac{r_{\!\phi}(c)}{\sqrt{2}}(0,1,1)$ in the minimum (\ref{eq:greenpoint}) corresponds to a point in the green stratum in FIG~\ref{fig:S41}.

\subsubsection{Obtaining minima in the blue stratum}
\label{sec:egfourb}

As in the previous example, let us consider a triplet $\phi$ transforming as (\ref{eq:s4gens}). To obtain minima in the blue stratum, we introduce a driving field $\rho=({\rho}_1, {\rho}_2)^\T$, which transforms in the same way as the doublet $\phi^{2\alpha}=(-2{\phi}_1^2+{\phi}_2^2+{\phi}_3^2, \sqrt{3}{\phi}_2^2-\sqrt{3}{\phi}_3^2)^\T$, i.e., as (\ref{eq:s4d6gens}). Let $\varphi=(\phi, \rho)^\T$. We construct the potentials in the following way,
\begin{align}
\V_\phi&=c_1 m^2 |\phi|^2+ c_2 |\phi|^4 + c_3 |\phi^{2\alpha}|^2,\label{eq:eg4b1}\\
\V_\rho&=c_4 m^2 |\rho|^2+ c_5 |\rho|^4 + c_6  m\,\rho^{2\alpha\,\T}\rho,\label{eq:eg4b2}\\
\Vb&=\V_\phi+\V_\rho,\label{eq:eg4b3}\\
\Vt&=\Vb+c_7 m\, \phi^{2\alpha\,\T}\rho,\label{eq:eg4b4}\\
\V&=\Vt+c_8 \phi^{2\alpha\,\T}\rho^{2\alpha}+c_9|\phi|^2|\rho|^2,\label{eq:eg4b5}
\end{align}
where $\rho^{2\alpha}=\left({\rho}^2_1-{\rho}^2_2, -2{\rho}_1{\rho}_2 \right)^\T$. $\V$ is the general renormalisable potential constructed using $\phi$ and $\rho$. We define $c_\phi=(c_1,c_2,c_3)$, $c_\rho=(c_4,c_5,c_6)$, $\cb=(c_1,..,c_6)$, $\ct=(c_1,..,c_7)$ and $c=(c_1,..,c_9)$.

$\V_\phi$ (\ref{eq:eg4b1}) has the same form as $\V$ (\ref{eq:eg42}). Similar to (\ref{eq:cubemin1}), $\V_\phi$ (\ref{eq:eg4b1}) is guaranteed to have the minimum,
\begin{equation}
\phio(c_\phi)=\frac{r_{\!\phi}(c_\phi)}{\sqrt{3}}\left(1,1,1\right)^\T,
\end{equation}
with the associated groups $\Ht_\phi=D_{6\phi}$ generated by $\matt_3$ and $-\matu_3$, and $\Tt_\phi=1$. The potential $\V_\rho$ (\ref{eq:eg4b2}) has the same form as (\ref{eq:dhpotb2}). It has the symmetry group $D_6$. Following the arguments similar to that given in Example~\ref{sec:egone}, we can show that $\V_\rho$ (\ref{eq:eg4b2}) is guaranteed to have the minimum,
\begin{equation}
\rhoo(c_\rho)=r_{\!\rho}(c_\rho)\left(1,0\right)^\T.
\end{equation}
We have the associated groups $\Ht_\rho=Z_{2\rho}$ generated by the group action $\rho\rightarrow\sigz\rho$, and $\Tt_\rho=1$.

The construction of $\Vb$ (\ref{eq:eg4b3}), $\Vt$ (\ref{eq:eg4b4}) and $\V$ (\ref{eq:eg4b5}) is in accordance with the middle route in FIG~\ref{fig:rerep}. The minimum of $\Vb$ is given by
\begin{equation}
\varphiob(\cb)=\big(\frac{r_{\!\phi}(c_\phi)}{\sqrt{3}}\left(1,1,1\right),r_{\!\rho}(c_\rho)\left(1,0\right)\big)^\T.
\end{equation}
The associated groups are $\Hb=\Ht_\phi\times\Ht_\rho=D_{6\phi}\times Z_{2\rho}$, and $\Tb=1$. 

We can show that the $7^\text{th}$ term (\ref{eq:eg4b4}) is not compatible with $\Hb=D_{6\phi}\times Z_{2\rho}$, and hence its addition causes bifurcation. Since $\varphiob(\cb)$ is a single point, we have $\mathcal{M}=\varphiob(\cb)$, and we are guaranteed to obtain a stationary point for $\Vt$ in relation to $\mathcal{M}$. Its HLICs are generated by $\Ht$, the subgroup of $\Hb$ with which the $7^\text{th}$ term is compatible. We have $\Ht=Z_{2\phi}\times Z_{2\rho}$, where $Z_{2\phi}$ is generated by the group action $\phi\rightarrow-\matu_3\,\phi$. We also have $\Tt=1$. With this information, we obtain the guaranteed stationary point of $\Vt$ as
\begin{equation}\label{eq:eg4bmin}
\varphiot(\ct)=\!\big(r_{\!\phi}(\ct)\big(\cos\theta(\ct),\frac{\sin\theta(\ct)}{\sqrt{2}},\frac{\sin\theta(\ct)}{\sqrt{2}}\big),  r_{\!\rho}(\ct)\left(1,0\right)\big)^{\!\T}\!\!,
\end{equation}
where $\theta(\ct)$ is a function of $\ct$ and an arbitrary constant like $r$'s and $k$'s. The HLICs in $\varphiot(\ct)$ correspond to $\text{fix}(Z_{2\phi}\times Z_{2\rho})^\perp$, which is the space spanned by $(\frac{1}{\sqrt{2}}(0,1,-1),(0,0))^\T$ and $((0,0,0),(0,1))^\T$. 

To explicitly show that $\varphiot(\ct)$ (\ref{eq:eg4bmin}) is a minimum, we rewrite $\Vt$ (\ref{eq:eg4b4}) as
\begin{equation}
\begin{split}
\Vt&\!=\!k_1(\ct) (|\phi|^2- r_{\!\phi}(\ct)^2)^2-k_1(\ct)r_{\!\phi}(\ct)^4\\
&\,\,\,\,+\!k_2(\ct) (|\rho|^2\!-\!r_{\!\rho}(\ct)^2)^2\!-\!k_2(\ct)r_{\!\rho}(\ct)^4\!+\!k_3(\ct)|\rho^{2\alpha}\!\!-\!r_{\!\rho}(\ct)\rho|^2\\
&\,\,\,\,+\!k_4(\ct)\big|\phi^{2\alpha}+\frac{r_{\!\phi}(\ct)^2}{r_{\!\rho}(\ct)}(2\cos^2 \theta(\ct)-\sin^2 \theta(\ct))\rho\big|^2.
\end{split}
\end{equation}
This potential contains seven arbitrary constants $r_{\!\phi}(\ct)$, $r_{\!\rho}(\ct)$, $\theta(\ct)$, $k_1(\ct), .., k_4(\ct)$, each of which is a function of the seven coefficients $\ct=\{c_1, .., c_7\}$.

Let us verify if the $8^\text{th}$ term $\phi^{2\alpha\,\T}\rho^{2\alpha}$ is compatible with $\Ht=Z_{2\phi}\times Z_{2\rho}$. The generators of $Z_{2\phi}$ and $Z_{2\rho}$ acting on $\phi$ are $-\matu_3$ and $1$, respectively, and acting on $\rho$ are $1$ and $\sigz$, respectively. Corresponding to $\phi^{2\alpha}$, they are $\sigz$ and $1$, respectively, and corresponding to $\rho^{2\alpha}$, they are $1$ and $\sigz$, respectively. Therefore, the set of representation matrices of $Z_{2\phi}\times Z_{2\rho}$ acting on $\phi^{2\alpha}$ and $\rho^{2\alpha}$ are the same, i.e., $\{1, \sigz\}$.  Therefore, by corollary A, the $8^\text{th}$ term $\phi^{2\alpha\,\T}\rho^{2\alpha}$ is compatible with $\Ht$. The $9^\text{th}$ term is invariant and hence, compatible. As a result, we obtain the minimum of $\V$ (\ref{eq:eg4b5}) as
\begin{equation}\label{eq:bluepoint}
\varphio(c)=\!\big(r_{\!\phi}(c)\big(\cos\theta(c),\frac{\sin\theta(c)}{\sqrt{2}},\frac{\sin\theta(c)}{\sqrt{2}}\big), r_{\!\rho}(c)\left(1,0\right)\big)^{\!\T}\!\!,
\end{equation}
with $\Ht=Z_{2\phi}\times Z_{2\rho}$ and $\Tt=1$.

The alignment $\phi=r_{\!\phi}(c)\big(\cos\theta(c),\frac{\sin\theta(c)}{\sqrt{2}},\frac{\sin\theta(c)}{\sqrt{2}}\big)$ in the minimum (\ref{eq:bluepoint}) corresponds to a point in the blue stratum in FIG~\ref{fig:S41}, i.e., we obtained a minimum in a non-maximal symmetry stratum in FIG~\ref{fig:S41} by using a driving field. Though we do not discuss it here, we state that the introduction of appropriate driving fields leads to minima in the red and grey strata as well. 

\vspace{-3.5mm}
\subsection{A complex representation\\ $\mathbb{G}=S_{4} \times Z_3$}
\label{sec:egfive}
\vspace{-3.5mm}

In this example, we complexify $S_4$ group using $Z_3$ group to construct $S_4\times Z_3$. Let $\varphi=({\varphi}_1,{\varphi}_2,{\varphi}_3)^\T$, where ${\varphi}_1={\varphi}_{1r}+i \varphi_{1i}$, ${\varphi}_2={\varphi}_{2r}+i {\varphi}_{2i}$ and ${\varphi}_3={\varphi}_{3r}+i {\varphi}_{3i}$ are complex numbers, be a triplet transforming under $S_4\times Z_3$. Under the generators of $S_4$, the multiplet $\varphi$ transforms in the same way as in the previous examples, while under the generator of $Z_3$, it transforms as $\varphi\rightarrow \omega \varphi$,
\begin{equation}
\mathbf S\equiv \mathbf s_3, \quad\mathbf T\equiv \mathbf t_3, \quad\mathbf U\equiv -\mathbf u_3, \quad\mathbf Z_3\equiv \om.
\end{equation}
We construct the following potentials using $\varphi$,
\begin{align}
\Vb&=c_1 m^2 |\varphi|^2+ c_2 |\varphi|^4,\label{eq:eq51}\\
\Vt&=\Vb+c_3 |\varphi^\T\varphi|^2+c_4 m \,\text{re}(\varphi^{2\beta\,\T}\varphi)+c_5 m \,\text{im}(\varphi^{2\beta\,\T}\varphi),\label{eq:eq52}\\
\V&=\Vt+ c_6 |\varphi^{2\alpha}|^2,\label{eq:eq53}
\end{align}
where $|\varphi^\T\!\varphi|^2\!=\!(\varphi^\T\!\varphi)^*(\varphi^\T\!\varphi)$, $\varphi^{2\beta}=({\varphi}_2 {\varphi}_3,{\varphi}_3 {\varphi}_1,{\varphi}_1 {\varphi}_2)^\T$, $\varphi^{2\alpha}=(-2|{\varphi}_1|^2+|{\varphi}_2|^2+|{\varphi}_3|^2, \sqrt{3}|{\varphi}_2|^2-\sqrt{3}|{\varphi}_3|^2)^{\T}$, $|\varphi^{2\alpha}|^2=\varphi^{2\alpha\,\T}\varphi^{2\alpha}$, and re and im represent the real part and imaginary part, respectively. $\V$ (\ref{eq:eq53}) is the general renormalizable potential of $\varphi$. We define $\cb=(c_1,c_2)$, $\ct=(c_1,..,c_5)$ and $c=(c_1,..,c_6)$.

The construction of $\Vb$ (\ref{eq:eq51}), $\Vt$ (\ref{eq:eq52}) and $\V$ (\ref{eq:eq53}) follows the middle route in FIG~\ref{fig:irrep}. $\Vb$ is $O(6)$ invariant. For $c_1<0$ and $c_2>0$, we obtain a $5$-sphere with radius $r(\cb)$ as the minima of $\Vb$,
\begin{equation}\label{eq:eg5sphere}
\varphiob(\cb)=5\text{-sphere}.
\end{equation}
We have $\Hb=1$ and $\Tb=O(6)$. The newly added terms (\ref{eq:eq52}) break $\Tb=O(6)$ to the discrete group $\Ga=S_4\times Z_3$. Rather than comprehensively studying the stratification of the $5$-sphere $\varphiob(\cb)$ under $S_4\times Z_3$, let us consider a maximal symmetry stratum that is interesting to us. 

Consider the one-dimensional subspace of $\varphiob(\cb)$ (\ref{eq:eg5sphere}),
\begin{equation}\label{eq:eg5circle}
\frac{r(\cb)}{\sqrt{3}}e^{i\theta}(1,\om,\ob).
\end{equation}
This subspace is nothing but a circle parameterized by $\theta$. Every point in this circle remains invariant under the action $\om \mathbf t_3^2$ since $\om\mathbf t_3^2(1,\om,\ob)=(1,\om,\ob)$. Here, $\om\mathbf t_3^2$ corresponds to the combined action of $\om\in Z_3$ and $\mathbf t_3\equiv \gent$. The element $\om\mathbf t^2_3$ generates a cyclic group, say, $Z_3'$, which is the pointwise stabilizer of the circle. No point exists in $\varphiob(\cb)$ infinitesimally close to the circle whose pointwise stabilizer is $Z_3'$ or a group larger than $Z_3'$. Therefore, the circle (\ref{eq:eg5circle}) forms a connected part of a maximal symmetry stratum in $\varphiob(\cb)$ (\ref{eq:eg5sphere}). 

It can be shown that (\ref{eq:eg5circle}) is one among the 8 circles,
\begin{equation}\label{eq:eg5circles}
\begin{split}
&\frac{r(\cb)}{\sqrt{3}}e^{i\theta}(1,\om,\pm\ob) , \,\,\frac{r(\cb)}{\sqrt{3}}e^{i\theta}(1,-\ob,\pm\om),\\
& \,\,\frac{r(\cb)}{\sqrt{3}}e^{i\theta}(1,\ob,\pm\om), \,\,\frac{r(\cb)}{\sqrt{3}}e^{i\theta}(1,-\ob,\pm\om), 
\end{split}
\end{equation}
which together form our maximal symmetry stratum in $\varphiob(\cb)$. The stabilizers of points in every one of these circles are conjugate to $Z_3'$. For example, points in the circle $\frac{r(\cb)}{\sqrt{3}}e^{i\theta}(1,\om,-\ob)$ have the stabilizer, say, $Z_3''$, generated by the group action $(\mathbf t_3\mathbf s_3\mathbf t_3^2)(\om \mathbf t_3^2)(\mathbf t_3\mathbf s_3\mathbf t_3^2)$. The action of $S_4\times Z_3$ on any point in this maximal symmetry stratum produces an orbit with $|S_4\times Z_3|/|Z_3'|=24$ points, with the 8 circles carrying 3 points each. For example, three points $\frac{r(\cb)}{\sqrt{3}}(1,\om,\ob)$, $\frac{r(\cb)}{\sqrt{3}}\om(1,\om,\ob)$ and $\frac{r(\cb)}{\sqrt{3}}\ob(1,\om,\ob)$ in (\ref{eq:eg5circle}) form a part of a $24$-point orbit. The maximal symmetry stratum (\ref{eq:eg5circles}) contains an infinite set of such $24$-point orbits. By Michel's theorem, we are guaranteed to obtain at least two stationary orbits of the potential $\Vt$ (\ref{eq:eq52}) corresponding to such a maximal symmetry stratum.

Let $\mathcal{M}$ be a point in the circle (\ref{eq:eg5circle}) with $\theta$ assigned to a specific value, say, $\theta=\boldsymbol \theta$,
\begin{equation}
\mathcal{M}=\frac{r(\cb)}{\sqrt{3}}e^{i\boldsymbol{\theta}}(1,\om,\ob).
\end{equation}
$\mathcal{M}$ breaks $\Ga=S_4\times Z_3$ into $\Ha=\Sa=Z_3'$ generated by $\om \mathbf t_3^2$. This leads to $\Ht=\Ha=Z_3'$ and $\Tt=\Sa/\Ha=1$. In relation to $\mathcal M$, we are guaranteed to obtain
\begin{align}\label{eq:eg5min}
\varphiot(\ct)=\frac{r(\ct)}{\sqrt{3}}e^{i\theta(\ct)}(1,\om,\ob),
\end{align}
as the stationary point of $\Vt$. As $(c_3, .., c_5)$ approaches zero, $\theta(\ct)$ approaches $\boldsymbol \theta$. Note that this is a path-dependent limit. To show that $\varphiot(\ct)$ can be obtained as a minimum, we rewrite $\Vt$ (\ref{eq:eq52}) as
\begin{equation}
\begin{split}
\Vt&=k_1(\ct) (|\varphi|^2- r(\ct)^2)^2-k_1(\ct)r(\ct)^4+k_2(\ct)|\varphi^\T\varphi|^2\\
&\quad +k_3(\ct)|\varphi^{2\beta}-\frac{r(\ct)e^{i3\theta(\ct)}}{\sqrt{3}}\varphi^*|^2+\frac{k_3(\ct)}{12}|\varphi^{2\alpha}|^2.
\end{split}
\end{equation}
This potential has $5$ arbitrary constants $r(\ct)$, $\theta(\ct)$, $k_1(\ct)$, $k_2(\ct)$ and $k_3(\ct)$, which are functions of the coefficients $\ct=(c_1, .., c_5)$.

The HLICs in $\varphiot(\ct)$ (\ref{eq:eg5min}) correspond to $\text{fix}(\Ht)^\perp=\text{fix}(Z_3')^\perp$. To explicitly obtain $\text{fix}(Z_3')^\perp$, consider the six-dimensional real space defined in terms of the components $({\phi}_{1r}, \phi_{1i}, {\phi}_{2r}, {\phi}_{2i}, {\phi}_{3r}, {\phi}_{3i})^\T$. In this basis, $\text{fix}(Z_3')^\perp$ is the space spanned by $\frac{1}{\sqrt{3}}(1,0,1,0,1,0)^\T$, $\frac{1}{\sqrt{3}}(0,1,0,1,0,1)^\T$, $\frac{1}{\sqrt{3}}(1,0,-\frac{1}{2},-\frac{\sqrt{3}}{2},-\frac{1}{2},\frac{\sqrt{3}}{2})^\T$ and $\frac{1}{\sqrt{3}}(0,1,\frac{\sqrt{3}}{2},-\frac{1}{2},-\frac{\sqrt{3}}{2},-\frac{1}{2})^\T$. Using complex numbers, $\text{fix}(Z_3')^\perp$ can be conveniently expressed as the space spanned by $\frac{1}{\sqrt{3}}(1,1,1)^\T$, $\frac{i}{\sqrt{3}}(1,1,1)^\T$, $\frac{1}{\sqrt{3}}(1,\ob,\om)^\T$ and $\frac{i}{\sqrt{3}}(1,\ob,\om)^\T$.

We add the $6^\text{th}$ term $|\varphi^{2\alpha}|^2$ to $\Vt$ to obtain $\V$. It can be shown that $\varphi^{2\alpha}$ vanishes at $\varphiot(\ct)$ (\ref{eq:eg5min}). Therefore, by corollary B, the $6^\text{th}$ term is compatible with $\Ht=Z_3'$. Thus we obtain the minimum of $\V$ in relation to $\varphiot(\ct)$ as
\begin{align}\label{eq:eg5min2}
\varphio(c)=\frac{r(c)}{\sqrt{3}}e^{i\theta(c)}(1,\om,\ob).
\end{align}
with $\Ht=Z_3'$, $\Tt=1$ and the HLICs preserved.

\subsection{Introducing complex conjugation\\ $\mathbb{G}=S_4\times Z_3\rtimes Z^c_2$}
\label{sec:egsix}

Let us modify Example~\ref{sec:egfive} by including complex conjugation, i.e., $\varphi\rightarrow\varphi^*$, as a group action. As a result, our symmetry group becomes $\mathbb{G}=S_4\times Z_3\rtimes Z^c_2$, where the generator of $Z^c_2$ is the complex conjugation. In this case, $\text{im}(\varphi^{2\beta\,\T}\varphi)$ is no longer an invariant. Hence, we obtain
\begin{align}
\Vb&=c_1 m^2 |\varphi|^2+ c_2 |\varphi|^4,\label{eq:eg61}\\
\Vt&=\Vb+c_3 |\varphi^\T\varphi|^2+c_4 m\,\text{re}(\varphi^{2\beta\,\T}\varphi),\label{eq:eg62}\\
\V&=\Vt+c_5 |\varphi^{2\alpha}|^2.\label{eq:eg63}
\end{align}
We define $\cb=(c_1,c_2)$, $\ct=(c_1,..,c_4)$ and $c=(c_1,..,c_5)$. 

The construction of $\Vb$ (\ref{eq:eg61}), $\Vt$ (\ref{eq:eg62}) and $\V$ (\ref{eq:eg63}) follows the middle route in FIG~\ref{fig:irrep}, similar to the analysis in Example~\ref{sec:egfive}. The minima of $\Vb$ (\ref{eq:eg61}) is obtained as
\begin{equation}\label{eq:eg6sphere}
\varphiob(\cb)=5\text{-sphere},
\end{equation}
having radius $r(\cb)$ with the associated groups $\Hb=1$ and $\Tb=O(6)$. 

The $O(6)$ symmetry of $\varphiob(\cb)$ is broken to $\Ga=S_4\times Z_3\rtimes Z^c_2$ by the newly added terms (\ref{eq:eg62}), which stratifies $\varphiob(\cb)$. Unlike in Example~\ref{sec:egfive}, the circle $\frac{r(\cb)}{\sqrt{3}}e^{i\theta}(1,\om,\ob)$ is no longer a part of a maximal symmetry stratum. We have six points in the circle, given by $\pm\frac{r(\cb)}{\sqrt{3}}(1,\om,\ob)$, $\pm\frac{r(\cb)}{\sqrt{3}}\om(1,\om,\ob)$ and $\pm\frac{r(\cb)}{\sqrt{3}}\ob(1,\om,\ob)$, that have higher symmetries. Let us assign one among them as $\mathcal M$, 
\begin{equation}\label{eq:eg6M}
\mathcal{M}=\frac{r(\cb)}{\sqrt{3}}(1,\om,\ob). 
\end{equation}
We have the symmetry $-\mathbf u \mathcal{M}^* = \mathcal{M}$, which corresponds to the action of abstract generator $\mathbf U$ together with complex conjugation. Let $Z_2'$ be the group generated by this group action. We also have the symmetry, $\om \mathbf t^2 \mathcal{M}= \mathcal{M}$, which is true for the whole circle (\ref{eq:eg5circle}) as described in Example~\ref{sec:egfive}, where we named the corresponding group $Z_3'$. Therefore, the stabilizer of $\mathcal M$ (\ref{eq:eg6M}) is given by $\Ha=\Sa=Z_3'\rtimes Z_2' = D_6'$. $\mathcal M$ is an example of a point in $\varphiob(\cb)$ (\ref{eq:eg6sphere}) whose every degree of freedom is determined by its stabilizer, i.e., it is a point isolated in its stratum. Therefore, Michel's theorem guarantees the existence of
\begin{equation}\label{eq:eg6min}
\varphiot(\ct)=\frac{r(\ct)}{\sqrt{3}}(1,\om,\ob),
\end{equation}
as the stationary point of $\Vt$ (\ref{eq:eg62}). We have $\Ht=\Ha=D_6'$ and $\Tt=\Sa/\Ha=1$. To obtain $\varphiot(\ct)$ as a minimum, we rewrite $\Vt$ in the form
\begin{equation}
\begin{split}
\Vt&=k_1(\ct) (|\varphi|^2- r(\ct)^2)^2-k_1(\ct)r(\ct)^4+k_2(\ct)|\varphi^\T\varphi|^2\\
&\quad +k_3(\ct)|\varphi^{2\beta}-\frac{r(\ct)}{\sqrt{3}}\varphi^*|^2+\frac{k_3(\ct)}{12}|\varphi^{2\alpha}|^2.
\end{split}
\end{equation}
This potential has $4$ arbitrary constants $r(\ct)$, $k_1(\ct)$, $k_2(\ct)$ and $k_3(\ct)$, which are functions of the four coefficients $\ct$.

The HLICs in $\varphiot(\ct)$ (\ref{eq:eg6min}) correspond to $\text{fix}(D_6')^\perp$ which is the space spanned by $\frac{1}{\sqrt{3}}(1,1,1)^\T$, $ \frac{i}{\sqrt{3}}(1,1,1)^\T$, $\frac{1}{\sqrt{3}}(1,\ob,\om)^\T$, $\frac{i}{\sqrt{3}}(1,\ob,\om)^\T$ and $\frac{i}{\sqrt{3}}(1,\om, \ob)^\T$, where we have conveniently used complex numbers to represent the six-component real vectors as we did in Example~\ref{sec:egfive}. There are five independent HLICs in (\ref{eq:eg6min}) as opposed to four in (\ref{eq:eg5min}) because the symmetries determine every degree of freedom, including the overall phase in (\ref{eq:eg6min}). Even though we have used real complex numbers in our analysis, real and complex representations are distinct from the point of view of the representation theory. The multiplet $\varphi$ appearing in Example~\ref{sec:egfive} is a three-dimensional complex representation of $S_4\times Z_3$ while that appearing here in Example~\ref{sec:egsix} is a six-dimensional real representation of $S_4\times Z_3\rtimes Z^c_2$.

The multiplet $\varphi^{2\alpha}$ vanishes at $\varphiot(\ct)$ (\ref{eq:eg6min}). This implies that, by corollary B, the term $|\varphi^{2\alpha}|^2$ is compatible with $\Ht=D_6'$. Therefore, we obtain 
\begin{equation}\label{eq:eg6min2}
\varphio(c)=\frac{r(c)}{\sqrt{3}}(1,\om,\ob),
\end{equation}
as the minimum of $\V$ (\ref{eq:eg63}) with $\Ht=D_6'$, $\Tt=1$ and the HLICs preserved.

\subsection{Unbroken accidental continuous symmetry\\ $\mathbb{G}=S_4\times Z_2\times Z_3\rtimes Z^c_2$}
\label{sec:egseven}

As in Example~\ref{sec:egsix}, let us use a triplet, $\varphi=({\varphi}_1,{\varphi}_2,{\varphi}_3)^\T$, transforming under $S_4$, $Z_3$ and $Z^c_2$. Besides these, let us introduce another $Z_2$ group generated by the action $\varphi\rightarrow -\varphi$. As a result, the term $\text{re}(\varphi^{2\beta\,\T}\varphi)$ is no longer an invariant. This results in the potentials,
\begin{align}
\Vb&=c_1 m^2 |\varphi|^2+ c_2 |\varphi|^4,\label{eq:eg71}\\
\V&=\Vb+c_3 |\varphi^\T\varphi|^2+ c_4 |\varphi^{2\alpha}|^2,\label{eq:eg72}
\end{align}
where $\V$ is the general renormalizable potential constructed with $\varphi$. We can see that $\V$ is invariant under the $U(1)$ transformation $\varphi\rightarrow e^{i\theta}\varphi$,  i.e., the symmetry of the potential got accidentally enlarged from $\mathbb{G}=S_4\times Z_2\times Z_3\rtimes Z^c_2$ to $G=S_4\times U(1) \rtimes Z^c_2=S_4\times O(2)$. For our analysis, we define $\cb=(c_1,c_2)$ and $c=(c_1,..,c_4)$. 

The construction of $\Vb$ (\ref{eq:eg71}) and $\V$ (\ref{eq:eg72}) follows the left route in FIG.~\ref{fig:irrep}. We obtain the minima of $\Vb$ as 
\begin{equation}\label{eq:eg7sphere}
\varphiob(\cb)=5\text{-sphere},
\end{equation}
having radius $r(\cb)$ with $\Hb=1$ and $\Tb=O(6)$. The added terms (\ref{eq:eg72}) break $\Tb=O(6)$ to $\Ga=S_4\times O(2)$. It is straightforward to see that eight circles similar to (\ref{eq:eg5circles}) form a single orbit in $\varphiob(\cb)$ (\ref{eq:eg7sphere}) under $\Ga=S_4\times O(2)$. Each circle forms a connected part (one-dimensional manifold) of the orbit. Let us consider one of them, say, 
\begin{align}\label{eq:eg7circle}
\mathcal{M}=\frac{r(\cb)}{\sqrt{3}}e^{i\theta}(1,\om,\ob).
\end{align}
Every point in it remains invariant under the action of $\om \matt_3^2$. The element $\om\mathbf t^2$ generates the cyclic group $Z_3'$, which forms the pointwise stabilizer of $\mathcal M$, i.e., $\Ha=Z_3'$. Every point in $\mathcal M$ moves within $\mathcal M$ under multiplication with a complex phase, i.e., under the action of the $U(1)$ group. Also, every point in $\mathcal M$ moves within $\mathcal M$ under the combined action of $-\mathbf u$ and complex conjugation $^*$. We call the group generated by this combined action $Z_2'$. The setwise stabilizer of $\mathcal M$ is generated by $Z_3'$, $U(1)$ and $Z_2'$, i.e., $\Sa=Z_3'\times U(1)\rtimes Z_2'$. By taking the quotient of $\Sa$ with $\Ha$, we obtain $\Tt=\Sa/\Ha=U(1)\rtimes Z_2'=O(2)'$. As we stated earlier, the pointwise stabilizer of $\mathcal M$ is $\Ha=Z_3'$. Since no point with the stabilizer $Z_3'$ exists infinitesimally close to $\mathcal{M}$, it is isolated in its stratum. Therefore, in relation to $\mathcal M$, we are guaranteed to obtain
\begin{align}\label{eq:eg7min}
\varphio(c)(\theta)=\frac{r(c)}{\sqrt{3}}e^{i\theta}(1,\om,\ob),
\end{align}
as the stationary manifold of $\V$ (\ref{eq:eg72}) with $\Ht=Z_3'$ and $\Tt=O(2)'$. 

To show that $\varphio(c)$ (\ref{eq:eg7min}) can be obtained as a minimum, we rewrite $\V$ (\ref{eq:eg72}) as
\begin{equation}
\begin{split}
\V&=k_1(c) (|\varphi|^2- r(c)^2)^2-k_1(c)r(c)^4\\
&\quad +k_2(c)|\varphi^\T\varphi|^2+k_3(c)|\varphi^{2\alpha}|^2.
\end{split}
\end{equation}
The four arbitrary constants $r(c)$, $k_1(c)$, $k_2(c)$ and $k_3(c)$ are functions of the four coefficients $c$. 

The HLICs in $\varphio(c)$ correspond to $\text{fix}(\Ht)^\perp=\text{fix}(Z_3')^\perp$, which is space spanned by $\frac{1}{\sqrt{3}}(1,1,1)^\T$, $\frac{i}{\sqrt{3}}(1,1,1)^\T$, $\frac{1}{\sqrt{3}}(1,\ob,\om)^\T$ and $\frac{i}{\sqrt{3}}(1,\ob,\om)^\T$. We urge the reader to note the differences among the points of minimum (\ref{eq:eg5min2}), (\ref{eq:eg6min2}) and the manifold of minima (\ref{eq:eg7min}). 

\subsection{Auxiliary symmetries and snowflakes}
\label{sec:egeight}

Let $\phia=({\phia}_1, {\phia}_2, {\phia}_3)^\T$, $\phib=({\phib}_1, {\phib}_2, {\phib}_3)^\T$ and $\phic=({\phic}_1, {\phic}_2, {\phic}_3)^\T$ be three complex triplets that transform as given in TABLE~\ref{tab:flavourcontent}. Our objective is to utilize these triplets to construct an effective doublet, say, $\phi$, that has $D_{12}$ symmetry as the doublet given in Example~\ref{sec:egtwo}, i.e., $\phi$ transforms as
\begin{equation}\label{eq:efftrof}
\begin{split}
&\qquad\quad\,\,\,\gent\equiv\matw_r,\quad\genu\equiv\sigz,\quad\genz\equiv-1,\\
&\genxs\equiv1,\,\,\,\,\genxd\equiv1,\,\,\,\,\genxa\equiv1,\,\,\,\,\genxb\equiv1,\,\,\,\,\genxc\equiv1.
\end{split}
\end{equation}
By inspecting TABLE~\ref{tab:flavourcontent}, we can infer that no quadratic combination of the triplets exists that results in the required transformations (\ref{eq:efftrof}). In the cubic order, the only possible combination is $\phia\otimes\phib\otimes\phic$. From $\phia\otimes\phib\otimes\phic$ we construct
\begin{equation}\label{eq:effflav}
\begin{split}
\phi&=\text{re}\big(\!-2{\phia}_1{\phib}_1{\phic}_1\!+{\phia}_2{\phib}_2{\phic}_2\!+{\phia}_3{\phib}_3{\phic}_3,\\
 &\qquad\qquad\qquad\quad\,\,\sqrt{3}{\phia}_2{\phib}_2{\phic}_2\!-\sqrt{3}{\phia}_3{\phib}_3{\phic}_3\big)^{\!\T}\!,\!\!\!
\end{split}
\end{equation}
which is the one and only leading order effective irrep that transforms as (\ref{eq:efftrof}).

{\renewcommand{\arraystretch}{1.2}
	\setlength{\tabcolsep}{5pt}
\begin{table}[]
	\begin{center}
	\begin{tabular}{|c|c c c c c c c c|}
\hline
	&$\gent$ & $\genu$	&$\genz$&$\genxs$&$\genxd$&$\genxa$&$\genxb$&$\genxc$ \\
\hline
$\phia$&$\matt_3$ & $-\matu_3$	&$-1$&$\matt_3^2\mats_3\matt_3$&$-1$&$\om$&$1$&$()^*$\\
$\phib$&$\matt_3$ & $-\matu_3$	&$1$&$\matt_3\mats_3\matt_3^2$&$-1$&$1$&$\om$&$()^*$\\
$\phic$&$\matt_3$ & $-\matu_3$	&$1$&$\mats_3$&$1$&$\ob$&$\ob$&$()^*$\\
\hline
$\etaa$&$1$ & $1$ &$-1$ &$1$&$-1$&$1$&$1$&$1$\\
$\etab$&$1$ & $1$ &$1$ &$1$&$-1$&$1$&$1$&$1$\\
\hline
	\end{tabular}
	\end{center}
	\caption{The transformation rules for the triplets $\phia$, $\phib$ and $\phic$, and the driving fields $\etaa$ and $\etab$.}
	\label{tab:flavourcontent}
\end{table}}

Before we construct the general renormalizable potential using our triplets, let us consider the following alignments,
\begin{equation}\label{eq:align}
\phia\propto\left(1,1,1\right)^{\T}\!\!, \,\,\,\phib\propto\left(-1,-1,1\right)^{\T}\!\!, \,\,\,\phic\propto\left(1,\om,\ob\right)^{\T}.
\end{equation}
We can show that each of these alignments is fully determined (up to its norm) by its stabilizer. Therefore, by symmetry arguments, each of these is guaranteed to be a stationary point of the corresponding potential (under the assumption that the potential does not possess accidental continuous symmetries). Let us verify if the cross terms among the triplets may spoil these stationary points. We can show that only six renormalizable cross terms exist, given by $\phia^{2\alpha\,\T}\phib^{2\alpha}$, $\phia^{2\alpha\,\T}\phic^{2\alpha}$, $\phib^{2\alpha\,\T}\phic^{2\alpha}$, $|\phia|^2|\phib|^2$, $|\phia|^2|\phic|^2$ and $|\phib|^2|\phic|^2$ where $\phia^{2\alpha}\!\!=\!(\!-2|{\phia}_1|^2\!+|{\phia}_2|^2\!+|{\phia}_3|^2\!,\! \sqrt{3}|{\phia}_2|^2\!-\sqrt{3}|{\phia}_3|^2)^{\T}$, and $\phib^{2\alpha}$ and $\phic^{2\alpha}$ defined in a similar manner. At (\ref{eq:align}), $\phia^{2\alpha}$, $\phib^{2\alpha}$ and $\phic^{2\alpha}$ vanish. Therefore, by corollary B, the first three cross terms are compatible with (\ref{eq:align}). The other three cross terms are products of squares of norms which are also compatible. Thus, we can conclude that the cross terms do not spoil our stationary points (\ref{eq:align}).

On further analysis, we can show that the renormalizable potentials of $\phia$ and $\phib$ contain accidental continuous symmetries. To break them, we introduce the driving fields $\etaa$ and $\etab$, respectively. They are singlets, and their transformation rules are provided in TABLE~\ref{tab:flavourcontent}. Let us define the following combined multiplets: $\varphia=(\phia,\etaa)^\T$, $\varphib=(\phib,\etab)^\T$ and $\varphi=(\varphia, \varphib, \phic)^\T$. With this field content, we construct the various potentials as
\begin{align}
\begin{split}
\V_{\varphia}&=c_{1} m^2 |\phia|^2+ c_{2} |\phia|^4+c_{3}m^2\etaa^2+c_{4}\etaa^4\\
&\quad\,\,+c_{5}|\phia^\T\phia|^2+c_{6} |\phia^{2\beta}|^2+c_{7} m\, \text{re}(\phia^{2\beta\,\T}\phia)\etaa,
\end{split}\\
\begin{split}
\V_{\varphib}&=c_{8} m^2 |\phib|^2+ c_{9} |\phib|^4+c_{10}m^2\etab^2+c_{11}\etab^4\\
&\quad\,\,+c_{12}|\phib^\T\phib|^2+c_{13} |\phib^{2\beta}|^2+c_{14} m\, \text{re}(\phib^{2\beta\,\T}\phib)\etab,
\end{split}\\
\begin{split}
\V_{\phic}&=c_{15} m^2 |\phic|^2+ c_{16} |\phic|^4+c_{17} |\phic^\T\phic|^2\\
&\qquad\qquad+c_{18} |\phic^{2\beta}|^2+c_{19} m\, \text{re}(\phic^{2\beta\,\T}\phic),
\end{split}\\
\Vt&=\V_{\varphia}+\V_{\varphib}+\V_{\phic},\label{eq:eg8sum}
\end{align}
\begin{align}
\begin{split}\notag
\V&=\Vt+c_{20}\phia^{2\alpha\,\T}\phib^{2\alpha}+c_{21}\phia^{2\alpha\,\T}\phic^{2\alpha}+c_{22}\phib^{2\alpha\,\T}\phic^{2\alpha}\\
&\quad\,\,+c_{23}|\phia|^2|\phib|^2+c_{24}|\phia|^2|\phic|^2+c_{25}|\phib|^2|\phic|^2
\end{split}\\
\begin{split}\label{eq:eg8gen}
&\quad\,\,+c_{26}\etaa^2\etab^2\!+c_{27}|\phia|^2\etaa^2\!+c_{28}|\phib|^2\etaa^2\!+c_{29}|\phic|^2\etaa^2\\
&\quad\,\,+c_{30}|\phia|^2\etab^2+c_{31}|\phib|^2\etab^2+c_{32}|\phic|^2\etab^2,
\end{split}
\end{align}
where $\phia^{2\beta}\!=\!({\phia}_2 {\phia}_3,{\phia}_3 {\phia}_1,{\phia}_1 {\phia}_2)^\T$\!\!, $|\phia^{2\beta}|^2\!=\!\phia^{2\beta\,\dagger}\phia^{2\beta}$, $|\phia^{2\alpha}|^2\!=\!\phia^{2\alpha\,\T}\phia^{2\alpha}$, and with similar definitions for the objects constructed with $\phib$ and $\phic$ as well. $\V$ (\ref{eq:eg8gen}) is the general renormalizable potential constructed with $\phia$, $\phib$, $\phic$, $\etaa$ and $\etab$.

Using the arguments developed in this paper, we assign
\begin{align}
\varphiao(c_{\varphia})&=\big(\frac{r_{\!\phia}(c_{\varphia})}{\sqrt{3}}\left(1,1,1\right),r_{\!\etaa}(c_{\varphia})\big)^{\!\T}\!\!,\\
\varphibo(c_{\varphib})&=\big(\frac{r_{\!\phib}(c_{\varphib})}{\sqrt{3}}\left(-1,-1,1\right),r_{\!\etab}(c_{\varphib})\big)^{\!\T}\!\!,\\
\phico(c_{\phic})&=\frac{r_{\!\phic}(c_{\phic})}{\sqrt{3}}\left(1,\om,\ob\right)^{\T}\!\!,
\end{align}
as the guaranteed stationary points of $\V_{\varphia}$, $\V_{\varphib}$ and $\V_{\phic}$, respectively. Corresponding to $\varphiao(c_{\varphia})$, $\varphibo(c_{\varphib})$ and $\phico(c_{\phic})$, we obtain the symmetry groups $\Ht_{\varphia}=(D_{6}\times Z^c_{2})_{\phia}$ generated by $\matt_3$, $-\matu_3$ and $()^*$ acting on $\phia$, $\Ht_{\varphib}=(D_{6}\times Z^c_{2})_{\phib}$ generated by $(\matt_3\mats_3\matt_3^2)(\matt_3)(\matt_3\mats_3\matt_3^2)$, $(\matt_3\mats_3\matt_3^2)(-\matu_3)(\matt_3\mats_3\matt_3^2)$ and $()^*$ acting on $\phib$, and $\Ht_{\phic}=D_{6\phic}$ generated by $\om\matt_3^2$ and $-\matu_3()^*$ acting on $\phic$, respectively. The corresponding HLICs are given by $\text{fix}(\Ht_{\varphia})$ which is the space spanned by $\frac{1}{\sqrt{3}}(1,\om,\ob)^\T$, $\frac{i}{\sqrt{3}}(1,\om,\ob)^\T$, $\frac{1}{\sqrt{3}}(1,\ob,\om)^\T$, $\frac{i}{\sqrt{3}}(1,\ob,\om)^\T$ and $\frac{i}{\sqrt{3}}(1,1,1)^\T$ in relation to $\phia$, $\text{fix}(\Ht_{\varphib})$ which is the space spanned by $\frac{1}{\sqrt{3}}(1,\om,-\ob)^\T$, $\frac{i}{\sqrt{3}}(1,\om,-\ob)^\T$, $\frac{1}{\sqrt{3}}(1,\ob,-\om)^\T$, $\frac{i}{\sqrt{3}}(1,\ob,-\om)^\T$ and $\frac{i}{\sqrt{3}}(1,1,-1)^\T$ in relation to $\phib$, and $\text{fix}(\Ht_{\phic})$ which is the space spanned by $\frac{1}{\sqrt{3}}(1,1,1)^\T$, $ \frac{i}{\sqrt{3}}(1,1,1)^\T$, $\frac{1}{\sqrt{3}}(1,\ob,\om)^\T$, $\frac{i}{\sqrt{3}}(1,\ob,\om)^\T$ and $\frac{i}{\sqrt{3}}(1,\om, \ob)^\T$ in relation to $\phic$. The stationary points $\varphiao(c_{\varphia})$, $\varphibo(c_{\varphib})$ and $\phico(c_{\phic})$ can be obtained as minima by rewriting the potentials, as we demonstrated in the previous examples.

The construction of $\Vt$ (\ref{eq:eg8sum}) and $\V$ (\ref{eq:eg8gen}) is in accordance with the right route in FIG~\ref{fig:rerep}. The resulting minimum of $\Vt$ is
\begin{equation}
\varphiot(\ct)=\big(\varphiao(c_{\varphia}), \varphibo(c_{\varphib}), \phic(c_{\phic})\big)^\T.
\end{equation}
The HLICs in $\varphiot(\ct)$ correspond to $\text{fix}(\Ht)$, where $\Ht=\Ht_{\varphia}\times\Ht_{\varphib}\times\Ht_{\phic}=(D_{6}\times Z^c_{2})_{\phia}\times (D_{6}\times Z^c_{2})_{\phib}\times D_{6\phic}$.

\begin{figure}[]
\subfloat[\label{fig:hexa}]{%
  \includegraphics[width=0.33\columnwidth]{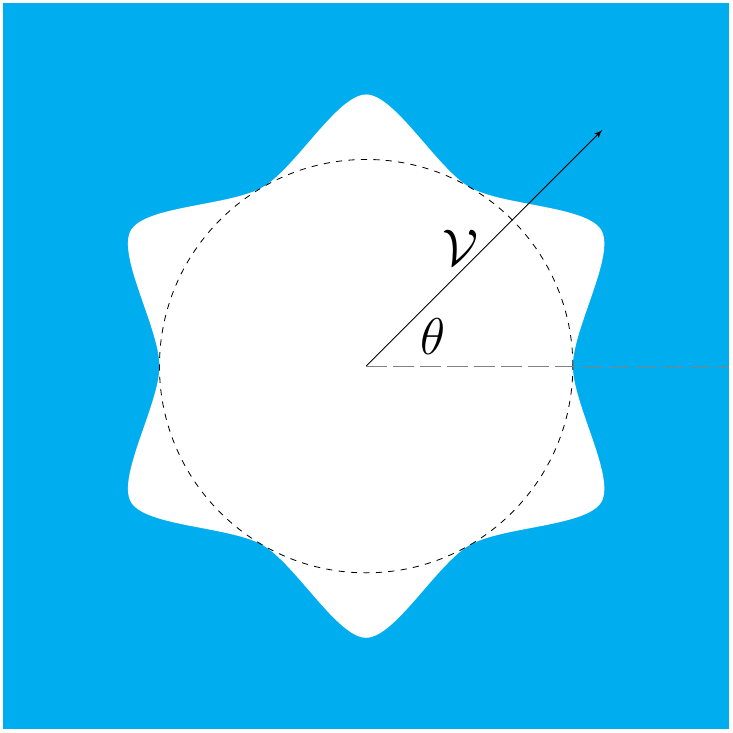}%
}
\subfloat[\label{fig:hexb}]{%
  \includegraphics[width=0.33\columnwidth]{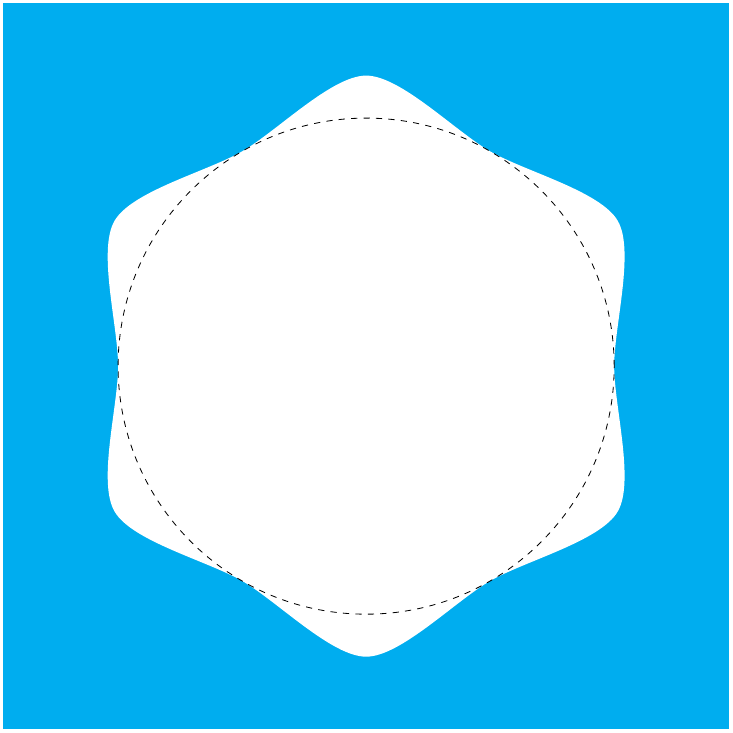}%
}
\subfloat[\label{fig:hexc}]{%
  \includegraphics[width=0.33\columnwidth]{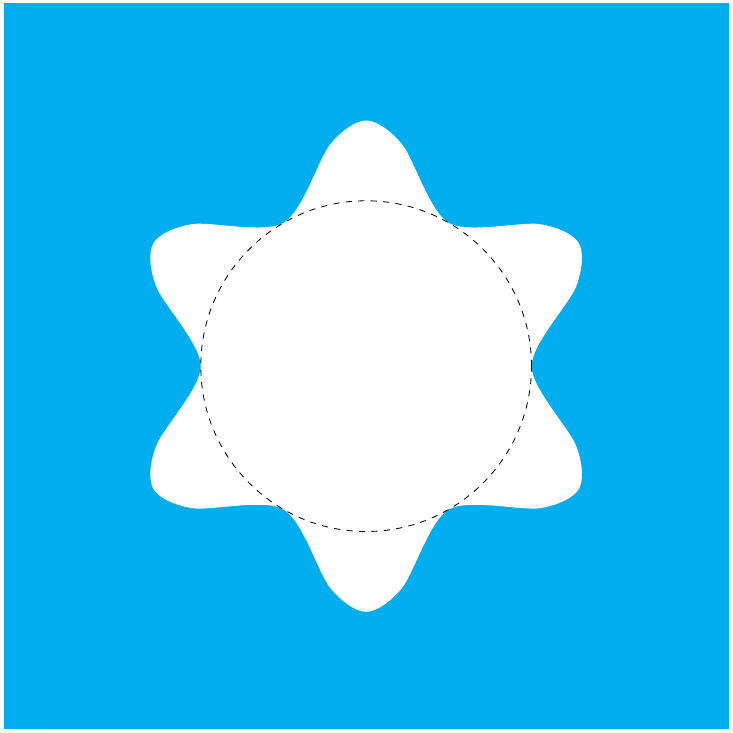}%
}\\
\subfloat[\label{fig:hexd}]{%
  \includegraphics[width=0.33\columnwidth]{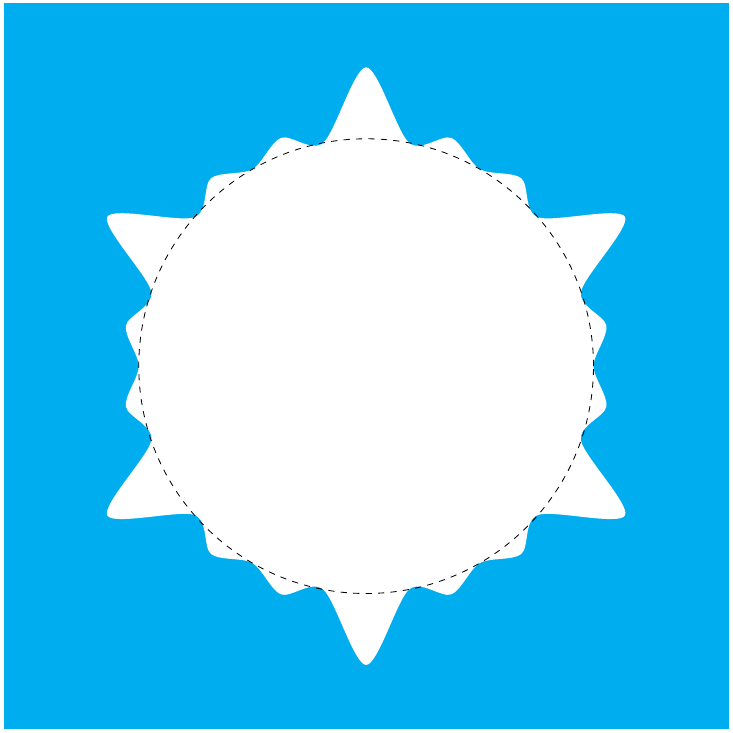}%
}
\subfloat[\label{fig:hexe}]{%
  \includegraphics[width=0.33\columnwidth]{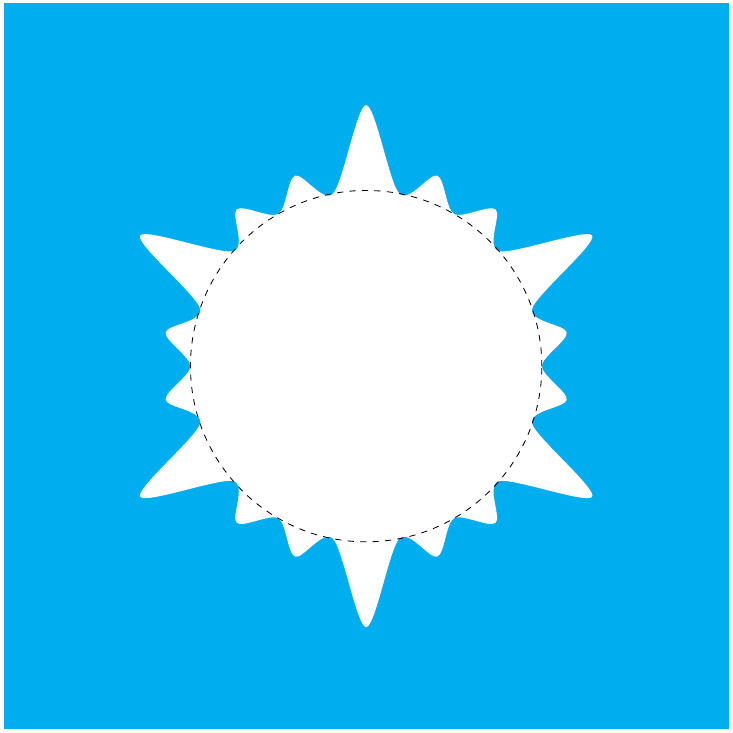}%
}
\subfloat[\label{fig:hexf}]{%
  \includegraphics[width=0.33\columnwidth]{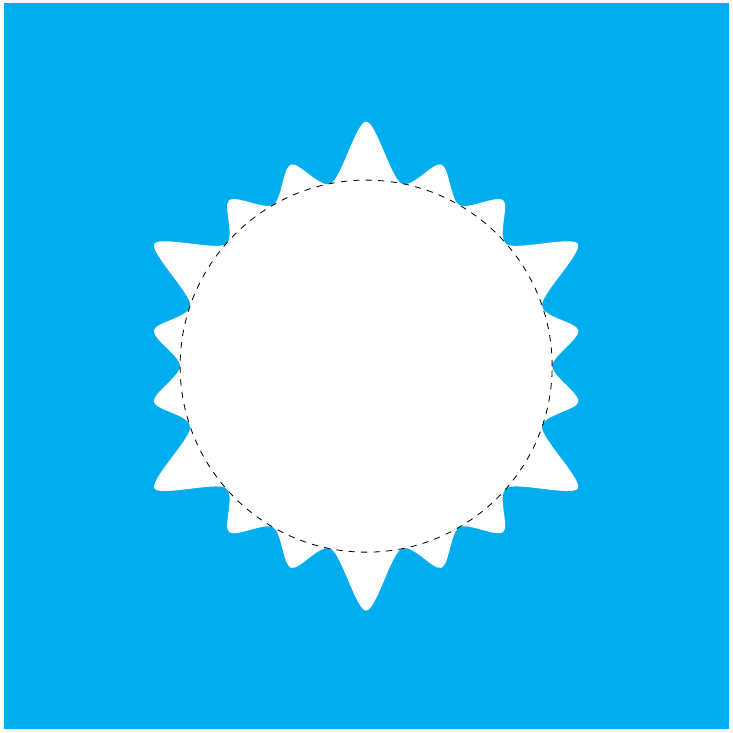}%
}
\caption{The top and bottom panels correspond to $\V$ constructed in Example~\ref{sec:egtwo} and Example~\ref{sec:egeight}, respectively. The radial distance denotes $\V$ and the angle $\theta$ denotes the angle between the components of $\phi$, i.e., $\theta=\tan^{-1}(\phi_2/\phi_1)$. The region where the potential exists is shaded in blue, so the minima correspond to the shortest radial distance (indicated by the dotted circle). (a), (b) and (c) are obtained with three random values for the set of six coefficients $(c_1,..,c_6)$ in (\ref{eq:eg25}). These figures have minima at $\theta=n\frac{\pi}{3}$ independently of the values of the coefficients, as expected. (d), (e) and (f) are obtained with three random values for the set of $32$ coefficients $(c_1,..,c_{32})$ in (\ref{eq:eg8gen}). Besides at $\theta=n\frac{\pi}{3}$, we have minima also at $\theta=n\frac{\pi}{3}\pm\tan^{-1}(\sqrt{3}/2)$, independently of the coefficients.}\label{fig:contrast}
\end{figure}

The potential $\V$ (\ref{eq:eg8gen}) is obtained by adding the terms numbering $20$ to $32$ to $\Vt$. Recalling our earlier analysis, we can see that all these terms are compatible with $\Ht$. This implies that $\V$ is guaranteed to have the minimum,
\begin{equation}\label{eq:finalmin}
\begin{split}
\varphio(c)&=\big(\frac{r_{\!\phia}(c)}{\sqrt{3}}\left(1,1,1\right),r_{\!\etaa}(c),\frac{r_{\!\phib}(c)}{\sqrt{3}}\left(-1,-1,1\right),r_{\!\etab}(c),\\
&\qquad\quad\frac{r_{\!\phic}(c)}{\sqrt{3}}\left(1,\om,\ob\right)\big)^{\!\T}\!\!,
\end{split}
\end{equation}
with $\Ht=(D_{6}\times Z^c_{2})_{\phia}\times (D_{6}\times Z^c_{2})_{\phib}\times D_{6\phic}$, and the HLICs preserved.

We can show that, even after we expanded the field content by including the driving fields $\etaa$ and $\etab$, the effective irrep $\phi$ as constructed in (\ref{eq:effflav}) remains the one and only leading order object that transforms as (\ref{eq:efftrof}). We substitute $\varphio(c)$ (\ref{eq:finalmin}) in (\ref{eq:effflav}) to obtain the minimum of the effective irrep $\phi$,
\begin{equation}\label{eq:effmin}
\phio(c)=r_{\!\phia}(c)\, r_{\!\phib}(c)\, r_{\!\phic}(c)\frac{1}{3\sqrt{3}}(2,\sqrt{3})^\T.
\end{equation}
This minimum contains one HLIC, namely, the ratio of its components being equal to $\frac{2}{\sqrt{3}}$, i.e., $\sqrt{3}\phio(c)_1=2\phio(c)_2$. This ratio remains independent of any general variation of the $32$ coefficients in the potential $\V$ (\ref{eq:eg8gen}). 

FIG.~\ref{fig:contrast} contrasts the minima of the potential obtained in Example~\ref{sec:egtwo} with the minima obtained here in Example~\ref{sec:egeight}. The irrep $\phi$ appearing in Example~\ref{sec:egtwo} as well as in Example~\ref{sec:egeight} transforms as the doublet of $D_{12}$. However, $\phi$ is an {it elementary} irrep in Example~\ref{sec:egtwo}, while it is constructed as an {\it effective} irrep using $\phia$, $\phib$ and $\phic$ in Example~\ref{sec:egeight}. These three multiplets transform not only under the generators of $D_{12}$ but also under $\genxs$, $\genxd$, $\genxa$, $\genxb$ and $\genxc$, TABLE~\ref{tab:flavourcontent}. We call them auxiliary generators. Even though the effective irrep $\phi$ remains invariant under the auxiliary generators, its alignments corresponding to the minima of $\V$ can contain HLICs that are determined by these generators, as is evident from FIG.~\ref{fig:contrast}. 

This example shows that a system that exhibits a set of discrete symmetries externally can also possess auxiliary symmetries, which relate to how the constituents of the system interact internally. Auxiliary symmetries are hidden in the sense that the external interactions do not possess these symmetries. However, the lowest energy states of the system, as observed externally, are determined not only by the external symmetries but also by the auxiliary symmetries. A rough analogy will be like observing hexagonal snowflakes that consistently possess features as shown in the lower panel of FIG.~\ref{fig:contrast}. The author is not aware of the existence of such physical systems, but this paper shows that we can construct them mathematically. In the companion paper\cite{2309.11542}, we argue that this mathematical framework could be underlying the flavour symmetries in particle physics. The fermions are assumed to transform under a set of discrete symmetries. We introduce several elementary flavons (irreps) which transform under them as well as under auxiliary symmetries. We construct effective flavons (irreps) that are invariants of the auxiliary symmetries and hence that couple with the fermions. The vacuum alignments of these effective flavons, which correspond to a minimum of the potential, are determined not only by the symmetries under which they interact with the fermions but also by the auxiliary symmetries. We utilize this framework in \cite{2309.11542} to construct the fermion mass matrices and obtain neutrino oscillation parameters consistent with the current experimental data.

$\gent$, $\genu$, $\genz$ and the auxiliary generators, TABLE~\ref{tab:flavourcontent}, generate the group $\mathbb G$. The renormalizable potential (\ref{eq:eg8gen}) possesses accidental symmetries, and hence $\mathbb G$ gets enlarged to $G$. Let $H$ be the stabilizer of $\varphio(c)$ (\ref{eq:finalmin}) under $G$. In our analysis, we determined neither $G$ nor $H$. Rather, we obtained $\Ht=(D_{6}\times Z^c_{2})_{\phia}\times (D_{6}\times Z^c_{2})_{\phib}\times D_{6\phic}$ and the resulting HLICs in $\varphio(c)$. It can be shown that $\text{fix}(H)^\perp=\text{fix}(\Ht)^\perp$ even though $H\subset \Ht$. In other words, we can obtain all the HLICs in $\varphio(c)$ (\ref{eq:finalmin}) from $H$ itself. This is true for every example provided in this paper. Please see \cite{2011.11653} for an example where $\text{fix}(H)^\perp$ is a subspace of $\text{fix}(\Ht)^\perp$, i.e.,~where we need to utilize $\Ht$ to obtain all the HLICs. We argue that even in cases where $\text{fix}(H)^\perp=\text{fix}(\Ht)^\perp$, our procedure of utilizing $\Ht$ is relevant as it lets us bypass analyzing the accidental symmetries and determining $H$. Obtaining $\Ht$ is rather straightforward since it is the direct product of the stabilizers corresponding to the individual irreps.

\bibliography{first_paper.bib, noninspire.bib}

\end{document}